\batchmode
\makeatletter
\def\input@path{{C:/Repozytoria/aldoktorat/}}
\makeatother
\documentclass[polish,b5paper,9pt,openany]{book}
\usepackage{mathpazo}
\usepackage[T1]{fontenc}
\usepackage[latin2]{inputenc}
\usepackage{mathrsfs}
\usepackage{amsmath}
\usepackage{amssymb}
\usepackage{graphicx}
\usepackage{esint}

\makeatletter
\@ifundefined{date}{}{\date{}}
\usepackage{geometry}
\usepackage{layouts}
\usepackage{dsfont}
\usepackage{cite}
\date{}
\usepackage{fancyhdr}
\usepackage{fancyhdr}
\makeatother

\usepackage{babel}
\begin{document}
\begin{titlepage}
	\setcounter{page}{0}
	\date{}
	\newcommand{\HRule}{\rule{\linewidth}{0.5mm}}
	\center
	\includegraphics[width=0.6\textwidth]{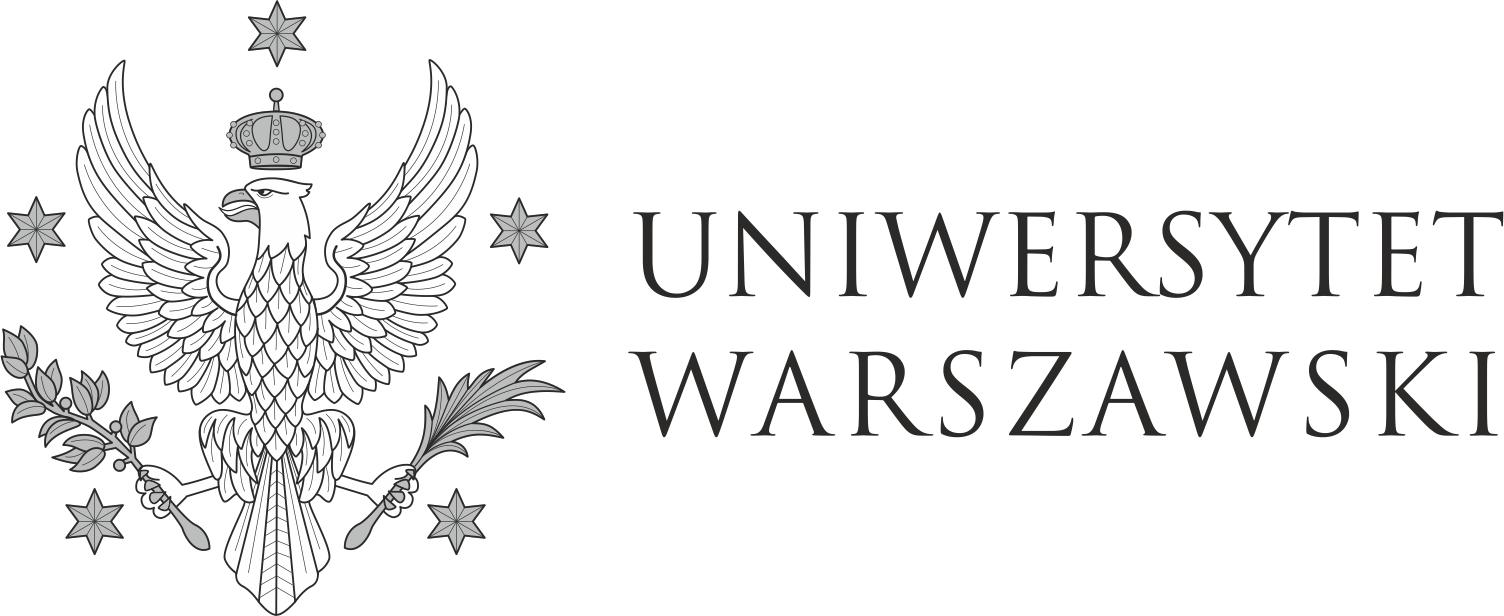}\\[1.5cm]
	\textsc{\Large Rozprawa doktorska}\\[0.5cm]
	\HRule\\[0.4cm] 
	{\huge\bfseries Przestrzenna modulacja fazy jako nastawny mechanizm transferu informacji między światłem, a pamięcią kwantową}\\[0.4cm] 
	\HRule\\[1.5cm] 
	\begin{minipage}{0.4\textwidth}\begin{flushleft}\large\textit{Autor}\\Adam Leszczyński\end{flushleft}\end{minipage}
	\begin{minipage}{0.4\textwidth}\begin{flushright}\large\textit{Promotor}\\dr hab. Wojciech Wasilewski\end{flushright}\end{minipage} 
	\vfill\vfill\vfill\vfill 
	Wydział Fizyki\\
	Uniwersytet Warszawski\\
	Warszawa 2021
\end{titlepage}

\begingroup   \pagestyle{empty}  \null   \newpage \endgroup
\sloppy
\begin{center}
\textbf{STRESZCZENIE ROZPRAWY DOKTORSKIEJ}\\
\textbf{Dopasowanie fazowe jako nastawny mechanizm transferu informacji między światłem, a pamięcią kwantową}
\end{center}
\addcontentsline{toc}{chapter}{Streszczenie}

Niniejsza rozprawa doktorska skupia się na zagadnieniu dopasowania
fazowego w pamięci kwantowej działającej na zimnej chmurze atomów
$^{87}$Rb, gdzie jako interfejs między światłem, a atomami wykorzystano
nierezonansowe rozpraszanie Ramana. Przedstawione zostały wyniki eksperymentów
oraz rozważania na temat sterowania procesem odczytu z pamięci poprzez
wykorzystanie efektu Zeemana, ac-Starka, kontrolę geometrii wiązek
laserowych, czy też wykorzystanie pierścieniowej wnęki optycznej.

W rozdziale 2 wprowadzona jest teoria oddziaływania światła z atomami
w trójpoziomowym układzie $\Lambda$. Analizowana jest wpływ czynników
takich jak natężenia wiązek, ich odstrojenia, geometria, czy też gęstość
optyczna, na szybkość i wydajność odczytu, a także straty związane
z dekoherencją, czy absorpcją światła.

Rozdział 3 prezentuje generację fikcyjnego pola magnetycznego za pomocą
efektu ac-Starka z rozdzielczością przestrzenną. Za jego pomocą modulowana
była przestrzenna faza precesji spinów oscylujących w zewnętrznym
polu magnetycznym. Uzyskane w eksperymencie wyniki pokazują, że kontrola
przestrzennej fazy umożliwia niejako włączanie lub wyłączanie odczytu
z pamięci kwantowej.

Rozdział 4 przedstawia eksperymentalną modulację przestrzennej fazy
fal spinowych w pamięci kwantowej przy wykorzystaniu efektu ac-Starka.
Zaprezentowana została możliwość kompensacji bezpośrednio na falach
spinowych dowolnych aberracji układu obrazującego. Za pomocą pomiarów
interferencyjnych na kamerze bliskiego pola oraz bezpośrednich pomiarów
na kamerze dalekiego pola została również dokładnie scharakteryzowana
praca przestrzennego modulatora fazy fal spinowych.

Rozdział 5 pokazuje połączenie pamięci typu GEM z przestrzenną modulacją
fazy. Zaprezentowano eksperymentalną realizację spektrometru o bardzo
wysokiej rozdzielczości (20 kHz \textasciitilde{} 83 peV \textasciitilde{}
$6\times10^{-7}$cm$^{-1}$), dostosowanego do wąskopasmowej emisji
atomowej. Przeanalizowana jest również zależność pomiędzy rozdzielczością,
szerokością pasma oraz wydajnością spektrometru.

Rozdział 6 proponuje możliwą realizację konwertera modów umożliwiającego
konwersję fal spinowych przechowywanych w różnych modach przestrzennych
naszej pamięci kwantowej na ciąg impulsów wprzęganych do światłowodu
jednomodowego. Zaprezentowane są wyniki symulacji numerycznych odczytu
z pamięci wewnątrz pierścieniowej wnęki optycznej oraz manipulacja
kierunkiem emisji fotonu za pomocą wiązki odczytującej o kontrolowanym
kącie padania na chmurę atomową.

\newpage

\begin{center}
\textbf{DOCTORAL THESIS ABSTRACT}\\
\textbf{Phase Matching as an adjustable mechanism of information transfer between light and quantum memory}
\end{center}
\addcontentsline{toc}{chapter}{Abstract}This doctoral thesis focuses on the issue of phase matching in a quantum
memory operating on a cold cloud of $^{87}$Rb atoms, where non-resonant
Raman scattering is used as an interface between light and atoms.
Experimental results and considerations for controlling the memory
readout process by using the Zeeman effect, ac-Stark effect, controlling
the laser beam geometry, or using an optical ring cavity are presented.

Chapter 2 introduces the theory of the interaction between light and
atoms in a three-level $\Lambda$ system. The influence of factors
such as beam intensities, beam offsets, beam geometry, or optical
density on the readout rate and efficiency, as well as losses due
to decoherence, or light absorption are analysed.

Chapter 3 presents the generation of a fictitious magnetic field using
the ac-Stark effect with spatial resolution. Using it, the spatial
precession phase of spins oscillating in an external magnetic field
was modulated. The results obtained in the experiment show that the
control of the spatial phase allow us to turn on or off the quantum
memory readout.

Chapter 4 presents the experimental modulation of the spatial phase
of spin waves in quantum memory using the ac-Stark effect. The possibility
of compensating directly on the spin waves for any aberrations of
the imaging system is presented. Using interference measurements on
a near-field camera and direct measurements on a far-field camera,
the operation of the spatial spin wave phase modulator is also characterised
in detail.

Chapter 5 shows the combination of a GEM with spatial phase modulation.
An experimental implementation of a very high resolution spectrometer
(20 kHz \textasciitilde{} 83 peV \textasciitilde{} $6\times10^{-7}$cm$^{-1}$)
adapted to narrowband atomic emission is presented. The relationship
between resolution, bandwidth and spectrometer efficiency is also
analysed.

Chapter 6 proposes a possible implementation of a mod converter enabling
the conversion of spin waves stored in different spatial modes of
our quantum memory onto a sequence of pulses coupled into a single-mode
optical fibre. Results of numerical simulations of the memory readout
inside an annular optical cavity are presented, as well as the manipulation
of the photon emission direction using a readout beam with a controlled
angle of incidence on the atomic cloud.

\newpage\begin{center}
\textbf{LISTA PUBLIKACJI}\\
\end{center}
\addcontentsline{toc}{chapter}{Lista publikacji}Publikacje bezpośrednio powiązane z niniejszą rozprawą:
\begin{enumerate}
\item \textbf{A. Leszczyński}, M. Mazelanik, M. Lipka, M. Parniak, M. Dąbrowski,
W. Wasilewski, ,,Spatially resolved control of fictitious magnetic
fields in a cold atomic ensemble'', Optics letters 43, 1147-1150
(2018)
\item M. Lipka, \textbf{A. Leszczyński}\footnote[2]{autor korespondencyjny},
M. Mazelanik, M. Parniak, W. Wasilewski, ,,Spatial spin-wave modulator
for quantum-memory-assisted adaptive measurements'', Physical Review
Applied 11, 034049 (2019)
\item M. Mazelanik\footnote[1]{\label{equal}równy wkład autorów}, \textbf{A.
Leszczyński}\footnotemark{\ref{equal}}, M. Lipka, M. Parniak, W.
Wasilewski, ,,Temporal imaging for ultra-narrowband few-photon states
of light'', Optica 7, 203-208 (2020)
\end{enumerate}
Pozostałe publikacje:
\begin{enumerate}
\item M. Mazelanik, \textbf{A. Leszczyński}, M. Lipka, W. Wasilewski, M.
Parniak, \textquotedblright Superradiant parametric conversion of
spin waves\textquotedblright , Physical Review A 100, 053850 (2019)
\item M. Mazelanik, M. Parniak, \textbf{A. Leszczyński}, M. Lipka, W. Wasilewski,
\textquotedblright Coherent spin-wave processor of stored optical
pulses\textquotedblright , npj Quantum Information 5, 1-9 (2019)
\item M. Parniak, M. Mazelanik, \textbf{A. Leszczyński}, M. Lipka, M. Dąbrowski,
W. Wasilewski, \textquotedblright Quantum optics of spin waves through
ac Stark modulation\textquotedblright , Physical review letters 122,
063604 (2019)
\item M. Dąbrowski, M. Mazelanik, M. Parniak, \textbf{A. Leszczyński}, M.
Lipka, W. Wasilewski, \textquotedblright Certification of high-dimensional
entanglement and Einstein-Podolsky-Rosen steering with cold atomic
quantum memory\textquotedblright , Physical Review A 98, 042126 (2018)
\item M. Parniak, M. Dąbrowski, M. Mazelanik, \textbf{A. Leszczyński}, M.
Lipka, W.Wasilewski, \textquotedblleft Wavevector multiplexed atomic
quantum memory via spatially-resolved single-photon detection\textquotedblright ,
Nature communications 8, 2140 (2017)
\item \textbf{A. Leszczyński}, M. Parniak, W. Wasilewski, \textquotedblleft Phase
matching alters spatial multiphoton processes in dense atomic ensembles\textquotedblright ,
Optics express 25, 284-295 (2017)
\item M. Parniak, \textbf{A. Leszczyński}, W. Wasilewski, \textquotedblleft Coupling
of four-wave mixing and Raman scattering by ground-state atomic coherence\textquotedblright ,
Physical Review A 93, 053821 (2016)
\item M. Parniak, \textbf{A. Leszczyński}, W. Wasilewski, \textquotedblleft Magneto-optical
polarization rotation in a ladder-type atomic system for tunable offset
locking\textquotedblright , Applied Physics Letters 108, 161103 (2016)
\item \textbf{A. Leszczyński}, W. Wasilewski, \textquotedblleft Calibration
of wavefront distortion in light modulator setup by Fourier analysis
of multibeam interference\textquotedblright , JOSA A 33, 683-688 (2016) 
\end{enumerate}
\newpage\begin{center}
\textbf{PODZIĘKOWANIA}\\
\end{center}
\addcontentsline{toc}{chapter}{Podziękowania}

Chciałem podziękować swojemu promotorowi, Wojtkowi Wasilewskiemu za
wsparcie okazane w trakcie całej mojej pracy w laboratorium pamięci
kwantowych. Zawsze służył mi radą i pomocą w najróżniejszych problemach.
Bez jego pomocy nie byłoby możliwe powstanie niniejszej rozprawy.

Chciałbym również podziękować całemu zespołowi, w szczególności Michałowi
Parniakowi, Mateuszowi Mazelanikowi oraz Michałowi Lipce. Wszystkie
osiągnięte wyniki są również rezultatem ich pracy oraz świetnych pomysłów,
których nigdy im nie brakowało.

Moja praca była współfinansowana przez Narodowe Centrum Nauki\newline(2017/25/N/ST2/00713)
oraz przez projekt ,,Kwantowe Technologie Optyczne'' działający
w ramach programu Międzynarodowych Agend Badawczych Fundacji Nauki
Polskiej, współfinansowany przez Unię Europejską w ramach Europejskiego
Funduszu Rozwoju Regionalnego.

\tableofcontents{}

\chapter{Wstęp}

Niniejsza rozprawa doktorska skupia się na analizie zagadnienia dopasowania
fazowego jako sposobu rozbudowy funkcji pamięci kwantowej. Prezentujemy
szereg prac, których wspólnym celem jest wypracowywanie nowych protokołów
przetwarzania i selekcji w wielomodowej pamięci kwantowej. W eksperymentach
wykorzystaliśmy pamięć opartą o nierezonansowe rozpraszanie Ramana,
działającą na zimnej chmurze atomów rubidu \cite{Parniak2017}. Prezentujemy
mechanizmy wykorzystujące gradient pola magnetycznego, geometrię użytych
wiązek laserowych, a także modulacji przestrzennej fazy fal spinowych
przechowywanych w pamięci.

\section{Motywacja}

Do dalszego badania oraz udoskonalania motywowała nas mnogość zastosowań
pamięci kwantowych. Do najważniejszych należą:

\paragraph{Dystrybucja splątania kwantowego.}

Przesyłanie informacji kwantowej jest tematem wielu prac naukowych
ostatnich lat. Kwantowa dystrybucja klucza kryptograficznego zapewnia
najbezpieczniejszy możliwy sposób przekazywania wiadomości, gdzie
bezpieczeństwo jest zapewniane przez zasady mechaniki kwantowej \cite{Pirandola2020,Gisin2002}.
Ze względu na relatywnie słabe oddziaływanie z otoczeniem, fotony
wydają się być najlepszym nośnikiem do przesyłu informacji na duże
odległości. Powszechnie używa się światłowodów telekomunikacyjnych
działających na długości fali około 1550 nm, które wykazują się niezwykle
niskim tłumieniem \cite{Gobby2004,Stucki2009,Korzh2015}.

Straty przy przesyle informacji za pomocą fotonów skalują się wykładniczą
wraz z odległością. Stanowi to poważne ograniczenie w maksymalnym
osiągalnym dystansie, na które wiadomość może zostać przekazana. Rozwiązaniem,
który zmniejsza ten problem, może być zastosowanie wtórników kwantowych
(ang. quantum repeater). Dzięki ich użyciu, zamiast wykładniczego
spadku prędkości transmisji wraz z odległością, skaluje się ona wielomianowo
\cite{Briegel1998}. W pracy \cite{Duan2001} zaproponowano protokół
(nazwany później DLCZ od pierwszych liter nazwisk jego twórców) wykorzystujący
pamięć kwantową, bazującą na nierezonansowym rozpraszaniu Ramana.
Z czasem powstało wiele prac dotyczących komunikacji kwantowej z użyciem
pamięci kwantowych \cite{Azuma2015,Zhang2017,Sangouard2011,Simon2010}

\paragraph{Przetwarzanie informacji kwantowej.}

W wielu przypadkach niezbędne jest posiadanie pamięci umożliwiającej
przechowywanie stanów kwantowych, a także ich interakcję między sobą
\cite{Furusawa2007,Nielsen2003,Parniak2019,Tao2015,Lu2013,Reimer2016,Brecht2015a,Humphreys2014,Lu2019,mazelanik_coherent_2019}.
Właściwościami pamięci kwantowych, które ograniczają możliwości jej
wykorzystania są wydajność przechowywania, wierność odtworzenia stanu
oraz czas przechowywania. Ważna jest też możliwość odczytania informacji
z pamięci na żądanie \cite{Bussieres2013}.

\paragraph{Wykorzystanie struktury czasowej światła.}

Czasowy stopień swobody zarówno klasycznych, jak i kwantowych stanów
światła umożliwia lub zwiększa szereg sposobów przetwarzania informacji
kwantowej \cite{Humphreys2014,Brecht2015a,Reimer2016,Lu2019}. W rozwoju
architektur sieci kwantowych oraz nowych rozwiązań w obliczeniach
kwantowych i metrologii, wiele wysiłku poświęcono pamięciom kwantowym
opartym na zespołach atomowych, oferującym wielomodowe przechowywanie
i przetwarzanie \cite{Pu2017,Parniak2017,mazelanik_coherent_2019,Seri2019},
wysoką wydajność \cite{Cho2016} lub długi czas przechowywania \cite{Bao2012}.
Możliwe do wykonania implementacje protokołów łączących elastyczność
systemów atomowych i możliwości przetwarzania czasowego z natury rzeczy
wymagają zdolności do manipulowania i wykrywania modów czasowych fotonów
z rozdzielczością spektralną i czasową dopasowaną do wąskopasmowej
emisji atomowej. Wszechstronne podejście wykorzystujące dwoistość
widmowo-czasową, polega na wykonaniu mapowania częstości na czas (transformacji
Fouriera) w analogii do obrazowania dalekiego pola w przestrzeni położeń
i pędów. W celu zachowania kwantowej struktury nieklasycznych stanów
światła stosuje się układy oparte na koncepcji soczewki czasowej \cite{kolner_temporal_1989,zhu_aberration-corrected_2013,patera_space-time_2018}.

\section{Koncepcje}

Omówmy krótko podstawowe koncepcje wokół których obraca się niniejsza
praca.

\paragraph{Pamięć kwantowa.}

Pamięć kwantowa dla światła składa się z układu materialnego zdolnego
przechować informacje w bezruchu oraz odwracalnego interfejsu światło-materia,
który pozwala przenieść informacje ze światła do materii lub z powrotem.
Pamięć zasługuje na miano kwantowej, jeśli wierność przechowywania
informacji jest lepsza, niż możliwa przy użyciu jakiegokolwiek protokołu
typu pomiar oraz ponowne wygenerowanie stanu kwantowego.

Istnieje niezwykle dużo implementacji pamięci kwantowych. Można je
klasyfikować ze względu na kilka kategorii.

\paragraph{Ośrodek materialny.}

Po pierwsze pamięci kwantowe mogą korzystać z różnych ośrodków materialnych,
w których przechowywana jest informacja. Są to między innymi: ciepłe
pary atomów \cite{Reim2010,Hosseini2011,Bashkansky2012}, zimne atomy
\cite{Bao2012,Stack2015,Parniak2017}, centra barwne w diamencie \cite{Maurer2012,Tsukanov2013,Pfender2017},
domieszki metali ziem rzadkich w ciele stałym \cite{Saglamyurek2011,DeRiedmatten2008},
czy też membrana \cite{Thomas2020}.

\paragraph{Rodzaj interfejsu.}

Po drugie pamięci kwantowe mogą posiadać różne interfejsy między światłem,
a materią. Należą do nich: rozpraszanie Ramana \cite{Reim2010,Parniak2017},
elektromagnetycznie indukowana przezroczystość \cite{Veissier2013,Ma2017},
atomowy grzebień częstości \cite{Hedges2010,Akhmedzhanov2016}, czy
też spójna oscylacja populacji \cite{DeAlmeida2015}.

\paragraph{Modowość.}

Część pamięci kwantowych potrafi zapamiętywać różne mody pola elektromagnetycznego.
Takie pamięci nazywamy wielomodowymi. Można je podzielić na grupy
ze względu na to z jakiego stopnia swobody korzystają. Pamięci wielomodowe
przestrzennie \cite{Vernaz-Gris2018,Parniak2017,Chen2016,Dai2012,Surmacz2008,Dabrowski2017_Optica,Pu2017,Lan2009}
rozróżniają fotony o różnych kierunkach wektora falowego. Pamięci
wielomodowe czasowo \cite{Seri2017,Cho2016,Kutluer2017,Tiranov2016,Tang2015,Campbell2014,Sparkes2013}
potrafią odtworzyć fotony, które zostały zapisane w różnych momentach
czasu, a wielomodowe spektralnie odróżniają natomiast fotony o różnej
częstości \cite{Sinclair2014,Seri2019}. Istnieją również pamięci
korzystające z więcej niż jednego stopnia swobody. Dla przykładu,
pamięć typ GEM (gradient echo memory) zapamiętuje zarówno informację
o modzie przestrzennym \cite{Higginbottom2012}, czasowym \cite{Hosseini2012}
jak i spektralnym \cite{Sparkes2013}.

\paragraph{Dopasowanie fazowe.}

Informacja o stanie zapisanym w pamięci kwantowej jest zawarta wewnątrz
(zazwyczaj całej) objętości ośrodka. Oznacza to, że aby w procesie
odczytu konwersja przechowywanego stanu na foton była wydajna, niezbędne
jest aby światło emitowane z różnych punktów w przestrzeni interferowało
konstruktywnie. Nazywamy to warunkiem dopasowania fazowego. Bez konstruktywnej
interferencji informacja nie może zostać odczytana. Wynika stąd, że
kontrolując dopasowanie fazowe dla każdego przechowywanego modu można
decydować o tym, który z nich zostanie odczytany, a który nie. Tego
typu zachowanie zostało zaprezentowane w pamięciach typu GEM \cite{Hosseini2012},
gdzie jako mechanizmu regulującego dopasowanie fazowe używa się pola
magnetycznego o stałym w przestrzeni gradiencie. W przypadku pamięci
wielomodowych przestrzennie jak na razie nie jest możliwe, by wybrać
mody przestrzenne, które chcemy odczytać. Odczytywane są wszystkie
na raz, co stanowi pewne ograniczenie.

\section{Zawartość rozprawy}

Początkiem prac było opracowanie metody pozwalającej na manipulowanie
atomami z rozdzielczością przestrzenną. Rozdział \ref{chap:Przestrzenna-modulacja-fazy}
opisuje eksperyment, w którym za pomocą efektu ac-Starka modulowana
była faza precesji spinów z rozdzielczością przestrzenną. Optyczna
modulacja fazy została przez nas wybrana \cite{Yale2013,Moriyasu2009,Park2001,Park2002},
ponieważ w przeciwieństwie do pola magnetycznego, czy elektrycznego,
światło oświetlające atomy może być łatwo kształtowane. Generowanie
fikcyjnych pól magnetycznych \cite{Cohen-Tannoudji1972} za pomocą
wektorowego efektu ac-Starka, ma już zastosowanie w magnetometrii
\cite{Zhivun2014,Lin2017,Sun2017}. Dodatkowa kontrola przestrzenna
takiego pola mogłaby otworzyć pole do rozwiązań takich jak magnetometria
z wysoką rozdzielczością przestrzenną \cite{Vengalattore2006}, gradiometria
magnetyczna \cite{Deb2013}, obrazowanie superrozdzielcze \cite{Hemmer2012}
czy generowanie przestrajalnych potencjałów w zimnych atomach \cite{Goldman2014}.
Precyzyjna kontrola przestrzenna mogłaby również umożliwić wydajne
echa fotonowe \cite{Rosatzin1990,Zielonkowski1998} stosowane w gradientowych
pamięciach kwantowych \cite{Sparkes2010,Chaneliere2015}, precyzyjne
operowanie atomami \cite{Park2001,Park2002}, jak również nowe techniki
pułapkowania atomów \cite{Schneeweiss2014,Albrecht2016}. Analogicznie
modulację światłem można również wykorzystać do kontroli przestrzennej
fazy fal spinowych przechowywanych w pamięci, co otwiera wiele nowych
możliwości z zastosowaniem dopasowania fazowego \cite{Parniak2018,mazelanik_coherent_2019}.
Wyniki zostały opublikowane w pracy\cite{Leszczynski2018}.

Dalszą pracą było zagadnienie kształtowania przestrzennej struktury
światła za pomocą pamięci kwantowej. Modulacja fazy fal spinowych
z rozdzielczością przestrzenną pozwala stworzyć efektywnie element
optyczny nadający światłu dowolnie zadaną fazę, co przedstawiliśmy
w publikacji \cite{Lipka2019}. Dzięki temu możliwe jest korygowanie
aberracji, czy też zniekształceń spowodowanych wadami układu obrazującego.
Pozwala to na zwiększenie liczby dostępnych modów przestrzennych,
a co za tym idzie wzrost wymiarowości przestrzeni Hilberta, co może
być wykorzystywane w optycznej komunikacji kwantowej \cite{Pu2017,Parniak2017,mazelanik_coherent_2019,Seri2019},
czy też przetwarzaniu informacji \cite{Andersen2015}. Posiadanie
elementu optycznego o kontrolowalnym profilu fazowym może być też
użyte w pomiarach adaptacyjnych \cite{Wiseman2009} czy też zmiany
bazy pomiarowej na żądanie, które może być przydatne w protokołach
opartych o paradoks Einsteina-Podolskiego-Rosena \cite{Edgar2012,Aspden2013,Dabrowski2018}.
Powyższe zagadnienie zostało opisane w rozdziale \ref{chap:Kompensacja-aberracji}
niniejszej rozprawy.

Kolejnym etapem było połączenie funkcjonalności przestrzennej modulacji
fazy za pomocą efektu ac-Starka oraz pamięci typu GEM, która zapamiętuje
czasowo-spektralne właściwości zapisywanego światła. Umożliwia to
tworzenie układów obrazujących w domenie czasowej, jak to opisaliśmy
w pracy \cite{Mazelanik2020}. Metody oparte na koncepcji soczewki
czasowej umożliwiają kształtowanie widm \cite{li_high-contrast_2015,Donohue2013,Lu2018},
dopasowanie szerokości pasma \cite{allgaier_highly_2017} dla fotonów
generowanych w różnych węzłach sieci kwantowej, a także obrazowanie
korelacyjne znane jako ghost imaging \cite{denis_temporal_2017,dong_long-distance_2016,ryczkowski_magnified_2017,wu_temporal_2019}.
Obecnie istniejące rozwiązania oparte o elektrooptyczną modulację
fazy \cite{kolner_active_1988,grischkowsky_optical_1974,karpinski_bandwidth_2017},
generację sumy częstości \cite{hernandez_104_2013,bennett_aberrations_2001,bennett_temporal_1994,bennett_upconversion_1999,agrawal_temporal_1989},
czy mieszanie czterech fal \cite{kuzucu_spectral_2009,okawachi_high-resolution_2009,foster_silicon-chip-based_2008,foster_ultrafast_2009}
w ciele stałym nadają się dobrze do przetwarzania światła o szerokim
widmie, na przykład impulsów piko- lub femtosekundowych. Rozwiązanie
oparte o pamięć kwantową jest natomiast dobrze dopasowane do wąskopasmowej
emisji atomowej o widmie szerokości zaledwie kilku MHz do kilkudziesięciu
kHz \cite{Zhao2014,Guo2017,Farrera2016} czy też jonów sprzężonych
ze światłem we wnęce (poniżej 100 kHz) \cite{Stute2012}, SPDC wzmocnionej
przez wnękę (poniżej 1 MHz) \cite{Rambach2016} lub układów opto-mechanicznych
\cite{Hong2017,Hill2012}. Dokładny opis konstrukcji oraz właściwości
zbudowanego spektrometru znajduje się w rozdziale \ref{chap:Czasowo-cz=000119stotliwosciowa-transf}
niniejszej rozprawy.

Ostatnim badanym zagadnieniem, opisanym w rozdziale \ref{chap:Sukcesywny-odczyt-konwerter-mod=0000F3},
jest zaprojektowanie konwertera modów przestrzennych pamięci kwantowej
na czasowe. Konstrukcja ta umożliwi selektywny odczyt fal spinowych
przechowywanych w różnych modach przestrzennych naszej pamięci kwantowej.
Daje to możliwość projektowania nowych protokołów przetwarzania informacji
kwantowej. Ponadto, daje sposobność do szerszego zastosowania w komunikacji
kwantowej, dzięki możliwości wprzęgania fotonów, wygenerowanych w
procesie odczytu z pamięci, bezpośrednio do światłowodu jednomodowego.

\section{Wkłady pracy}

Wyniki zaprezentowane w niniejszej rozprawie są efektem pracy całego
zespołu pracującego w Laboratorium Pamięci Kwantowych. Poniżej zaprezentowany
jest wkład poszczególnych jego członków: Wojciech Wasilewski (WW),
Michał Parniak (MP), Mateusz Mazelanik (MM), Michał Lipka (ML) i Adam
Leszczyński (AL). 

WW napisał w języku Python system komunikacji umożliwiający zdalne
sterowanie wszystkimi programami w LabView podpiętymi do serwera.
Narzędzie to znacząco ułatwiało tworzenie sekwencji pomiarowych oraz
zbieranie danych, gdzie niezbędne było skoordynowanie programów działających
na kilku komputerach. AL i MM napisali oprogramowanie do sterowania
oraz kalibracji przestrzennego modulatora fazy (SLM) w języku Python.
AL, MM oraz MP zbudowali układ do kształtowania przestrzennego profilu
natężenia wiązki laserowej za pomocą SLM oraz oświetlający chmurę
atomową. MM oraz MP zoptymalizowali gęstość optyczną atomów w pułapce
magnetooptycznej. AL, MM i MP zbudowali układ do pomiaru precesji
średniego spinu atomów rubidu oraz wykonali pomiary badając wpływ
fikcyjnych pól magnetycznych generowanych przez efekt ac-Starka. AL
zajmował się analizą otrzymanych danych.

MP razem z MM zbudowali układ modulujący wiązkę lasera na częstości
6,8 GHz, niezbędne do generowania wymuszonego rozpraszania Ramana
w atomach $^{87}$Rb. AL wraz z ML dokonali charakteryzacji modulacji
fal spinowych efektem ac-Starka za pomocą bezpośrednich pomiarów sygnału
odczytanego z pamięci kwantowej na kamerze dalekiego pola oraz pomiaru
interferencyjnego na kamerze bliskiego pola. ML zaprojektował i zbudował
efektywną słabą soczewkę cylindryczną skonstruowaną z dwóch soczewek
sferycznych, która służyła do wprowadzenia sztucznych aberracji do
układu obrazującego. AL zajął się analizą i opracowaniem danych zebranych
za pomocą kamery dalekiego pola, natomiast ML danych z kamery bliskiego
pola. 

MP wraz z MM nawinęli cewki generujące gradient pola magnetycznego
w obrębie pułapki magnetooptycznej oraz skonstruowali przełącznik,
umożliwiający zmianę kierunku przepływającego przez nie prądu w czasie
rzędu kilku mikrosekund. AL oraz MM zajęli się konstrukcją spektrometru
opartego na pamięci GEM oraz ac-Starkowskiego modulatora fal spinowych.
AL zmodyfikował program WW sterujący DDS, by generować sygnał o liniowo
zmieniającej się częstości. MM udoskonalił interfejs do generowania
sekwencji pomiarowych. AL wykonał symulacje numeryczne odtwarzające
procesy zachodzące w pamięci kwantowej. MM opracował dane pod kątem
wpływu szerokości pasma i rozdzielczości spektrometru na jego wydajność.
MM dokonał również pomiaru i analizy szumu w funkcji mocy wiązki oświetlającej
atomy.

AL dokonał analizy zachowania dopasowania fazowego od konfiguracji
przestrzennej wiązek biorących udział w procesie zapisu i odczytu
z pamięci kwantowej. AL wraz z pomocą WW stworzył symulację oddziaływania
światła z atomami wewnątrz pierścieniowej wnęki rezonansowej. 

\chapter{Interfejs światło-atomy\label{chap:interfejs-=00015Bwiat=000142o-atomy}}

Rozdział przedstawia wprowadzenie do teorii oddziaływania między światłem,
a atomami. Wykorzystuje przykład trójpoziomowego układu typu $\Lambda$.
Analizowany jest wpływ czynników takich jak natężenia wiązek, ich
odstrojenia, ich geometria, czy też gęstość optyczna, na szybkość
i wydajność odczytu, a także straty związane z dekoherencją, czy absorpcją
światła. Wyprowadzone zależności będą wykorzystywane we wszystkich
kolejnych częściach niniejszej rozprawy. Struktura rozdziału jest
następująca:
\begin{itemize}
\item Sekcja 2.1 - Wyprowadzony zostanie hamiltonian atomu trójpoziomowego
oddziałującego ze światłem.
\item Sekcja 2.2 - Wprowadzone zostanie równanie Lindblada opisujące ewolucję
macierzy gęstości atomu trójpoziomowego.
\item Sekcja 2.3 - Wyprowadzone zostanie równanie na ewolucję światła w
układzie $\Lambda$ w obecności pola sprzęgającego oraz spójności
atomowej.
\item Sekcja 2.4 - Uogólnienie równań wyprowadzonych w poprzednich sekcjach
do układu $\Lambda$ z wieloma poziomami wzbudzonymi.
\item Sekcja 2.5 - Przeanalizowane zostaną równania wyprowadzone w sekcjach
2.2 i 2.3. Pokazane zostaną właściwości oddziaływania światła z atomami
takie jak zachowanie liczby wzbudzeń, przedstawienie oddziaływania
jako złożenie obrotów na małych odcinkach czasoprzestrzennych, czy
wpływ geometrii na dopasowanie fazowe.
\item Sekcja 2.6 - Omówione zostaną mechanizmy strat wynikające z dekoherencji
i absorpcji światła.
\end{itemize}

\section{Trójpoziomowy układ $\Lambda$}

\begin{figure}
\centering \includegraphics{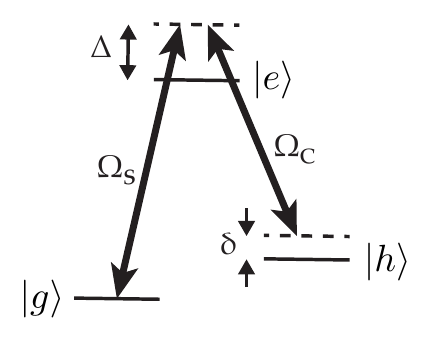}

\caption{Schemat układu $\Lambda$\label{fig:Schemat-ukladu}}
\end{figure}

W jednej z najprostszych konfiguracji pamięć kwantową można zbudować
z zespołu atomów trójpoziomowych, w których dozwolone są dwa przejścia
optyczne o wspólnym poziomie wzbudzonym $|e\rangle$ i dwóch długożyjących
poziomach podstawowych $|g\rangle$ i $|h\rangle$, jak to przedstawia
rysunek \ref{fig:Schemat-ukladu}. Jest to tak zwany układ $\Lambda$.
Czysta pamięć to taka w której wszystkie atomy są na poziomie początkowym
$|g\rangle$, czyli łącznie cała pamięć jest w stanie $|g\rangle\otimes|g\rangle\dots\otimes|g\rangle$.
Spójna superpozycja tego stanu i stanu, w którym część atomów jest
w $|h\rangle$posłuży nam do zapamiętania światła.

Zapamiętanie (czyli interfejs) polega na absorpcji fotonów słabego
światła sygnałowego na przejściu $|g\rangle\rightarrow|e\rangle$
i jednocześnie wymuszonej emisji fotonów $|e\rangle\rightarrow|h\rangle$
do silnej wiązki laserowej tzw. sprzęgającej (ang. \textit{coupling}).
Po absorpcji należy wyłączyć wiązkę sprzęgającą. Jej ponowne włączenie
spowoduje odwrócenie działania interfejsu: przejście $|h\rangle\rightarrow|e\rangle\rightarrow|g\rangle$
i emisję sygnału.

\paragraph{Hamiltonian.}

Hamiltonian nieoddziałującego atomu zapiszemy następująco:
\begin{equation}
\hat{H}_{0}=\hbar(\omega_{g}|g\rangle\langle g|+\omega_{h}|h\rangle\langle h|+\omega_{e}|e\rangle\langle e|)=\hbar\left(\begin{array}{ccc}
\omega_{g} & 0 & 0\\
0 & \omega_{h} & 0\\
0 & 0 & \omega_{e}
\end{array}\right),
\end{equation}
gdzie $\omega_{g}$, $\omega_{h}$ i $\omega_{e}$ są częstościami
odpowiadającymi energiom stanów $|g\rangle$, $|h\rangle$ i $|e\rangle$.Przy
oddziaływaniu atomu ze światłem przyjmujemy założenie, że moment dipolowy
pomiędzy poziomami $|g\rangle$ i $|h\rangle$ jest zerowy. Zakładamy
ponadto, że przerwa energetyczne pomiędzy stanami podstawowymi jest
tak duża, że światło oddziałuje tylko z jednym przejściem na raz,
$|g\rangle\rightarrow|e\rangle$, albo $|h\rangle\rightarrow|e\rangle$.
W takiej sytuacji atom oświetlany jest dwoma polami: sygnałowym o
amplitudzie $A_{s}$ na przejściu $|g\rangle\rightarrow|e\rangle$
oraz sprzęgającym $A_{c}$ na przejściu $|h\rangle\rightarrow|e\rangle$.
Pole elektryczne sygnałowe o częstości $\omega_{s}$ i potrzebnym
nam w kolejnych sekcjach wektorze falowym $\mathbf{k}_{s}$ ma postać
$E_{s}=\Re\left(A_{s}\exp(ik_{s}z)\exp(-i\omega_{s}t)\right)$. Wygląda
ona analogicznie dla pola sprzęgającego. Dodając sprzężenia dipolowe
do hamiltonaniu dostaniemy:
\begin{equation}
\hat{H}=\frac{\hbar}{2}\left(\begin{array}{ccc}
\omega_{g} & 0 & -\Omega_{s}^{*}\mathrm{e}^{-i\omega_{s}t}\\
0 & \omega_{h} & -\Omega_{c}^{*}\mathrm{e}^{-i\omega_{c}t}\\
-\Omega_{\text{s}}\mathrm{e}^{i\omega_{s}t} & -\Omega_{\text{c}}\mathrm{e}^{i\omega_{c}t} & \omega_{e}
\end{array}\right)+\mathrm{c.c.},
\end{equation}
gdzie $\Omega_{\text{s}}=d_{g,e}A_{\text{s}}\exp(ik_{s}z)/\hbar$,
$\Omega_{\text{c}}=d_{h,e}A_{\text{c}}\exp(ik_{c}z)/\hbar$ to częstości
Rabiego, a $d_{ge}$ i $d_{he}$ są momentami dipolowymi na przejściach
$|g\rangle\rightarrow|e\rangle$ oraz $|h\rangle\rightarrow|e\rangle$.

Przechodzimy następnie do obrazu oddziaływania tak dobranego, aby
uzyskać Hamiltonian niezależny od czasu. Jako część niezaburzoną weźmiemy
hamiltonian atomu, który miałby poziomy w miejscach linii przerywanych
na rysunku \ref{fig:Schemat-ukladu} $H_{b}=\hbar\omega_{s}|e\rangle\langle e|+\hbar(\omega_{s}-\omega_{c})|h\rangle\langle h|$.
Różnice energii pomiędzy założonymi, a rzeczywistymi położeniami poziomów
będą dalej grały role odstrojeń $\Delta=\omega_{s}-\omega_{g,e}$
oraz $\delta=(\omega_{s}-\omega_{c})-(\omega_{g,e}-\omega_{h,e})$.
Możemy obliczyć hamiltonian oddziaływania: 
\begin{equation}
\begin{array}{llc}
\hat{H}_{I}\!\!\!\!\!\! & =\!\!\!\!\!\!\!\!\!\!\!\!\! & \mathrm{e^{-\frac{iH_{b}t}{\hbar}}}(\hat{H}-\hat{H}_{b})\mathrm{e^{\frac{iH_{b}t}{\hbar}}}\\
 & = & -\frac{\hbar}{2}\left(\begin{array}{ccc}
0 & 0 & \Omega_{\text{s}}^{*}\\
0 & 2\delta & \Omega_{\text{c}}^{*}\\
\Omega_{\text{s}} & \Omega_{\text{c}} & 2\Delta
\end{array}\right),
\end{array}
\end{equation}
Zastosowano tu tzw. przybliżenie wirującej fali (RWA) \cite{Allen1975}.
Człony oscylujące z częstościami rzędu setek THz, jak to ma miejsce
w przypadku atomów rubidu, można zaniedbać, jako że w trakcie oddziaływania
o czasie trwania rzędu mikrosekund, uśrednią się one do zera.

\section{Ewolucja macierzy gęstości}

W układzie izolowanym ewolucja jest opisywana równaniem Liouville'a.
W przypadku opisanego wyżej układu $\Lambda$ należy jednak uwzględnić
emisję spontaniczną z poziomu wzbudzonego $|e\rangle$. Wtedy ewolucja
macierzy gęstości jest opisywana równaniem Lindblada \cite{Manzano2020}:
\begin{equation}
\begin{array}{c}
\frac{\partial\hat{\rho}}{\partial t}=\overbrace{-\frac{i}{\hbar}[\hat{H}_{I},\hat{\rho}]}^{\mathrm{Liouville}}-\overbrace{\frac{1}{2}\{\hat{\Gamma},\hat{\rho}\}}^{\mathrm{dekoherencja}}+\overbrace{\hat{\Upsilon}}^{\mathrm{repopulacja}},\\
\hat{\Gamma}=\Gamma|e\rangle\langle e|,\\
\hat{\Upsilon}=\Gamma\rho_{e,e}(|g\rangle\langle g|+|h\rangle\langle h|)/2,
\end{array}\label{eq:liouville}
\end{equation}
gdzie $\Gamma$ to stała zaniku dla poziomu wzbudzonego. $\hat{\Upsilon}$
jest tak zwaną macierzą repopulacji, która obsadza w sposób niespójny
poziomy podstawowe $|g\rangle$i $|h\rangle$ w wyniku emisji spontanicznej
z poziomu $|e\rangle$.

Powyższe równanie na ewolucję macierzy gęstości można uprościć zakładając
reżim, w którym można pominąć populację stanu wzbudzonego $\rho_{e,e}\simeq0$.
Ponadto stosujemy przybliżenie adiabatyczne, które mówi że spójności
optyczne podążają natychmiast za zmianami pól optycznych $\Omega_{\text{s}}$,
$\Omega_{\text{c}}$ czyli $\frac{\mathrm{d}\rho_{g,e}}{\mathrm{d}t}=\frac{\mathrm{d}\rho_{h,e}}{\mathrm{d}t}\approx0$.
Uzyskamy stąd:
\begin{equation}
\begin{gathered}\rho_{g,e}=i\frac{\Omega_{\text{s}}^{*}+\Omega_{\text{c}}^{*}\rho_{g,h}}{2\Delta-i\Gamma}\\
\rho_{h,e}=i\frac{\Omega_{\text{s}}^{*}\rho_{g,h}^{*}}{2\Delta-i\Gamma}
\end{gathered}
\label{eq: rh_ge}
\end{equation}
Podstawiając \ref{eq: rh_ge} do \ref{eq:liouville} uzyskamy uproszczone
równania ewolucji.

\paragraph{Stany bliskie początkowego.}

Będzie nas interesował właściwie tylko przypadek, gdy praktycznie
wszystkie atomy znajdują się w stanie podstawowym $|g\rangle$, czyli
$\rho_{g,g}\simeq1$, zaś populacje pozostałych stanów są zaniedbywalne
$\rho_{h,h}\simeq0$. Ponadto załóżmy, że pole $\Omega_{\text{s}}$
jest słabym sygnałem, a $\Omega_{\text{c}}$ silnym polem sprzęgającym,
tzn. $|\Omega_{\text{s}}|^{2}\ll|\Omega_{\text{c}}|^{2}.$ Cała informacja
o układzie zawiera się wtedy w spójności między długo żyjącymi stanami
podstawowymi $\rho_{g,h}$.

\paragraph{Ewolucja spójności.}

Ewolucja spójności w czasie dana jest wówczas równaniem:

\begin{equation}
\begin{array}{c}
\frac{\partial\rho_{g,h}}{\partial t}=(-i(\delta-\delta_{\text{acS}})-\gamma)\rho_{g,h}+\frac{i}{2}\frac{\Omega_{\text{s}}^{*}\Omega_{\text{c}}}{2\Delta-i\Gamma},\\
\delta_{\text{acS}}=-\Delta\frac{|\Omega_{\text{c}}|^{2}}{\Gamma^{2}+4\Delta^{2}},\\
\gamma=\frac{\Gamma}{2}\frac{|\Omega_{\text{c}}|^{2}}{\Gamma^{2}+4\Delta^{2}}.
\end{array}\label{eq:spojnosc}
\end{equation}
Człon $\delta-\delta_{\text{acS}}$ oznacza całkowite odstrojenie
od rezonansu dwufotonowego z uwzględnieniem przesunięcia ac-Starka
wywołanego polem sprzęgającym $\Omega_{\text{c}}$. Czynnik $\gamma$
odpowiada za dekoherencję spójności atomowej pod wpływem $\Omega_{\text{c}}$.
Jest to tzw. poszerzenie natężeniowe linii. Człon proporcjonalny do
$\Omega_{\text{s}}^{*}\Omega_{\text{c}}$ odpowiada za nieliniową
konwersję światła na spójność $\rho_{g,h}$,

\paragraph{Zapis fali płaskiej.}

Jeżeli przyjmiemy, że pola sygnałowe i laser sprzęgający mają postać
fal płaskich o wektorach falowych $\mathbf{k}_{\text{s}}$ i $\mathbf{k}_{\text{c}}$
oraz stałych w czasie amplitudach to człon opisujący nieliniową konwersję
$\Omega_{\text{s}}^{*}\Omega_{\text{c}}$ będzie miał zależność przestrzenną\footnote{aby pozostać w zgodzie z konwencją śledzenia jedynie wolnozmiennej
obwiedni pól względem fali nośnej $\exp(ik_{0}z-\omega_{0}t)$ dla
fal płaskich przestrzenna zależność częstości Rabiego przyjmie postać
$\Omega_{\text{s}}\sim\exp(i\mathbf{k}_{\text{s}}\mathbf{r}-ik_{0}z)$
oraz $\Omega_{\text{c}}\sim\exp(i\mathbf{k}_{\text{c}}\mathbf{r}-ik_{0}z)$.
Odjecie wektora falowego $k_{0}$ nie wypływa na człon $\Omega_{\text{s}}^{*}\Omega_{\text{c}}$
w równaniu \ref{eq:spojnosc}.} typu $\exp(i\mathbf{k}_{\text{c}}\mathbf{r}-i\mathbf{k}_{\text{s}}\mathbf{r})$.
Równanie \ref{eq:spojnosc} można odcałkować. Zapisana spójność również
będzie miała postać fali płaskiej z wektorem falowym $\mathbf{k}_{\rho}$:
\begin{equation}
\rho_{g,h}\sim\exp(i\mathbf{k}_{\rho}\mathbf{r})\qquad\mathbf{k}_{\rho}=\mathbf{k}_{\text{c}}-\mathbf{k}_{\text{s}}.\label{eq:rho=00003Dexp(ikr)}
\end{equation}
Wzór ten uzasadnia przekonanie, że w procesie rozpraszania Ramana
absorpcja fotonu z wiązki sygnałowej i emisja do wiązki sprzęgającej
prowadzi do przekazu pędu do atomów, który przyjmuje postać przestrzennie
zmiennej fazy spójności atomowej, a charakter tej zmienności ma postać
fali płaskiej.

\section{Ewolucja pola}

Ewolucja pola elektrycznego w ośrodku atomowym określona jest równaniem
falowym z członem źródłowym:
\begin{equation}
\nabla^{2}\mathbf{E}-\frac{1}{c^{2}}\frac{\partial^{2}}{\partial t^{2}}\mathbf{E}=\frac{1}{c^{2}\epsilon_{0}}\frac{\partial^{2}}{\partial t^{2}}\mathbf{P},
\end{equation}
gdzie $\mathbf{E}$ jest polem elektrycznym, $\mathbf{P}$ polaryzacją
atomową, $c$ prędkością światła, $\epsilon_{0}$ przenikalnością
elektryczną próżni. Do powyższego równania można podstawić $\mathbf{E}=\Re\left(A\exp(ik_{0}z-\omega_{0}t)\right)\hat{e}_{x}$
oraz $\mathbf{P}=\Re\left(P\exp(ik_{0}z-\omega_{0}t)\right)\hat{e}_{x}$,
gdzie $A$ to obwiednia pola, a $P$ to obwiednia polaryzacji atomowej,
$\omega_{0}$ częstością fali elektromagnetycznej, a $k_{0}=\omega_{0}/c$
liczbą falową. Stosując przybliżenie wolno zmiennej obwiedni, w którym
$\partial_{z}^{2}A=0$, $\partial_{t}^{2}A=0$, $\partial_{t}^{2}P=0$,
można uzyskać z równania falowego równanie na obwiednie pola:

\begin{equation}
\frac{\partial}{\partial z}A+\frac{1}{c}\frac{\partial}{\partial t}A=i\frac{k_{0}}{2\epsilon_{0}}P+\frac{i}{2k_{0}}\nabla_{\perp}^{2}A,\label{eq:ewolucja pola}
\end{equation}
gdzie $\nabla_{\perp}^{2}=\frac{\partial^{2}}{\partial x^{2}}+\frac{\partial^{2}}{\partial y^{2}}$.

Lewa strona otrzymanego równania opisuje falę biegnącą z prędkością
światła. Równanie upraszcza się w układzie współrzędnych w którym
mierzymy czas względem środka impulsu świetlnego ($t'=t-z/c$). Po
przekształceniu współrzędnych z powyższego równania znika pochodna
czasowa i przyjmuje ono postać:
\begin{equation}
\frac{\partial}{\partial z}A=\frac{i}{2k_{0}}\nabla_{\perp}^{2}A+i\frac{k_{0}}{2\epsilon_{0}}P\label{eq:=00015Bwiatlo_biegnace}
\end{equation}

Otrzymane równanie opisuje jak zmienia się obwiednia pola $A$ po
przejściu przez plasterek ośrodka materialnego $dz$. Po prawej stronie
mamy człon opisujący dyfrakcję oraz polaryzacje ośrodka, która może
powstać za sprawą innych pól.

\paragraph{Wiele pól.}

Jeśli pole elektryczne składała się z szeregu składowych 
\begin{equation}
\mathbf{E}(\mathbf{r},t)=\sum_{j}\Re\left(A_{j}(\mathbf{r},t-z/c)\exp(ik_{j}z-\omega_{j}t)\right)\hat{e}_{j},\label{eq:E=00003DsumAeikz-iwt}
\end{equation}
oraz analogicznie polaryzacja $\mathbf{P}=\sum_{j}\Re\left(P_{j}\exp(ik_{j}z-\omega_{j}t)\right)\hat{e}_{j}$,
to równanie \ref{eq:=00015Bwiatlo_biegnace} można rozdzielić na szereg
równań niezależnych dla każdej polaryzacji $\hat{e}_{j}$, częstości
$\omega_{j}$ i wektora falowego $k_{j}$.

\paragraph{Fale płaskie}

Zauważmy, że jeśli tylko jedno $A_{j}$ jest niezerowe i nie zależy
od położenia, to wzór \ref{eq:E=00003DsumAeikz-iwt} opisuje fale
płaską o wektorze falowym $k_{j}\hat{e}_{z}$ i częstości $\omega_{j}$.
W dalszej analize będziemy potrzebowali rozważać jednocześnie fale
płaskie o wektorach falowych lekko odchylonych oraz zmienionych częstościach.
Zauważmy że podstawiając amplitudę w postaci bardzo długiej fali płaskiej
$A_{j}\sim\exp(i\mathbf{d}\mathbf{\cdot r}-i\delta t)$ reprezentującej
małe odchylenie od centralnego wektora falowego i centralnej częstości,
uzyskamy zależność przestrzenną pola $\mathbf{\mathbf{E}(\mathbf{r},t)}$
z wektorem falowym $\mathbf{k}=\mathbf{d}+k_{j}\hat{e}_{z}$ oraz
czasową z częstością $\omega=\omega_{j}+\delta+d_{z}/c$. Odwrotnie,
pragnąc opisać falę płaską o wektorze falowym $\mathbf{k}$ i częstości
$\omega$ musimy użyć amplitudy typu $A_{j}\sim\exp\left[i\left(\mathbf{k}-k_{j}\hat{e}_{z}\right)\mathbf{\cdot r}-i\left(\omega-\omega_{j}-d_{z}/c\right)t\right]$.

\subsection{Równania Maxwella-Blocha}

Rozważmy chmurę atomów trójpoziomowych, z których każdy ewoluuje zgodnie
z opisanymi wcześniej prawami. W niniejszej rozprawie interesuje nas
sytuacja kiedy praktycznie wszystkie atomy są w stanie $|g\rangle$
a pole sprzężone z tym stanem $\Omega_{s}$ jest słabe (obwiednia
$A_{s}=\hbar\Omega_{s}/d_{g,e}$). Na drugim przejściu $|h\rangle\rightarrow|e\rangle$
atomy oddziałują z silną, przychodzącą z zewnątrz wiązką laserową
$\Omega_{\text{c}}$ której amplituda praktycznie nie zmienia się
przy przejściu przez chmurę --- w stanie $|h\rangle$ jest rzędy
wielkości mniej atomów niż fotonów w wiązce laserowej.

Pole $\Omega_{s}$ w przyjętym modelu atomu trójpoziomowego oddziałuje
ze spójnością atomową za pośrednictwem przejścia $|g\rangle\rightarrow|e\rangle$.
Polaryzacja na tym przejściu z definicji jest równa gęstości objętościowej
momentu dipolowego: 
\begin{equation}
P=2nd_{g,e}\rho_{g,e}^{*},\label{eq:polaryzacja}
\end{equation}
gdzie $n$ jest koncentracją atomów. Podstawiając powyższą formułę,
wartość spójności $\rho_{g,e}^{*}$ uzyskaną w eliminacji adiabatycznej
\ref{eq: rh_ge} oraz $A_{\text{s}}=\hbar\Omega_{\text{s}}/d_{g,e}$
do równania \ref{eq:=00015Bwiatlo_biegnace} uzyskujemy:
\begin{equation}
\frac{\partial}{\partial z}\Omega_{\text{s}}=-i\frac{k_{s}}{\hbar\epsilon_{0}}\frac{nd_{g,e}^{2}}{2\Delta+i\Gamma}(\!\!\!\!\!\!\!\!\!\underbrace{\Omega_{\text{s}}}_{\mathrm{\begin{array}{c}
\textrm{dyspersja}\\
\textrm{absorpcja}
\end{array}}}\!\!\!+\!\!\!\!\!\!\underbrace{\Omega_{\text{c}}\rho_{g,h}^{*}}_{\mathrm{\begin{array}{c}
\textrm{konwersja}\\
\textrm{światło-spójność}
\end{array}}}\!\!\!\!\!\!\!\!\!\!\!\!\!\!\!)+\frac{i}{2k_{s}}\nabla_{\perp}^{2}\Omega_{\text{s}}\label{eq:omega_fin}
\end{equation}
Pod nieobecność $\Omega_{\text{c}}$ równanie to opisuje dyspersję
oraz absorpcję pola $\Omega_{\text{s}}$ pod wpływem oddziaływania
na przejściu $|g\rangle\rightarrow|e\rangle$, a także dyfrakcję związaną
z propagacją. Dodatkowy człon $\Omega_{\text{c}}$ opisuje konwersje
pomiędzy spójnością $\rho_{g,h}$, a polem $\Omega_{\text{s}}$.

Zestaw równań \ref{eq:spojnosc} oraz \ref{eq:omega_fin} stanowią
razem tzw. równania Maxwella-Blocha opisujące interakcję światła z
atomami.

\paragraph{Wiele stanów wzbudzonych.}

W eksperymentach opisywanych w pracy, w atomie $^{87}$Rb, spójność
jest przechowywana między poziomami stanu podstawowego $|g\rangle$i
$|h\rangle$ o $F=1$ oraz $F=2$, gdzie $F$ jest liczbą kwantową
całkowitego atomowego momentem pędu. Oznacza to, że w przejściach
Ramanowskich mogą uczestniczyć poziomy wzbudzone $|e\rangle$ o $F'=1$
oraz $F'=2$. Jeżeli odstrojenia od poszczególnych poziomów wzbudzonych
mają porównywalne wartości, należy uwzględnić rozszerzony model z
dwoma poziomami wzbudzonymi. Przyjmujmy nazwę drugiego poziomu wzbudzonego
$|f\rangle$ oraz $\Omega_{i,j}$ - częstości Rabiego na przejściach
$|i\rangle\rightarrow|j\rangle\text{, }i\in\{g,h\}\text{, }j\in\{e,f\}$.
Ponadto niech $\Delta_{e}$ oraz $\Delta_{f}$ oznaczają odpowiednio
odstrojenia od poziomów $|e\rangle$ i $|f\rangle$. Postępując analogicznie
jak przy wyprowadzaniu formuły dla atomu trójpoziomowego można pokazać,
że wkłady od poszczególnych poziomów wzbudzonych dodają się.

Podstawiając do równania \ref{eq:ewolucja pola} polaryzację atomową
$P=2n(d_{g,e}\rho_{g,e}^{*}+d_{g,f}\rho_{g,f}^{*}$), również można
zauważyć, że wkłady policzone niezależnie dla każdego poziomu wzbudzonego
dodają się.

\section{Dopasowanie fazowe\label{sec:Dopasowanie-fazowe}}

Rozważania o ewolucji macierzy gęstości zakończyliśmy analizą zależności
przestrzennych w trakcie zapisu fali sygnałowej do pamięci przewidując,
że zapisana spójność będzie miała postać fali płaskiej $\rho_{g,h}\sim\exp(i\mathbf{k}_{\rho}\mathbf{r})$
z wektorem falowym $\mathbf{k}_{\rho}=\mathbf{k}_{\text{c}}-\mathbf{k}_{\text{s}}$
równym różnicy wektorów falowych fali sprzęgającej i sygnałowej. Rozważmy
teraz proces odczytu, całkując równanie \ref{eq:omega_fin} z zapisaną
spójnością atomową przy płaskiej fali sprzęgającej.

\paragraph{Sygnałowe fale płaskie.}

Rozważmy najpierw rozwiązanie jednorodne równania \ref{eq:omega_fin}
pod nieobecność lasera $\Omega_{\text{c}}=0$. Rozwiazania te to fale
płaskie, a równanie \ref{eq:omega_fin} nakłada jeden warunek wiążący
częstość i wektor falowy. W próżni $c|k|=\omega$, co można przepisać
jako $k_{z}\simeq k_{0}-k_{\perp}^{2}/2k_{0}$ natomiast w obecności
atomów pojawia sie korekta od współczynnika załamana. Aby odzyskać
skorygowany wektor falowy opisujący oscylacje pola elektrycznego $\mathbf{k}_{\text{s}}$
podstawiamy $\Omega_{\text{s}}\sim\exp(i\mathbf{k}_{\text{s}}\mathbf{r}-ik_{s}z)$
do równania \ref{eq:omega_fin} i uzyskujemy wówczas:
\begin{equation}
k_{s,z}=k_{s}^{at}-\frac{1}{2k_{0}}\mathbf{k}_{\text{s,\ensuremath{\perp}}}^{2},\qquad k_{s}^{at}=k_{s}\left(1-\frac{1}{\hbar\epsilon_{0}}\frac{nd_{g,e}^{2}}{2\Delta+i\Gamma}\right)\label{eq:ksz=00003D}
\end{equation}
gdzie $\mathbf{k}_{\text{s,\ensuremath{\perp}}}^{2}$ to poprzeczna
cześć wektora falowego. Warunek ten reprezentuje paraboloidę obrotową
dopuszczalnych wektorów falowych, czyli przyosiowy fragment sfery
o promieniu równym długości wektora falowego $k_{s}^{at}$ skorygowanym
na dyspersję atomów.

\paragraph{Odczyt fali płaskiej i dopasowanie fazowe.}

Aby przekonwertować spójność atomową na światło, atomy są oświetlane
wiązką sprzęgającą. Przyjmijmy, że pole elektryczne lasera ma postać
fali płaskiej o wektorze falowym $\mathbf{k}_{c2}$, czyli $\Omega_{\text{c2}}\sim\exp(i\mathbf{k}_{\text{c2}}\mathbf{r}-ik_{s}z)$.
Niech początkowa spójność ma postać fali płaskiej, jaką otrzymaliśmy
po modelowym zapisie $\rho_{g,h}\sim\exp(i\mathbf{k}_{\rho}\mathbf{r})$.
Prześledźmy co odczytalibyśmy, gdyby oba wspomiane pola nie ulegały
osłabieniu. Tworzą one wspólnie człon źródłowy w równaniu \ref{eq:omega_fin}
o zależności przestrzennej typu $\Omega_{\text{c}}\rho_{g,h}^{*}\sim\exp\left[i\left(\mathbf{k}_{\text{c2}}-\mathbf{k}_{\rho}\right)\mathbf{r}-ik_{s}z\right]$.
Zastosujmy wobec tego ansatz rozwiązania zgodny z równaniem bez nieliniowości
postaci $\Omega_{\text{s}}(\mathbf{r})=s(z)\exp(i\mathbf{k}_{\text{s2}}\mathbf{r}-ik_{s}z)$
gdzie składową $k_{s2,z}$ wyznaczamy z równania \ref{eq:ksz=00003D}.
Zauważmy, że spełnienie \ref{eq:omega_fin} jednocześnie dla wszystkich
$x$/$y$ wymaga zgodności poprzecznej składowej przyjętego ansatzu
z członem źródłowym, czyli $\mathbf{k}_{\text{s2,\ensuremath{\perp}}}=\mathbf{k}_{\text{c2,\ensuremath{\perp}}}-\mathbf{k}_{\rho,\ensuremath{\perp}}$.
W tym momencie zdajmy sobie sprawę, że oscylacje członu źródłowego
wzdłuż $z$ mają potencjalnie inny wektor falowy $k_{\text{c2,z}}-k_{\rho,z}-k_{s}$
niż oscylacje ansatzu z wektorem falowym $k_{s2,z}-k_{s}$. Różnica
wektorów falowych wynosi: 
\[
\delta k_{z}=k_{\text{c2,z}}-k_{\rho,z}-k_{s2,z}=k_{\text{c2,z}}-k_{\rho,z}-k_{s}^{at}+\frac{\left(\mathbf{k}_{\text{c2,\ensuremath{\perp}}}-\mathbf{k}_{\rho,\ensuremath{\perp}}\right)^{2}}{2k_{s}}.
\]

Podstawiając podany ansatz do równania \ref{eq:omega_fin} i wykorzystując
podane zależności uzyskamy proste równanie na narastanie amplitudy
fali odczytywanej wzdłuż $z$ $\partial s/\partial z\sim n\exp(i\delta k_{z}z)$,
gdzie dla uproszczenia pominęliśmy szereg stałych multiplikatywnych
pozostawiając jedynie gęstość atomów $n$, która może się zmieniać
wzdłuż $z$. Dla jednorodnego zespołu atomów o długości L (szklana
komórka z gazem) możemy podane równanie odcałkować uzyskując znane
skalowanie wydajności z różnicą wektorów falowych $\text{\textasciitilde sinc}\left(\delta k_{z}L/2\right)$.
Dla gaussowskiego w przestrzeni rozkładu gęstości (zimne atomy) uzyskamy
gładką zależność gaussowską $\sim\exp(-\delta k_{z}^{2}L^{2})$. Widzimy
więc, że wydajny odczyt wymaga dopasowania fazowego $|\delta k_{z}L|\lesssim1$.

\paragraph{Rzeczywisty zapis i odczyt.}

Dobra pamięć kwantowa powinna cechować się wysoką sprawnością zapisu
i odczytu. Innymi słowy pole sygnałowe na etapie zapisu powinno być
absorbowane i nie opuszczać zespołu atomów. W efekcie zależność przestrzenna
amplitudy wytworzonej spójności atomowej będzie inna niż otrzymaliśmy
w najprostszym de facto pierwszym rzędzie rozwinięcia ze względu na
sprzężenie atomy-światło. Z kolei proces odczytu będziemy prowadzić,
aż do zaniku spójności atomowej. Dlatego przedstawione zależności
należy traktować jedynie jako pierwsze przybliżenie. Poniżej zasugerujemy
sposób konstrukcji prostej i niezawodnej symulacji numerycznej.

\section{Zachowanie liczby wzbudzeń\label{subsec:Zachowanie-liczby-wzbudze=000144}}

Zgodnie z fizycznym opisem zagadnienia przedstawionym na początku
spodziewamy się, że absorpcja jednego fotonu sygnałowego spowoduje
przeniesienie jednego atomu do stanu $|h\rangle$. Pokażmy, że tak
jest. Dla uproszczenia zaniedbajmy wymiary poprzeczne $x$, $y$.

Wartość $|\rho_{g,h}|^{2}$ mówi nam o tym jaki odsetek atomów znajduje
się w stanie $|h\rangle$. Oznacza to, że całkowita liczba atomów
przeniesionych do stanu $|h\rangle$ wynosi:
\begin{equation}
n_{h}=\intop_{-\infty}^{\infty}n|\rho_{g,h}|^{2}\text{d}z=\intop_{-\infty}^{\infty}n|\widetilde{\rho}_{g,h}|^{2}\text{d}k_{z},\label{eq:n_h}
\end{equation}
gdzie tylda oznacza transformacje Fouriera względem $z$. Jednocześnie
całkowita liczba fotonów sygnałowych $n_{ph}$ równa jest energii
pola $\int\epsilon_{0}|A_{1}|^{2}/2$ podzielonej przez energię fotonu
$\hbar\omega_{0}$ i wynosi: 
\begin{equation}
n_{\text{ph}}=\frac{\epsilon_{0}}{2\hbar\omega_{0}}\intop_{-\infty}^{\infty}|A_{\text{s}}|^{2}\text{d}z=\frac{\epsilon_{0}}{2\hbar\omega_{0}}\intop_{-\infty}^{\infty}|\widetilde{A_{\text{s}}}|^{2}\text{d}k_{z}.\label{eq:n_ph}
\end{equation}
Chcemy pokazać, że całkowita liczba wzbudzeń nie zmienia się w czasie:
\begin{equation}
\frac{\partial}{\partial t}\left(n_{\text{at}}+n_{\text{ph}}\right)=0.
\end{equation}
W tym celu podstawimy podane wyżej definicje (\ref{eq:n_h},\ref{eq:n_ph}),
przejdziemy z pochodną czasową pod całki i wykorzystamy poznane równania
na pochodne czasowe ewoluujących pól. Wykorzystując równanie \ref{eq:spojnosc}
i zaniedbując emisję spontaniczną $\Gamma$ można obliczyć, że: 
\begin{equation}
\frac{\partial}{\partial t}(|\widetilde{\rho}_{g,h}|^{2})=\frac{\partial}{\partial t}(\widetilde{\rho}_{g,h}\widetilde{\rho}_{g,h}^{*})=2\Re\left(i\frac{\Omega_{c}}{4\Delta}\widetilde{\Omega}_{\text{s}}^{*}\widetilde{\rho}_{g,h}^{*}\right).
\end{equation}
 Z kolei z równań \ref{eq:ewolucja pola} i \ref{eq:polaryzacja}
otrzymujemy po zaniedbaniu dyspersji oraz absorpcji:
\begin{equation}
\frac{\partial}{\partial t}(|\widetilde{A_{\text{s}}}|^{2})=\frac{\partial}{\partial t}(\widetilde{A_{\text{s}}}\widetilde{A_{\text{s}}}^{*})=2\Re\left(i\frac{ck_{o}n\hbar\Omega_{\text{c}}^{*}}{2\epsilon_{0}\Delta}\widetilde{\Omega}_{\text{s}}\widetilde{\rho}_{g,h}\right).
\end{equation}
Obliczmy pochodną czasową całkowitej liczby wzbudzeń $n_{at}+n_{ph}$
wykorzystując otrzymane wyrażenia. Uzyskamy:
\begin{equation}
\frac{\partial}{\partial t}\left(n_{\text{at}}+n_{\text{ph}}\right)=\intop_{-\infty}^{\infty}2\Re\left(i\frac{n}{4\Delta}\left(\Omega_{\text{c}}\widetilde{\Omega}_{\text{s}}^{*}\widetilde{\rho}_{g,h}^{*}+\Omega_{\text{c}}^{*}\widetilde{\Omega}_{\text{s}}\widetilde{\rho}_{g,h}\right)\right)\text{d}k_{z}.
\end{equation}

Zauważmy że suma iloczynów pól pod całką jest czysto rzeczywista ---
ma postać $\varpi+\varpi^{*}$. Wobec tego pod całką jest równe 0.
Zatem całkowita liczby wzbudzeń $n_{at}+n_{ph}$ nie zmienia się w
czasie po zaniedbaniu emisji spontanicznej $\Gamma\rightarrow0$.

\section{Interfejs światło-atomy jako obroty\label{subsec:Oddzia=000142ywanie-=00015Bwiat=000142o-atomy-jako} }

\begin{figure}
\centering\includegraphics{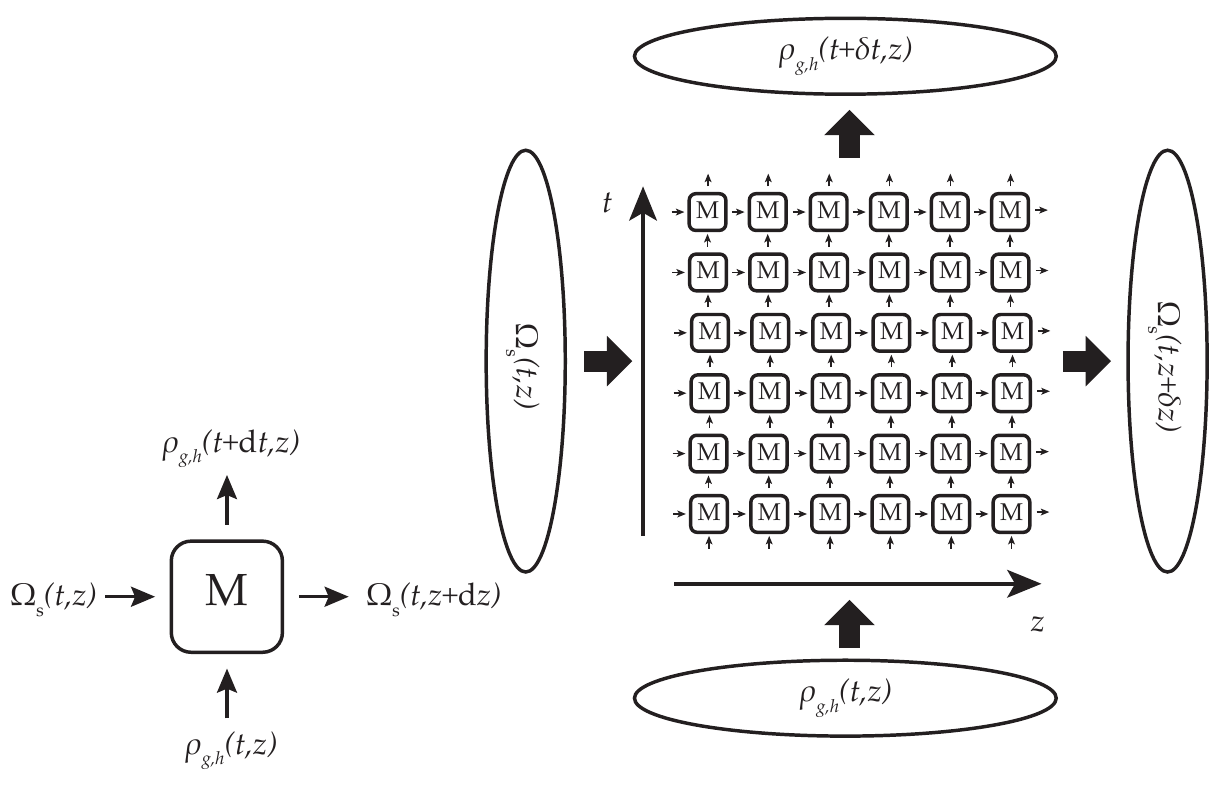}

\caption{Ilustracja interfejsu światło $\Omega_{\mathrm{s}}$ --- spójność
atomowa $\rho_{g,h}$ jako obrotów. Pojedynczy element nazwany M przedstawia
unitarną operację pokazaną w formule \ref{eq:obroty_prawdziwe}, która
zachowuje sumę liczby wzbudzeń atomowych i liczby fotonów i propaguje
światło w przestrzeni o infinitezymalną odległość d$z$ oraz spójność
w czasie o infinitezymalną odległość d$t$. Pełną transformację można
przedstawić jako złożenie operacji elementarnych M.\label{fig:Ilustracja-oddzia=000142ywania-=00015Bwiat=000142a}}
\end{figure}
Przyjmując niewielki krok czasowy $\mathrm{d}t$ i przestrzenny $\mathrm{d}z$,
pochodne $\partial_{t}\rho{}_{g,h}$ oraz $\partial_{z}\Omega_{\mathrm{s}}$
można przybliżyć poprzez ilorazy
\begin{equation}
\begin{gathered}\frac{\partial\rho{}_{g,h}}{\partial t}\approx\frac{\rho{}_{g,h}(t+\mathrm{d}t)-\rho{}_{g,h}(t)}{dt}\\
\frac{\partial\Omega_{\text{s}}}{\partial z}\approx\frac{\Omega_{\text{s}}(z+\mathrm{d}z)-\Omega_{\text{s}}(t)}{dz}
\end{gathered}
\end{equation}
Podstawiając powyższe przybliżenie do równań \ref{eq:spojnosc} oraz
\ref{eq:omega_fin} całe oddziaływanie światła i atomów można zapisać
w postaci macierzowej 
\begin{equation}
\left(\begin{array}{c}
\rho'{}_{g,h}\\
\Omega'{}_{\text{s}}^{*}
\end{array}\right)=\left(\begin{array}{cc}
1 & a\\
b & 1
\end{array}\right)\left(\begin{array}{c}
\rho{}_{g,h}\\
\Omega{}_{\text{s}}^{*}
\end{array}\right),\label{eq:rho-omega}
\end{equation}
gdzie: 
\begin{equation}
\begin{array}{c}
a=\frac{i}{2}\frac{\Omega_{\text{c}}}{2\Delta}\mathrm{d}t,\\
b=ik_{0}\frac{nd_{g,e}^{2}}{\hbar\epsilon_{0}}\frac{\Omega_{\text{c}}^{*}}{2\Delta}\mathrm{d}z,\\
\rho'{}_{g,h}=\rho{}_{g,h}(t+\mathrm{d}t),\\
\Omega'{}_{\text{s}}=\Omega_{\text{s}}(z+\mathrm{d}z).
\end{array}
\end{equation}
Dokonajmy teraz przeskalowania spójności i amplitudy pola wprowadzając
zmienne $u_{\mathrm{at}}=c_{\text{at}}\rho_{g,h}$ oraz $u_{\mathrm{ph}}=c_{\text{ph}}\Omega_{\text{s}}^{*}$.
Współczynniki skalowania $c$ dobierzemy tak, żeby $|u_{\text{at}}|^{2}$
było ilością wzbudzeń atomowych na odcinku $[z,z+\text{d}z]$, zaś
$|u_{\text{ph}}|^{2}$ ilością fotonów na odcinku czasowym $[t,t+\text{d}t]$:
\begin{equation}
\begin{array}{c}
c_{\mathrm{at}}=e^{-i\frac{\pi}{4}}\sqrt{n\mathrm{d}z},\\
c_{\mathrm{ph}}=e^{i\frac{\pi}{4}}\sqrt{\frac{\hbar}{d_{g,e}^{2}}\frac{\epsilon_{0}c}{2\omega_{0}}\mathrm{d}t}.
\end{array}
\end{equation}
równanie \ref{eq:rho-omega} przyjmuje po przeskalowaniu postać:
\begin{equation}
\left(\!\!\!\begin{array}{c}
u'_{\mathrm{at}}\\
u'_{\mathrm{ph}}
\end{array}\!\!\!\right)\!\!\!=\!\!\!\left(\!\!\!\begin{array}{cc}
1 & \frac{\Omega_{\text{c}}}{2\sqrt{2}\Delta}d_{g,e}\sqrt{\frac{nk_{0}}{\hbar\epsilon_{0}}}\sqrt{dt\mathrm{d}z}\\
-\frac{\Omega_{\text{c}}^{*}}{2\sqrt{2}\Delta}d_{g,e}\sqrt{\frac{nk_{0}}{\hbar\epsilon_{0}}}\sqrt{dt\mathrm{d}z} & 1
\end{array}\!\!\!\right)\!\!\!\left(\!\!\!\begin{array}{c}
u_{\mathrm{at}}\\
u_{\mathrm{ph}}
\end{array}\!\!\!\right).\label{eq:obroty}
\end{equation}
Przeskalujmy także oś $z$ na gęstość optyczną OD. Na każdy przyrost
d$z$ przypada przyrost gęstości optycznej:
\begin{equation}
\text{d}\text{OD}=2\frac{nk_{0}d_{g,e}^{*2}}{\hbar\epsilon_{0}\Gamma}\text{d}z,
\end{equation}
Wobec tego równanie \ref{eq:obroty} można przepisać w postaci
\begin{equation}
\begin{gathered}\left(\!\!\!\begin{array}{c}
u'_{\mathrm{at}}\\
u'_{\mathrm{ph}}
\end{array}\!\!\!\right)\!\!\!=\!\!\!\left(\!\!\!\begin{array}{cc}
1 & \alpha\\
-\alpha & 1
\end{array}\!\!\!\right)\!\!\!\left(\!\!\!\begin{array}{c}
u_{\mathrm{at}}\\
u_{\mathrm{ph}}
\end{array}\!\!\!\right),\\
\alpha=\sqrt{\Gamma}\frac{\Omega_{\text{c}}}{4\Delta}\sqrt{\text{dOD}\text{d}t}.
\end{gathered}
\label{eq:alphascaling}
\end{equation}
Jak wynika z analizy przeprowadzonej w podsekcji \ref{subsec:Zachowanie-liczby-wzbudze=000144}
całkowita liczba wzbudzeń $n_{\text{at}}+n_{\text{ph}}=|u_{\mathrm{at}}|^{2}+|u_{\mathrm{ph}}|^{2}$
musi być zachowana. Powyższa transformacja jest pierwszym rzędem rozwinięcia
unitarnej transformacji obrotu. Ostatecznie więc oddziaływanie między
spójnością i polem w małym wycinku przestrzennoczasowym $[z,z+\text{d}z]$,
$[t,t+\text{d}t]$ musi być opisywane przez 
\begin{equation}
\left(\!\!\!\begin{array}{c}
u'_{\mathrm{at}}\\
u'_{\mathrm{ph}}
\end{array}\!\!\!\right)\!\!\!=\!\!\!\left(\!\!\!\begin{array}{cc}
\cos(\alpha) & \sin(\alpha)\\
-\sin(\alpha) & \cos(\alpha)
\end{array}\!\!\!\right)\!\!\!\left(\!\!\!\begin{array}{c}
u_{\mathrm{at}}\\
u_{\mathrm{ph}}
\end{array}\!\!\!.\right).\label{eq:obroty_prawdziwe}
\end{equation}

\paragraph{Skalowanie.}

Przedstawienie oddziaływania światła z atomami w powyższej postaci
pozwala również jasno pokazać, że szybkość konwersji skaluje się proporcjonalnie
do pierwiastka z koncentracji atomów $\sqrt{n},$ pierwiastka z natężenia
wiązki sprzęgającej $\sqrt{I}\sim|\Omega_{2}|$ oraz odwrotnie proporcjonalnie
do odstrojenia $\Delta$. Rysunek \ref{fig:Ilustracja-oddzia=000142ywania-=00015Bwiat=000142a}
pokazuje ideę oddziaływania światła z atomami jako złożenie elementarnych
transformacji z wyrażenia \ref{eq:obroty_prawdziwe}.

\section{Straty \label{sec:wydajno=00015B=000107-i-straty}}

Istnieją dwa mechanizmy strat podczas konwersji światła i spójności
atomowej. Pierwszy jest związany z poszerzeniem natężeniowym powodowanym
przez wiązkę sprzęgającą, które niszczy spójność atomową. Zapominając
na chwilę o konwersji na światło, zgodnie z równaniem \ref{eq:spojnosc}
spójność atomowa będzie zanikać wykładniczo w czasie: 
\begin{equation}
\rho_{g,h}(t)=\rho_{g,h}(0)\mathrm{e}^{-\gamma t},\qquad\gamma=\frac{\Gamma}{2}\frac{|\Omega_{\text{c}}|^{2}}{\Gamma^{2}+4\Delta^{2}}.\label{eq:natezeniowe}
\end{equation}
Zauważmy, że dla dużych odstrojeń $\gamma\mathrm{d}t=2|\alpha|^{2}/\mathrm{OD}$,
gdzie $\alpha$ jest kątem uogólnionego obrotu przeprowadzającego
wzbudzenia ze światła do atomów lub odwrotnie, wprowadzonym w równaniu
\ref{eq:obroty_prawdziwe}. Wobec tego o ile jest prawdą, że straty
$\gamma$ można redukować poprzez zwiększanie odstrojenia $\Delta$
lub zmniejszenia natężenia wiązki sprzęgającej proporcjonalnego do
$|\Omega_{c}|^{2}$, to uzyskanie małych start przy wydajnym zapisie
lub odczycie jest wyłącznie kwestią dużej gęstości optycznej. Zaawansowane
analizy teoretyczne potwierdzają tę konkluzję \cite{Gorshkov2008}.

Drugim mechanizmem strat jest absorpcja jednofotonowa światła sygnałowego
$\Omega_{\text{s}}$ przez chmurę atomową. Amplituda światła, które
przepropagowało się przez chmurę atomową, zakładając brak wiązki sprzęgającej
$\Omega_{\text{c}}$, spada wykładniczo:
\begin{equation}
\begin{gathered}\begin{array}{c}
A_{\mathrm{f}}=A_{\mathrm{i}}\mathrm{e}^{-\Gamma^{2}\frac{\mathrm{OD}}{2\Gamma^{2}+8\Delta^{2}}}\end{array},\\
\mathrm{OD}=2\frac{k_{0}d_{g,e}^{*2}N}{\hbar\epsilon_{0}\Gamma},
\end{gathered}
\label{eq:absorpcja}
\end{equation}
gdzie $N$ to liczba atomów. Straty te zależą od gęstości optycznej
$\mathrm{OD}$ oraz od odstrojenia $\Delta$. 

W obu przypadkach straty mogą zostać zmniejszone poprzez zwiększanie
odstrojenia $\Delta$. Należy jednak pamiętać, że dzieje się to kosztem
spowolnienia konwersji między spójnością atomową, a polem. Uwzględniając
analizę przeprowadzoną w sekcji \ref{subsec:Oddzia=000142ywanie-=00015Bwiat=000142o-atomy-jako}
zwiększanie natężenia wiązki sprzęgającej $I$ oraz gęstości optycznej
OD proporcjonalnie do odstrojenia $\Delta$ zmniejszy straty, a jednocześnie
w granicy dużych $\Delta$, zachowa stałą szybkość oddziaływania.
Ponieważ odstrojenie, jak i natężenie są łatwo regulowalne, ważne
jest, by budując układ, optymalizować go pod względem jak największej
gęstości optycznej.

\section{Podsumowanie}

W tym rozdziale wyprowadzone zostały równania opisujące oddziaływanie
światła z atomami w modelu atomu trójpoziomowego w układzie $\Lambda$.
Najważniejsze wyniki uzyskane w tej części rozprawy to:
\begin{itemize}
\item Zinterpretowanie oddziaływania światła z atomami jako złożenie transformacji
obrotów na niewielkich odcinkach czasoprzestrzennych, które konwertują
fotony sygnału na wzbudzenia atomowe i odwrotnie. 
\item Demonstracja wpływu niedopasowania fazowego na amplitudę odczytu.
Szerokość zakresu niedopasowania $\delta k_{z}$ w którym możliwy
jest wydajny odczyt, jest odwrotnie proporcjonalna do długości chmury
atomowej.
\item Opisanie mechanizmów strat podczas konwersji światło-spójność atomowa.
Jest to dekoherencja spójności spowodowana oddziaływaniem z wiązką
z sprzęgającą, czyli tzw. poszerzenie natężeniowe oraz absorpcja jednofotonowa
światła przez chmurę atomową.
\end{itemize}

\chapter{Przestrzenna modulacja fazy \label{chap:Przestrzenna-modulacja-fazy}}

Pierwszym krokiem prezentowanej rozprawy jest eksperymentalna weryfikacja
faktycznej możliwości przestrzennego modulowania fazy fal spinowych.
Pierwszy eksperyment powinien prowadzić do uzyskania łatwego do detekcji
i jednoznacznego efektu. Jako jedną z najpowszechniej stosowanych
opcji wybraliśmy detekcję precesji spinu atomowego za pomocą pomiaru
skręcenia polaryzacji. Po ustawieniu wszystkich spinów wzdłuż osi
wiązki próbkującej, a prostopadle do pola magnetycznego wynikiem pomiaru
jest oscylujący sygnał z detektora różnicowego, który można na żywo
obserwować na oscyloskopie upewniając się natychmiast o postępie eksperymentu.
Wiązka próbkująca przechodzi przez eliptyczną chmurę atomową wzdłuż
jej najdłuższej osi ($z)$, natomiast modulacja optyczna ma postać
wyświetlanego z boku za pomocą projektora laserowego obrazu. W takim
układzie łatwo można zaobserwować zmianę częstości precesji przy jednorodnym
oświetleniu. Towarzyszy jej jednak znaczne przyspieszenie zaniku.
W niniejszym rozdziale przedstawiamy wyniki pomiaru tych efektów.

Manipulowanie spinem atomowym za pomocą efektu ac-Starka posiada dwie
zasadnicze przewagi na polem magnetycznym. Po pierwsze, wiązka Starkowska
może zostać szybko włączona lub wyłączona. W naszym przypadku za pomocą
modulatorów akustooptycznych w czasie rzędu kilkudziesięciu nanosekund.
W przypadku pola magnetycznego jest to niemożliwe do osiągnięcia,
nawet w przy użyciu cewek o niskiej indukcyjności. Po drugie, za pomocą
przestrzennego modulatora fazy możliwe jest wygenerowanie wiązki o
prawie dowolnym przestrzennym rozkładzie natężenia. Oznacza to, że
faza może być modulowana z rozdzielczością przestrzenną. Wytworzenie
takich pól magnetycznych jest praktycznie niewykonalne.

Byliśmy w stanie odróżnić zanik sygnału wywołany destruktywną interferencją
wkładów od poszczególnych części chmury od lokalnej dekoherencji.
W szczególności przeprowadziliśmy prosty eksperyment, w którym chmura
podzielona jest na kilka spójnie interferujących fragmentów. Analiza
pozwoliła ustalić, że głównym powodem dekoherencji są niedoskonałości
wyświetlanego obrazu modulującego --- speckle. Powodują one ograniczenie
maksymalnej możliwej do nałożenia fazy bez zniszczenia spójności atomowej.

Struktura Rozdziału jest następująca:
\begin{itemize}
\item Sekcja \ref{sec:Efekt-ac-Starka} - Omówiony zostanie model przesunięcia
ac-Starka. Pokazane zostanie, że przy kołowej polaryzacji i w granicy
dużego odstrojenia hamiltonian oddziaływania między światłem a atomami
przyjmuje formę analogiczną jak dla hamiltonian atomu w zewnętrznym
polu magnetycznym.
\item Sekcja \ref{sec:Uk=000142ad-do=00015Bwiadczalny} - Zaprezentowana
zostanie budowa układu doświadczalnego, który umożliwia obserwowanie
oscylacji średniego rzutu spinu atomowego na oś $z$ oraz wytwarzanie
fikcyjnego pola magnetycznego z rozdzielczością przestrzenną.
\item Sekcja \ref{sec:Charakteryzacja-fikcyjnego-pola} - Przedstawione
zostaną pomiary charakteryzujące działanie fikcyjnego pola magnetycznego
w funkcji parametrów wiązki oświetlającej atomy rubidu.
\item Sekcja \ref{sec:Revival-koncepcja-i-wyniki} - Pokazana zostanie dynamika
średniego spinu przy schodkowych rozkładach fikcyjnego pola magnetycznego.
Pokazana zostaje również analogia obserwowanych oscylacji do wydajności
odczytu z pamięci kwantowej.
\end{itemize}
Wyniki zostały opublikowane w pracy \cite{Leszczynski2018}.

\section{Efekt ac-Starka\label{sec:Efekt-ac-Starka}}

Hamiltonian poziomu nadsubtelnego o spinie atomowym $F$ i całkowitym
elektronowym momencie pędu $J$, oddziałującego ze światłem propagującym
się wzdłuż osi $z$, składa się z trzech komponentów \cite{DeEchaniz2008,Geremia2006,Colangelo2013}:
\begin{equation}
\hat{H}_{s}=\underbrace{\hat{H}_{s}^{(0)}}_{\text{skalarny}}+\underbrace{\hat{H}_{s}^{(1)}}_{\text{wektorowy}}+\underbrace{\hat{H}_{s}^{(2)}}_{\text{tensorowy}},
\end{equation}
gdzie
\begin{equation}
\begin{gathered}\hat{H}_{s}^{(0)}=\frac{2}{3}g\alpha_{FJJ'I}^{(0)}(\Delta_{s})\hat{S}_{0},\\
\hat{H}_{s}^{(1)}=g\alpha_{FJJ'I}^{(1)}(\Delta_{s})\hat{S}_{z}\hat{F}_{z},\\
\hat{H}_{s}^{(2)}=g\alpha_{FJJ'I}^{(2)}(\Delta_{s})\!\!\left(\frac{1}{3}\hat{S}_{0}\left(3\hat{F}_{z}^{2}-2\hat{\mathbb{1}}\right)+\hat{S}_{x}\left(\hat{F}_{x}^{2}-\hat{F}_{y}^{2}\right)+\hat{S}_{y}\left(\hat{F}_{x}\hat{F}_{y}+\hat{F}_{y}\hat{F}_{x}\right)\!\!\right).
\end{gathered}
\label{eq:stark-ogolny}
\end{equation}
$\alpha_{FJJ'I}^{(i)}(\Delta_{s})$ jest tensorem polaryzacji atomowej
zależnym od całkowitego elektronowego momentu pędu stanu wzbudzonego
$J'$, spinu jądrowego $I$ oraz odstrojenia światła $\Delta_{s}$
od centroidu linii. Współczynnik $g=\omega_{0}/(2\epsilon_{0}V)$,
gdzie $\omega_{0}$ jest częstością rezonansową linii, a $V$ objętością
oddziaływania. $\hat{S}_{i}$ oraz $\hat{F}_{i}$ są odpowiednio operatorami
Stokesa oraz operatorami spinu atomowego.

Przy odstrojeniu $\Delta_{s}$ znacznie większym niż rozszczepienie
nadsubtelne poziomu wzbudzonego, czynnik $\alpha^{(2)}$ skaluje się
jak $1/\Delta_{s}^{2}$, podczas gdy $\alpha^{(0)}$ oraz $\alpha^{(1)}$
jak $1/\Delta_{s}$ \cite{DeEchaniz2008}. Ponadto, gdy polaryzacja
światła jest kołowa, $|\langle\hat{S}_{z}\rangle|\gg|\langle\hat{S}_{x}\rangle|,|\langle\hat{S}_{y}\rangle|$.
W takich warunkach tensorowa część hamiltonianu $\hat{H}_{s}^{(2)}$
może zostać zaniedbana. Skalarna część hamiltonianu $\hat{H}_{s}^{(0)}$
daje jedynie stały wkład do energii wszystkich poziomów energetycznych,
co nie wpływa na dynamikę spinu atomowego. Wobec tego ten człon również
może zostać pominięty. 

\paragraph{Fikcyjne pole magnetyczne.}

Ostatecznie jedynym nietrywialnym członem hamiltonianu $\hat{H}_{s}$
jest człon wektorowy $\hat{H}_{s}^{(1)}$. W przybliżeniu klasycznego
elektromagnetyzmu przyjmuje on postać
\begin{equation}
\hat{H}_{s}^{(1)}=q\frac{\kappa}{\Delta_{s}}\frac{I_{s}}{2\hbar\epsilon_{0}c}\hat{F}_{z},
\end{equation}
gdzie $\kappa=\alpha^{(1)}\Delta_{s}$ jest stałe w granicy dużych
$\Delta_{s}$, $q$ przyjmuje wartości $\pm1$ odpowiednio dla polaryzacji
$\sigma_{\pm}$, a $I_{s}$ jest natężeniem światła. Powyższy hamiltonian
ma formę analogiczną do hamiltonianu oddziaływania atomu z polem magnetycznym.
Pozwala to zdefiniować tzw. fikcyjne pole magnetyczne \cite{Cohen-Tannoudji1972}
\begin{equation}
B_{\mathrm{f}}=q\frac{1}{g_{F}\mu_{B}}\frac{\kappa}{\Delta_{s}}\frac{I_{s}}{2\hbar\epsilon_{0}c},\label{eq:b_f}
\end{equation}
 Jeżeli atomy znajdą się również pod wpływem prawdziwego pola magnetycznego
$\mathbf{B}$ to całkowity hamiltonian przyjmie postać
\begin{equation}
\hat{H}=g_{F}\mu_{B}\mathbf{B}_{\mathrm{eff}}\mathbf{\hat{F}},\label{eq:hamiltonian_fikcyjne_pole}
\end{equation}
gdzie $\mathbf{B}_{\mathrm{eff}}=\mathbf{B}+\mathbf{B}_{\mathrm{f}}$
jest efektywnym polem magnetycznym, $\mathbf{B}_{\mathrm{f}}=(0,0,B_{\text{f}})$
jest wektorem fikcyjnego pola magnetycznego, a $\hat{\mathbf{F}}=(\hat{F}_{x}\hat{,F}_{y},\hat{F}_{z})$.
W konsekwencji można obserwować precesję spinów z częstością Larmora
\begin{equation}
\omega_{\mathrm{L}}=g_{F}\mu_{B}\mathbf{|B}_{\mathrm{eff}}|/\hbar.\label{eq:larmor}
\end{equation}
Hamiltonian przedstawiony w formule \ref{eq:hamiltonian_fikcyjne_pole}
został wyprowadzony zakładając, że światło generujące fikcyjne pole
magnetyczne jest skierowane wzdłuż osi $z$. Ponieważ iloczyn skalarny
$\mathbf{B}_{\mathrm{eff}}\mathbf{\hat{F}}$ nie zależy od wyboru
bazy, wektory $\mathbf{B}_{\mathrm{eff}}$ oraz $\mathbf{\hat{F}}$
można zapisać w dowolnym innym, obróconym układzie współrzędnych.
W takim przypadku $\mathbf{B}_{\mathrm{f}}=B_{\text{f}}\hat{e}_{s}$,
gdzie $\hat{e}_{s}$ jest wersorem wskazującym kierunek fikcyjnego
pola magnetycznego w nowej bazie. By zachować spójność z pozostałymi
rozdziałami niniejszej rozprawy, w których kierunek wiązki Starkowskiej
modulującej atomy jest wyznaczony przez oś $x$, kierunek fikcyjnego
pola magnetycznego opisanego w tym rozdziale również będzie pokrywał
się z osią $x$, a nie osią $z$, jak ma to miejsce w wyprowadzeniu.

\section{Układ doświadczalny \label{sec:Uk=000142ad-do=00015Bwiadczalny}}

\begin{figure}

\centering\includegraphics{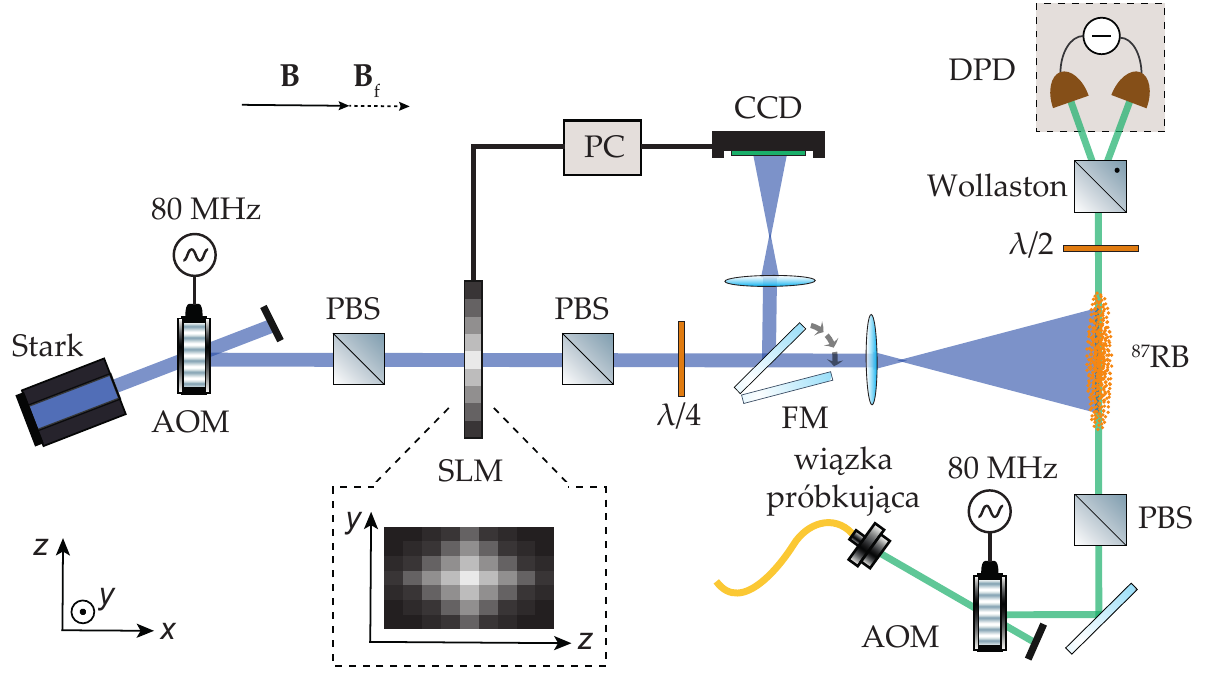}\caption{Schemat układu doświadczalnego. Atomy $^{87}$Rb, po wypuszczeniu
z pułapki magnetooptycznej są oświetlane przez kołowo spolaryzowaną
wiązkę Starkowską oraz liniowo spolaryzowaną wiązkę próbkującą. Wiązka
Starkowska jest kształtowana za pomocą odbiciowego SLM, natomiast
dla większej czytelności schemat przedstawia SLM transmisyjny. FM
przełącza obraz pomiędzy atomami $^{87}$Rb, a kamerą CCD. Obrót polaryzacji
wiązki próbkującej jest rejestrowany za pomocą polaryzatora Wollastona
oraz DPD. Użyte skróty: AOM - modulator akustooptyczny, PBS - polaryzująca
kostka światłodzieląca, FM - przestawne lustro, DPD - fotodioda różnicowa.\label{fig:Schemat-uk=000142adu-fid.}}

\end{figure}
Centralnym elementem układu eksperymentalnego przedstawionego na rysunku
\ref{fig:Schemat-uk=000142adu-fid.} jest chmura atomów $^{87}$Rb.
Za pomocą pułapki magnetooptycznej jest ona uformowana w kształt cygara
o długości około 1 cm i średnicy około 0,6 mm. Po pułapkowaniu przez
około 19 ms chmura atomów jest dodatkowo chłodzona przez 300 $\mu$s
za pomocą melasy optycznej, co pozwala zmniejszyć jej temperaturę
do 22 $\mu$K \cite{Parniak2017}. Po uformowaniu atomy są przez 15$\mu$s
pompowane do stanu $|g\rangle=5S_{1/2},F=1,m_{F}=1$ za pomocą dwóch
wiązek: niespolaryzowanej dostrojonej do przejścia $5S_{1/2},F=2\rightarrow5P_{1/2},F=2$
oraz wiązki o polaryzacji $\sigma_{+}$ propagującej się wzdłuż osi
$z$, dostrojonej do przejścia $5S_{1/2},F=1\rightarrow5P_{3/2},F=1$,
gdzie oś kwantyzacji $z$ jest skierowana wzdłuż chmury. Po zakończeniu
pompowania następuje pomiar precesji. Atomy oświetlane są liniowo
spolaryzowaną wiązką próbkującą, świecącą wzdłuż osi $z$, odstrojoną
o 100 MHz od przejścia $5S_{1/2},F=1\rightarrow5P_{3/2},F=2$. Skręcenie
jej polaryzacji, które zależy od średniego rzutu spinu atomowego $\langle\hat{F}_{z}\rangle$
na oś $z$, jest mierzone przy pomocy polaryzatora Wollastona oraz
fotodiody różnicowej \cite{Geremia2006,Behbood2013a}. Dzięki temu
możliwa jest obserwacja oscylacji Larmora pod wpływem efektywnego
pola magnetycznego. Po około 100 $\mu$s swobodnych oscylacji załączamy
modulację. Cały eksperyment odbywa się cyklicznie z częstością 50
Hz, synchronizowanymi do częstości napięcia w sieci. Dzięki temu zewnętrzne
pole magnetyczne pochodzące od wszystkich urządzeń elektrycznych znajdujących
się wokół układu doświadczalnego jest dokładnie takie samo w każdym
cyklu.

Do wytwarzania fikcyjnego pola magnetycznego użyta została silna,
kołowo spolaryzowana wiązka laserowa (zwana dalej wiązką Starkowską)
daleko odstrojona od linii $5S_{1/2},F=1\rightarrow5P_{3/2}$ i oświetlająca
atomy wzdłuż osi $x$. Dodatkowo został wybudowany układ kształtujący
jej przestrzenny rozkład natężenia. Składa się on z odbiciowego przestrzennego
modulatora fazy światła (ang. spatial light modulator, SLM), znajdującego
się w płaszczyźnie obrazowania chmury atomowej oraz polaryzującej
kostki światłodzielącej. Całość działa na dokładnie takiej samej zasadzie
jak powszechnie używane ekrany ciekłokrystaliczne. Obraz powierzchni
SLM jest przełączany pomiędzy atomy, a kamerę CCD. Umożliwia to bezpośrednią
kalibrację przestrzennego rozkładu natężenia światła \cite{Leszczynski2015}
padającego na chmurę atomów. Dodatkowo cała pułapka otoczona jest
zestawem 6 cewek, które umożliwiają precyzyjną kontrolę prawdziwego
pola magnetycznego, w jakim znajdują się atomy. Zostały one skonfigurowane
tak, by wytworzyć stałe w przestrzeni pole magnetyczne $\boldsymbol{\mathbf{B}}$
o natężeniu $B=$100 mG wzdłuż osi $x$, czyli tak by pokrywało się
z kierunkiem fikcyjnego pola $\mathbf{B}_{\mathrm{f}}$.

\section{Charakteryzacja fikcyjnego pola magnetycznego\label{sec:Charakteryzacja-fikcyjnego-pola}}

\begin{figure}
\centering\includegraphics{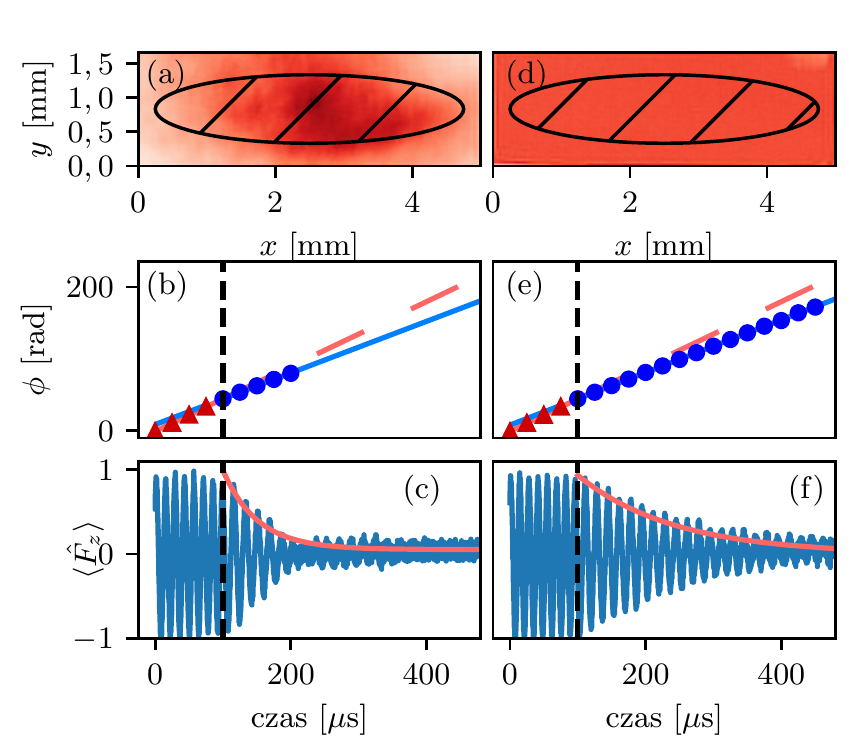}

\caption{Wpływ wiązki Starkowskiej na precesję spinów. Lewa oraz prawa kolumna
przedstawiają wyniki pomiarów odpowiednio dla wiązki o niejednorodnym
(a) oraz jednorodnym (b) rozkładzie natężenia. W obu przypadkach $\Delta_{s}=2\pi\times30$
GHz, a średnie natężenie wynosi $I_{s}=160$ mW/cm$^{2}$, co odpowiada
fikcyjnemu polu magnetycznemu o natężeniu $B_{\mathrm{f}}=20$ mG.
Panele (c) i (f) przedstawiają oscylacje Larmora wraz z dopasowanym
zanikiem wykładniczym. Panele (b) i (e) prezentują skumulowaną fazę
tych oscylacji. Czerwona przerywana linia przedstawia fazę spodziewaną
przy braku modulacji wiązką starkowską, natomiast kolor niebieski
przedstawia zmierzoną fazę. Przerywana linia pionowa oznacza moment
w którym wiązka starkowska została włączona. \label{fig:dekoherencja}}

\end{figure}

Rysunek \ref{fig:dekoherencja} przedstawia wpływ wiązki Starkowskiej,
o polaryzacji $\sigma_{-}$, odstrojeniu $\Delta_{s}=2\pi\times30$
GHz i średnim natężeniu $I_{s}=160$ mW/cm$^{2}$, na precesję spinu.
Odpowiada to fikcyjnemu polu magnetycznemu $\mathbf{B}_{\mathrm{f}}$
o natężeniu $B_{\mathrm{f}}=20$ mG. Jego zwrot jest przeciwny do
pola zewnętrznego $\mathbf{B}$. Lewa (a-c) oraz prawa (d-f) kolumna
przedstawiają wyniki pomiaru dla wiązek Starkowskich o odpowiednio
niejednorodnym rozkładzie natężenia (a) oraz o natężeniu wyrównanym
za pomocą SLM (d). Panele (c) i (f) pokazują mierzone na detektorze
różnicowym oscylacje $\langle\hat{F}_{z}\rangle(t)$ w czasie. 

Ponieważ oscylacje są równomierne i wolnozmienne, to stosując transformację
Hilberta na zmierzonym sygnale $\langle\hat{F}_{z}\rangle(t)$ można
odzyskać czasowy przebieg amplitudy i fazy oscylacji. Moduł transformaty
Hilberta stanowi obwiednia, która została narysowana czerwoną linią
na panelach (c) i (d). Natomiast faza transformaty Hilberta (czyli
skumulowana faza oscylacji) została narysowana na panelach (b) i (e).

Jak widać z pomiaru, czas zaniku $\tau_{\mathrm{s}}$ obserwowanych
oscylacji $\langle\hat{F}_{z}\rangle$ po wyrównaniu natężenia światła
znacząco się wydłużył.

Zanik oscylacji $\langle\hat{F}_{z}\rangle_{z}$ może być spowodowany
utratą napompowania (dekoherencją) lub przestrzennym rozfazowaniem
spinów. Utrata napompowania może być wywołana absorpcją wiązki próbującej
lub Starkowskiej. Absorpcja wiązki Starkowskiej jest jednak proporcjonalna
do czynnika $1/\Delta_{s}^{2}$ \cite{Sparkes2010}, co przy odstrojeniu
$\Delta_{s}=2\pi\cdot30$ GHz jest całkowicie zaniedbywalne. Zmniejszenie
mocy wiązki próbującej do poziomu kilku $\mu$W pozwala osiągnąć czas
zaniku oscylacji $\tau_{s}=700$ $\mu$s.

Największy wkład do zaniku pochodzi od przestrzennej nierównomierności
natężenia wiązki Starkowskiej. Przy oświetleniu chmury atomów wiązką
o rozkładzie natężenia przedstawionym na rysunku \ref{fig:dekoherencja}
(a), czas zaniku oscylacji Larmora wyniósł zaledwie $\tau_{s}=50$
$\mu$s (rysunek \ref{fig:dekoherencja} (c)). Wyrównanie go za pomocą
SLM (rysunek \ref{fig:dekoherencja} (d)) zwiększyło go do około $\tau_{s}=250$
$\mu$s (rysunek \ref{fig:dekoherencja} (f)).
\begin{figure}
\centering\includegraphics{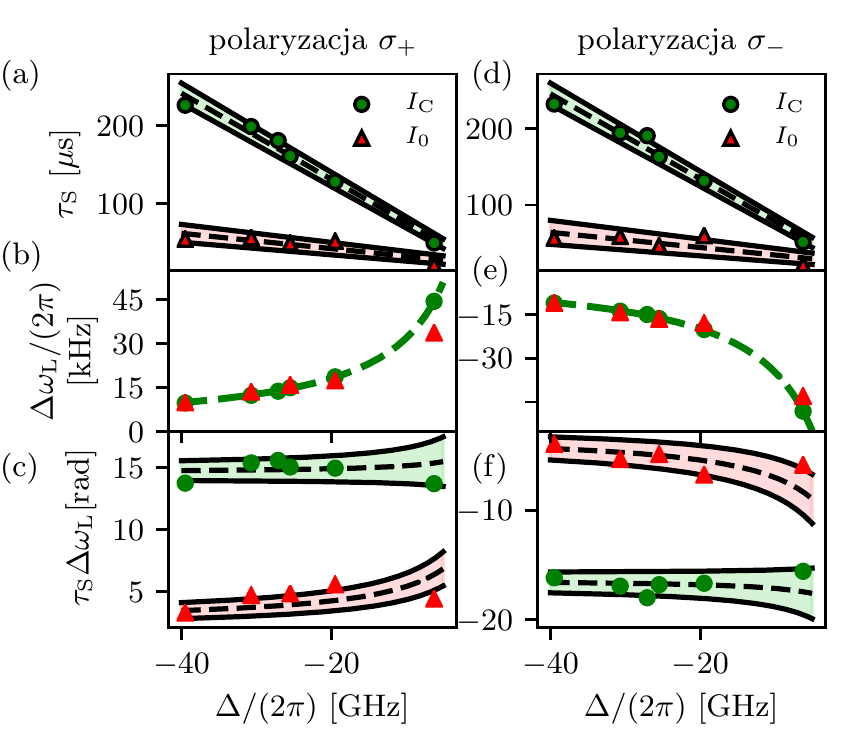}

\caption{Wpływ odstrojenia $\Delta$ na oscylacje Larmora z ($I_{C}$) oraz
bez ($I_{0}$) korekcji natężenia za pomocą SLM. (a, d) Czas życia
oscylacji $\tau_{s}$ wraz z dopasowaną prostą. (b, e) Zmiana częstości
Larmora $\Delta\omega_{\mathrm{L}}$ wywołana wiązką strarkowską wraz
z dopasowaną krzywą teoretyczną (równanie \ref{eq:larmor}) odpowiadającą
natężeniu $I_{s}=160$ mW/cm$^{2}$. (c, f) Całkowita faza zakumulowana
w czasie zaniku oscylacji Larmora $\phi_{s}=\tau_{s}\Delta\omega_{\mathrm{L}}$.
Lewa oraz prawa kolumna odpowiadają dwóm polaryzacjom kołowym wiązki
Starkowskiej, odpowiednio $\sigma_{+}$ oraz $\sigma_{-}$.\label{fig:Wp=000142yw-odstrojenia-}}

\end{figure}

Rysunek \ref{fig:Wp=000142yw-odstrojenia-} przedstawia, w jaki sposób
zachowują się oscylacje Larmora pod wpływem wiązki Starkowskiej o
różnych odstrojeniach $\Delta$. Lewa oraz prawa kolumna odpowiadają
polaryzacjom kołowym $\sigma_{+}$ oraz $\sigma_{-}$ wiązki Starkowskiej,
co odpowiada dwóm przeciwnym zwrotom generowanego fikcyjnego pola
magnetycznego. Czas życia oscylacji Larmora $\tau_{s}$ skaluje się
liniowo (a, d) z odstrojeniem $\Delta$. Jednocześnie, zgodnie z równaniami
\ref{eq:b_f} i \ref{eq:larmor}, zmiana częstości Larmora $\Delta\omega_{\mathrm{L}}$
wywołana fikcyjnym polem magnetycznym skaluje się jak odwrotność odstrojenia
$\Delta^{-1}$ (b, e). Oznacza to, że całkowita faza nadana spinom
w czasie zaniku $\tau_{s}$ wynosi $\phi_{s}=\tau_{s}\Delta\omega_{\mathrm{L}}$.
W granicy dużych odstrojeń $\Delta$ faza $\phi_{s}$ nie zależy od
$\Delta$ a jedynie od stopnia niejednorodności przestrzennego rozkładu
natężenia wiązki Starkowskiej. W zaprezentowanym eksperymencie korekcja
za pomocą SLM pozwoliła podnieść osiągalną fazę do $\phi_{s}=15$
rad. Szczegółowa analiza dekoherencji spowodowanej niejednorodnością
nadanej przestrzennej fazy znajduje się w następnym rozdziale poświęconym
przestrzennej modulacji fazy fali spinowej.

\section{Zanik i odrodzenie\label{sec:Revival-koncepcja-i-wyniki}}

\begin{figure}
\centering\includegraphics{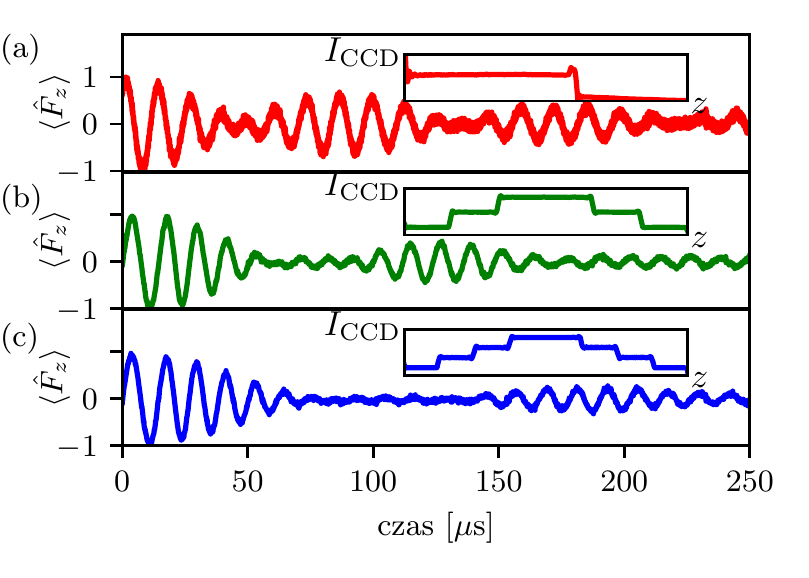}\caption{Dynamika spinu dla schodkowych profili przestrzennych natężenia wiązki
Starkowskiej. Małe wykresy wewnętrzne pokazują przestrzenny rozkład
natężenia wzdłuż osi $z$ zarejestrowany przez kamerę CCD. Zaobserwowano
dudnienia dla natężenia dwustopniowego (a) oraz zanik i odrodzenie
sygnału dla profili trzy (b) i czterostopniowego (c).\label{fig:Dynamika-spinu-}}
\end{figure}

Rysunek \ref{fig:Dynamika-spinu-} przedstawia oscylacje średniego
rzutu spinu $\langle\hat{F}_{z}\rangle(t)$ przy modulacji za pomocą
wiązki Starkowskiej o schodkowym przestrzennym rozkładzie natężenia
$I_{CCD}(z)$. Taka modulacja odpowiada podzieleniu atomów na kilka
grup z których każda ma inną częstość precesji $\omega_{L}$. Dla
dwóch stopni (a) obserwujemy dudnienia. Przy zwiększaniu liczby stopni,
okres wygaszenia oscylacji $\langle\hat{F}_{z}\rangle$ staje się
coraz wyraźniejszy. Przy czterech poziomach (c) oscylacje znikają
niemal całkowicie na okres około 80 $\mu$s. Po tym czasie pojawiają
się ponownie. Widzimy więc, że za pomocą SLM można nadać spinom w
poszczególnych miejscach przestrzeni takie fazy, by całkowite $\langle\hat{F}_{z}\rangle$
wynosiło zero. Dalsze działanie wiązki Starkowskiej może przywrócić
konstruktywną interferencję wkładów od poszczególnych fragmentów chmury.
Wyniki takiego eksperymentu jednoznacznie wykazują, że rzeczywiście
dokonaliśmy modulacji fazy precesji z rozdzielczością przestrzenną
a nie po prostu zniszczenia stanu atomów lub ich wydmuchnięcia z pułapki.

Sygnał obserwowany na fotodiodzie $S(t)$ można przepisać w postaci
analogicznej, jak w przy rozważaniu dopasowania fazowego przy odczycie
fali spinowej (sekcja \ref{sec:Dopasowanie-fazowe}). Faza nabyta
przez atomy $\phi$ jest zależna od położenia $z$ i wynosi $\phi(z)=\omega_{L}(z)t$
gdzie $\omega_{L}(z)$ jest częstością precesji w obecności modulacji.
Łatwo napisać:
\begin{equation}
S(t)\sim\intop n\mathrm{e}^{i\phi(z)}\text{d}z.\label{eq:S=00003Dcolapse_revival}
\end{equation}

W przypadku podziału atomów na kilka różnie modulowanych części uzyskamy
$S(t)\sim\sum n_{j}\exp(i\omega_{j}t)$, gdzie $n_{j}$ jeśli liczbą
atomów na grupie $j$.

\section{Podsumowanie}

W tym rozdziale przedstawiliśmy pierwszą eksperymentalną weryfikację
możliwości modulacji fazy spójności atomowej z rozdzielczością przestrzenną.
Konkretnie pokazaliśmy wprost możliwość zmieniania częstości precesji
Larmora spinu atomów $\hat{F}$ poprzez wyświetlanie obrazów z projektora
laserowego na chmurze. Do tego celu zbudowaliśmy układ umożliwiający
dowolne kształtowanie przestrzennego rozkładu natężenia wiązki światła
oświetlającego atomy. Z przedstawionych pomiarów można wyciągnąć kilka
najważniejszych wniosków:
\begin{itemize}
\item Lasery i modulatory którymi dysponujemy umożliwiły modulowanie fazy
przestrzennej precesujących spinów
\item Możliwe jest uzyskanie efektów interferencyjnych pomiędzy różnymi
fragmentami chmury - sygnał precesji można odwracalnie wygasić - co
jest zupełnie alogiczne do zniszczenia i następnie naprawienia dopasowania
fazowego
\item Maksymalna osiągalna modulacja jest ograniczona przez niejednorodności
przestrzenne wiązki Starkowskiej
\end{itemize}
Koncepcje pokazane w tym rozdziale stanowią preludium do modulacji
fazy fal spinowych w pamięci kwantowej. Rzut spinów $\langle\hat{F}_{z}\rangle$
zostanie zastąpiony przez spójność pomiędzy poziomami stanu podstawowego
$\rho_{g,h}$. Obserwowana amplituda sygnału średniego rzutu spinu
na oś $z$ wykazuje analogię z wydajnością odczytu z pamięci kwantowej,
jak to zaargumentowaliśmy w ostatnim podrozdziale. Efekt ac-Starka
wykorzystany do wytworzenia fikcyjnego pola magnetycznego może również
posłużyć do zmiany rozsunięcia poziomów energetycznych stanu podstawowego,
co po upływie czasu $t$ w konsekwencji prowadzi do nadania spójności
$\rho_{g,h}$ dodatkowej fazy. 

\chapter{Kompensacja aberracji \label{chap:Kompensacja-aberracji}}

Po zaprezentowaniu modulacji przestrzennej fazy precesujących spinów,
kolejnym krokiem jest modulacja przestrzennej fazy fal spinowych przechowywanych
w pamięci kwantowej i weryfikacja nałożonego profilu fazowego. Do
tego celu obmyśliliśmy eksperyment w którym wprowadzamy i kompensujemy
zniekształcenie fazy fali spinowej. Weryfikacja poprawności procesu
przeprowadzana jest na dwa sposoby: w polu dalekim sprawdzamy poprawność
ogniskowania poprawionej wiązki, zaś w polu bliskim dokonujemy interferencyjnego
pomiaru frontu falowego. Pomiar interferencyjny udało nam się rozszerzyć
i pokazać, że dwa kolejne odczyty z pamięci maja taką samą fazę, niezależnie
od powolnego dryfu interferometru pomiarowego.

W tym rozdziale pokazane zostanie działanie przestrzennego modulatora
fazy fali spinowej. Zaprezentowana będzie możliwość kompensacji aberracji
układu obrazującego, bezpośrednio na falach spinowych. Omówione zostaną
również zagadnienia związane z niszczeniem spójności atomowej spowodowanej
niejednorodnością nadrukowanej fazy. Wyniki te zostały opublikowane
w \cite{Lipka2019}.

Struktura rozdziału jest następująca:
\begin{itemize}
\item Sekcja \ref{sec:Interfejs-=00015Bwiat=000142o-atomy} - Omówiona jest
sekwencja przygotowująca atomy w odpowiednim stanie początkowym oraz
sekwencja pomiarowa.
\item Sekcja \ref{sec:ac-Starkowski-przestrzenny-modul} - Przedstawiona
jest budowa i działanie przestrzennego modulatora fazy fali spinowej. 
\item Sekcja \ref{sec:Charakteryzacja-dalekopolowa} - Zaprezentowane są
wyniki pomiarów wykonane za pomocą kamery dalekiego pola. Pokazana
jest kompensacja fazy wprowadzonej przez dodatkową soczewkę cylindryczną.
\item Sekcja \ref{sec:charakteryzacja-interferometrycz} - Przedstawione
są pomiary interferencyjne za pomocą kamery bliskiego pola, które
umożliwiają dokładne odtworzenie fazy nadanej falom spinowym oraz
dekoherencji spójności atomowej nią spowodowaną.
\item Sekcja \ref{sec:SSM---dekoherencja} - Wyprowadzona została formuła
określająca w jaki sposób dekoherencja fal spinowych skaluje się z
nadaną im fazą przestrzenną. Wykonane pomiary potwierdzają jej zgodność
z doświadczeniem.
\item Sekcja \ref{sec:Stabilno=00015B=000107-fazy} - Omówione jest zagadnienie
stabilności fazy fal spinowych przechowywanych w pamięci kwantowej. 
\end{itemize}

\section{Eksperyment\label{sec:Interfejs-=00015Bwiat=000142o-atomy}}

\begin{figure}

\centering\includegraphics{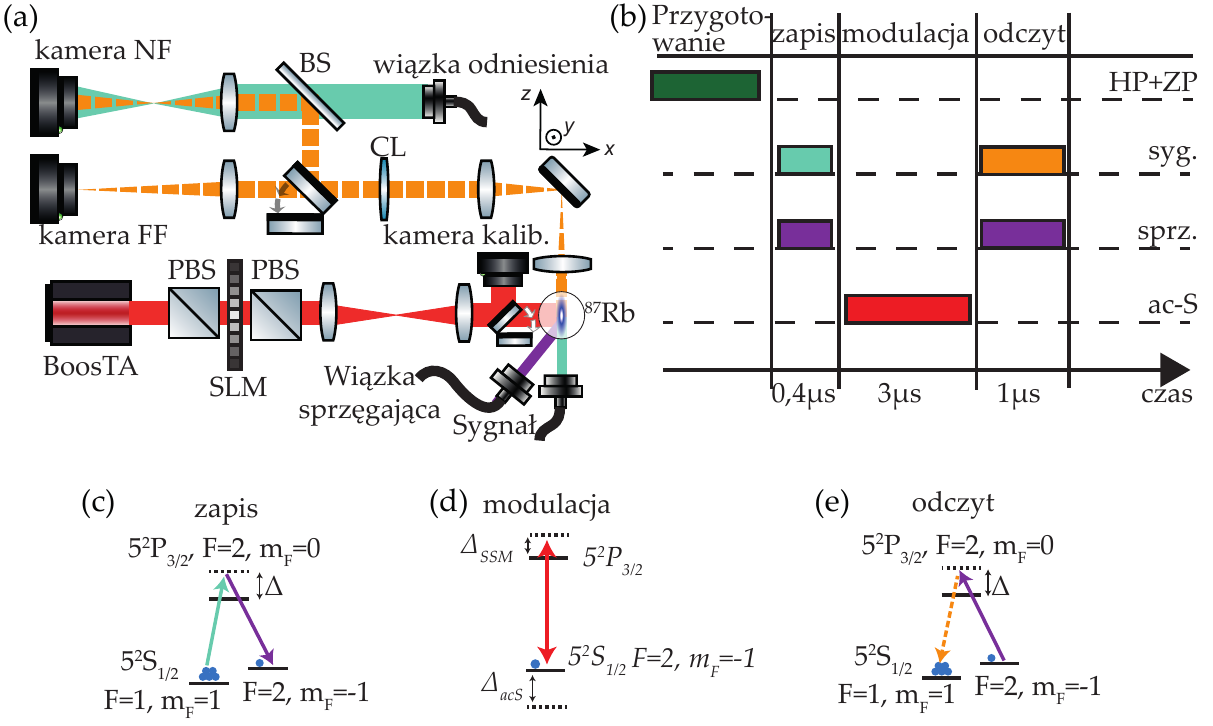}\caption{(a) Uproszczony schemat eksperymentu. Przestrzenny modulator fazy
fali spinowej składa się ze wzmacniacza laserowego (BoosTA) oraz przestrzennego
modulatora światła (SLM) skonfigurowanych do kształtowania profilu
natężenia wiązki Starkowskiej i zobrazowania jej na zespole atomów
rubidu. (b) Sekwencja czasowa impulsów używanych w eksperymencie.
(c-e) Schemat poziomów energetycznych używanych w procesie zapisu,
modulacji wiązką Starkowską oraz odczytu. Użyte skróty: NF ---bliskie
pole, FF --- dalekie pole, PBS --- polaryzująca kostka światłodzieląca,
BS --- płytka światłodzieląca, HP+ZP --- Pompy nadsubtelna i Zeemanowska.
\label{fig:a)-Uproszczona-schemat}}
\end{figure}
Do zademonstrowania modulacji przestrzennej fazy fal spinowych wykorzystujemy
wielomodową przestrzennie pamięć kwantową realizowaną w zimnym zespole
atomów $^{87}\mathrm{Rb}$ uwięzionych w pułapce magnetooptycznej
(MOT). Rysunek \ref{fig:a)-Uproszczona-schemat} koncepcyjnie przedstawia
układ eksperymentalny oraz sekwencję wykonywanych operacji. Sama pamięć
działa w układzie $\Lambda$ przedstawionym na rysunku \ref{fig:a)-Uproszczona-schemat}
(c) i (e). Wiązka sprzęgająca $\Omega_{\text{c}}$ (silne pole laserowe)
jest odstrojona o $\Delta=2\pi\cdot20\;\mathrm{MHz}$ od przejścia
$|h\rangle=5^{2}\mathrm{S}_{1/2},\;F=2,\;m_{F}=1\rightarrow|e\rangle=5^{2}\mathrm{P}_{3/2},\;F'=2,\;m_{F}=0$
natomiast słaby sygnał $\Omega_{\text{s}}$ jest z nią w rezonansie
dwufotonowym na przejściu $|h\rangle\rightarrow|g\rangle=5^{2}\mathrm{S}_{1/2},\;F=2,\;m_{F}=-1$.

Ponieważ rezonans dwufotonowy jest bardzo wąski (w praktyce tak szeroki
jak poszerzenie natężeniowe spowodowane wiązką sprzęgającą), ważne
jest by różnica częstości między wiązką sprzęgającą, a sygnałem była
stabilna. Z tego powodu aby utworzyć wiązkę sygnałową, niewielka część
lasera wiązki sprzęgającej jest podebrana i zmodulowana za pomocą
modulatora elektrooptycznego na częstości bliskiej rozszczepieniu
poziomów $|h\rangle\rightarrow|g\rangle$, syntezowanej jako $6,835\;\mathrm{GHz}$
względem generatora kwarcowego. Boczne pasmo modulacji jest izolowane
przez aktywnie stabilizowaną wnękę Fabry-Pérot \cite{Parniak2017}.
Wiązki sprzęgająca i sygnałowa są dwukrotnie uginane na modulatorach
akustooptycznych, które są niezależnie sterowane z generatorów DDS
działających na częstości około 80 MHz. Poprzez drobne przestrajanie
jednego z tych generatorów możliwe jest bardzo dokładne ustawienie
różnicy częstości między sygnałem i wiązką sprzęgającą, z precyzją
dużo lepszą niż wymaga tego czas prowadzenia przejścia dwufotonowego.
W praktyce znajdujemy rezonans dwufotonowy obserwując absorpcję wiązki
sygnałowej w chmurze atomów.

\paragraph{Sekwencja eksperymentalna.}

Rysunek \ref{fig:a)-Uproszczona-schemat} przedstawia sekwencję czasową
używanych impulsów laserowych. Na początku atomy są pompowane do stanu
$|g\rangle$ za pomocą dwóch laserów: tzw. pompy nadsubtelnej (HP)
dostrojonej do przejścia $5^{2}\mathrm{S}_{1/2},\;F=2\rightarrow5^{2}\mathrm{P}_{1/2},\;F'=2$
oświetlającej atomy z kilku stron oraz kołowo spolaryzowanej pompy
Zeemanowskiej (ZP) dostrojonej do przejścia $5^{2}\mathrm{S}_{1/2},\;F=1\rightarrow5^{2}\mathrm{P}_{3/2},\;F'=1$.
Pompa Zeemanowska propaguje się wzdłuż osi $z$, która tym sposobem
staje się osią kwantyzacji. Następnie przez około 300 ns wiązki sprzęgająca
oraz sygnałowa wytwarzają w atomach spójność $\rho_{g,h}$ pomiędzy
stanami $|g\rangle$ i $|h\rangle$. Obie wiązki maja postać fal płaskich
na obszarze chmury. Pomiędzy zapisem, a odczytem, przez okres 3 $\mu$s,
fale spinowe są modulowane za pomocą przestrzennego modulatora fazy
fali spinowej opisanego w następnym punkcie. Ponowne oświetlenie atomów
wiązką sprzęgającą powoduje konwersję spójności atomowej na sygnał
odczytu.

\paragraph{Modulator przestrzenny.\label{sec:ac-Starkowski-przestrzenny-modul}}

Przestrzenny modulator fazy fali spinowej opiera się na przestrzennie
zmiennym efekcie ac-Starka uzyskanym za pomocą silnej wiązki Starkowskiej,
odstrojonej o $\Delta_{\mathrm{SSM}}\approx1.5\;\mathrm{GHz}$ od
przejścia $|h\rangle\rightarrow5^{2}\mathrm{P}_{3/2}$ (rysunek \ref{fig:a)-Uproszczona-schemat}
(d)). Budowa modulatora jest dokładnie taka jak przedstawiona w sekcji
\ref{sec:Uk=000142ad-do=00015Bwiadczalny}, z tą różnicą, że światło
oświetlające atomy ma polaryzację liniową $\pi$ wzdłuż osi kwantyzacji
$z$ (rysunek \ref{fig:a)-Uproszczona-schemat} (a)). Indukuje ono
przestrzennie zmienne przesunięcie ac-Starka $\Delta_{\mathrm{acS}}$,
zmieniające rozsunięcie poziomów $|g\rangle$ i $|h\rangle$. Przesunięcie
zmienia tempo akumulacji fazy przechowywanej spójności atomowej $\rho_{g,h}$
. Z równań \ref{eq:stark-ogolny} znajdujących się w poprzednim rozdziale
można pokazać, że przesunięcie $\Delta_{\mathrm{acS}}$ jest proporcjonalne
do natężenia światła $I(y)$. W ten sposób można nadać falom spinowym
dodatkową fazę $\rho_{g,h}\rightarrow\rho_{g,h}\exp[i\varphi(y)]$
, gdzie $\varphi(y)\propto I(y)T$, a T jest czasem przez jaki atomy
są oświetlane przez wiązkę Starkowską \cite{DeEchaniz2008}.

\paragraph{Możliwości pomiarowe.}

Światło odczytane z atomów jest odwzorowywane przy pomocy teleskopu
o powiększeniu $M=4$ na płaszczyznę w której można wstawić soczewkę
cylindryczną o ogniskowej $f_{\mathrm{ph}}=-2000\;\mathrm{mm}$, aby
zakrzywić front falowy lub skompensować jego uprzednie zakrzywienie.
Następnie światło może biec albo do kamery obserwacji natężenia w
dalekim polu (FF), albo alternatywnie do interferometru z kamerą w
bliskim polu (NF).

\section{Charakteryzacja dalekopolowa\label{sec:Charakteryzacja-dalekopolowa}}

\begin{figure}
\centering\includegraphics{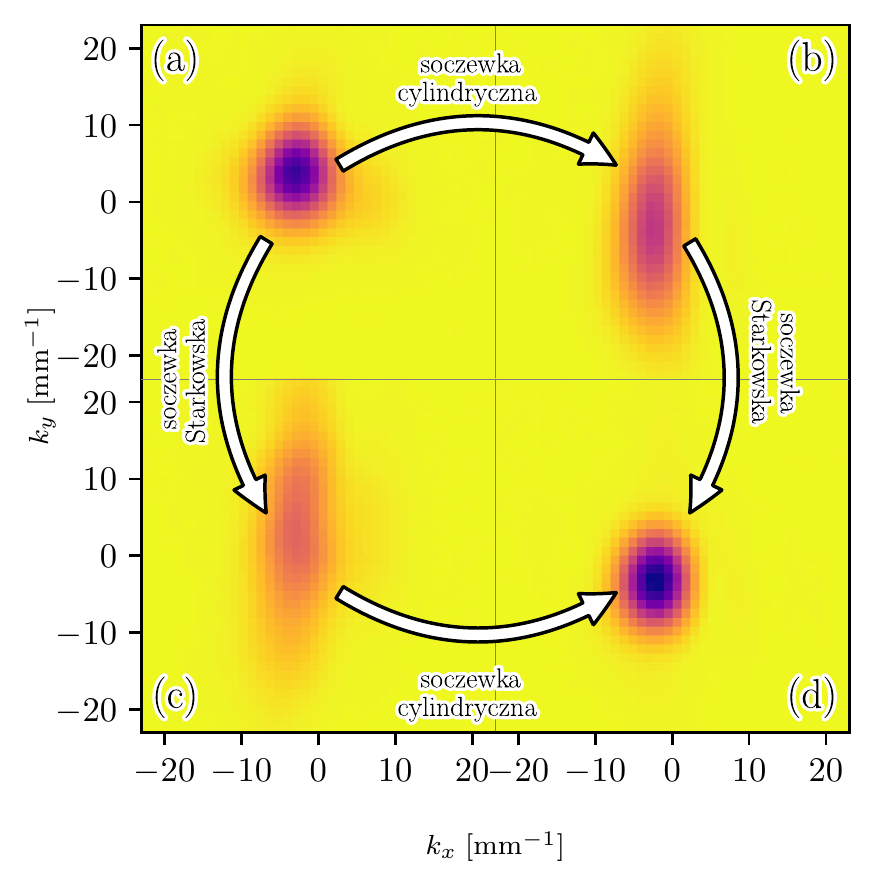}

\caption{Obraz sygnału odczytu zarejestrowany przez kamerę dalekiego pola (a)
bez modulacji, (b) z soczewką cylindryczną w układzie obrazującym,
(c) z paraboliczną fazą nałożoną na spójność atomową, (d) zarówno
z soczewką cylindryczną jak i modulacją fazy kompensującą jej efekt.
Modulacja niszczy około 20\% spójności. Zgodność kształtu (a) i (d)
mierzona jako wierność wynosi $\mathscr{F}\approx96\%$.\label{fig:Obraz-sygna=000142u-odczytu}}
\end{figure}
Eksperyment można skonfigurować na 4 podstawowe sposoby (a) bez żadnej
modulacji, (b) z soczewką cylindryczną w układzie obrazującym, (c)
z paraboliczną fazą nałożoną na spójność atomową, (d) zarówno z soczewką
cylindryczną jak i modulacją fazy kompensującą jej efekt. Przestrzenny
rozkład natężenia światła odczytywanego w dalekim polu we wszystkich
tych przypadkach przedstawia rysunek \ref{fig:Obraz-sygna=000142u-odczytu}.
Wydajność kompensacji, czyli stosunek całkowitego natężenia światła
odczytanego po zastosowaniu modulacji i soczewki do natężenia bez
nich wynosi $\eta=I_{d}/I_{a}\approx80\%$.

\paragraph{Wierność.}

Dodatkowo zdefiniować można wierność kompensacji
\begin{equation}
\mathscr{F}=\frac{\langle\sqrt{I_{d}(x',y')}\sqrt{I_{a}(x',y')}\rangle_{x',y'}}{\sqrt{\langle I_{d}(x',y')\rangle_{x',y'}\langle I_{a}(x',y')\rangle_{x',y'}}},\label{eq:fidelity}
\end{equation}
która sięga około $96\%$. Należy wspomnieć, że powyższa formuła działa
dobrze jedynie, gdy poza obszarem wiązek nie znajduje się żaden podkład,
tzn. wartość natężenia wynosi 0. W przeciwnym razie otrzymana wartość
wierności będzie zależała od wielkości obszaru na którym uśredniany
był sygnał. Aby warunek ten był spełniony została zmierzona średnia
wartość tła obrazu na kamerze, a następnie odjęta od właściwego pomiaru.
Następnie ustalony został próg, poniżej którego zmierzona wartość
natężenia była przyjmowana jako 0. Obserwowane natężenia $I_{d}(x',y')$
oraz $I_{a}(x',y')$ są przedstawione na rysunku \ref{fig:Obraz-sygna=000142u-odczytu}
(d) oraz \ref{fig:Obraz-sygna=000142u-odczytu} (a).

\begin{figure}
\centering\includegraphics{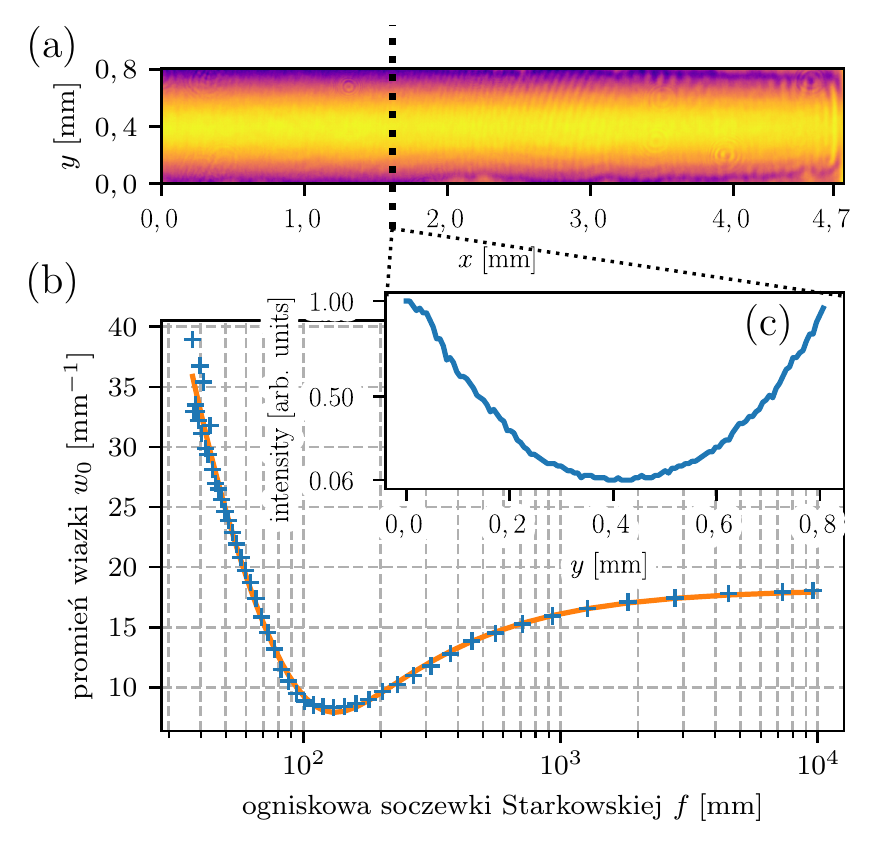}

\caption{Kompensacja soczewki cylindrycznej za pomocą modulacji Starkowskiej.
(a), (c) przedstawia przestrzenny profil natężenia wiązki modulującej,
zarejestrowany przez kamerę kalibracyjną, który jest paraboliczny
wzdłuż osi $y$ oraz stały wzdłuż osi $z$. (b) Promień wiązki $w_{0}$
sygnału odczytu zarejestrowany przez kamerę dalekiego pola w funkcji
mocy $1/f$ soczewki Starkowskiej, proporcjonalnej do natężenia wiązki
modulującej. Linia ciągła jest dopasowaniem modelu teoretycznego.\label{fig:Kompensacja-soczewki-cylindryczn}}
\end{figure}

\paragraph{Modulacja jako soczewka o zmiennej mocy.}

Aby upewnić się co do poprawności działania modulacji przeprowadziliśmy
pomiar średnicy pionowej wiązki odczytywanej w dalekim polu w funkcji
natężenia wiązki modulującej przy obecności soczewki cylindrycznej.
Minimalna średnica wiązki odczytywanej w dalekim polu jest ograniczona
przez poprzeczny rozmiar chmury atomowej $w_{sw}$. Konkretnie spodziewamy
się, że amplituda światła odczytywanego w płaszczyźnie chmury atomowej
$A(y)$ ma postać gaussowską $\exp(-y^{2}/w_{sw}^{2})$, na którą
nałożone są fazy pochodzące od soczewki $\varphi_{\mathrm{ph}}(y)$,
od modulacji $\varphi(y)$, a także tłumienie wynikające z dekoherencji
wywołanej modulacją. Kamera umieszczona w dalekim polu rejestruje
transformatę Fouriera opisanej amplitudy $\tilde{A}(k_{y})$. Można
w ten sposób obliczyć średnicę plamki rejestrowanej w dalekim polu
w funkcji mocy modulacji. Rysunek \ref{fig:Kompensacja-soczewki-cylindryczn}
przedstawia porównanie wyników eksperymentalnych z przewidywaniem
analitycznym. Stosując modulację przez czas $T=3\;\mu\mathrm{s}$,
najkrótsza uzyskana ogniskowa wynosiła $f=40\;\mathrm{mm}$.

\section{Charakteryzacja interferometryczna\label{sec:charakteryzacja-interferometrycz}}

\begin{figure}
\centering\includegraphics{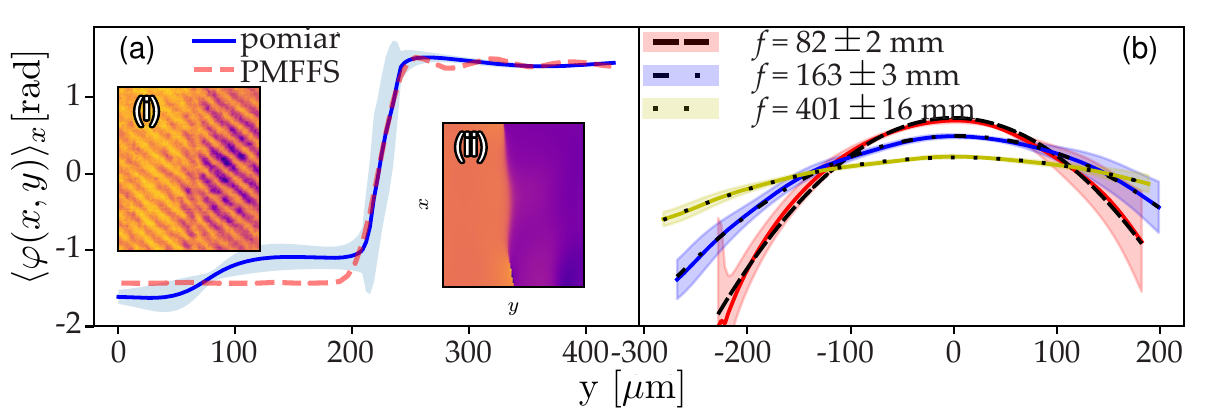}\caption{Interferencyjny pomiar fazy przestrzennej nadanej spójności atomowej.
(i) Przykładowy obraz prążków interferencyjnych zarejestrowanych przez
kamerę bliskiego pola oraz (ii) odpowiadająca mu zrekonstruowana przestrzenna
faza. (a) Faza ze schodkowym skokiem o $\pi$ rad, uśredniony wzdłuż
osi $x$. Niebieska linia jest wartością zmierzoną, natomiast przerywana
czerwona wartością spodziewaną na podstawie obrazu wiązki Starkowskiej
z kamery kalibracyjnej. (b) Paraboliczne profile fazy nadrukowane
na fale spinowe.\label{fig:Interferencyjny-pomiar-fazy}}
\end{figure}
Aby precyzyjnie scharakteryzować przestrzenną modulację fazy spójności
atomowej, sygnał odczytany z pamięci jest interferowany z dużą, skolimowaną
wiązką odniesienia, nachyloną pod kątem 22 mrad, a następnie rejestrowany
na kamerze bliskiego pola (rysunek \ref{fig:Interferencyjny-pomiar-fazy}).
Przyjmijmy, że $h(\mathbf{r_{\bot}})=|h(\mathbf{r_{\bot}})|\exp\left[i\varphi(\mathbf{r_{\bot}})\right]$
jest przestrzennym rozkładem amplitudy sygnału odczytu. Wtedy, przyjmując
amplitudę wiązki odniesienia jako $A_{0}$, natężenie rejestrowane
przez kamerę będzie wynosić
\begin{equation}
\begin{alignedat}{1}I_{\text{inter}}(\mathbf{r_{\bot}}) & =|h(\mathbf{r_{\bot}})+A_{0}e^{i\mathbf{K}_{0}\mathbf{r}_{\perp}}|^{2}\\
 & =|h(\mathbf{r_{\bot}})|^{2}+A_{0}^{2}+A_{0}|h(\mathbf{r_{\bot}})|\exp(i\mathbf{K}_{0}\cdot\mathbf{r_{\bot}}+i\varphi(\mathbf{r_{\bot}}))+c.c.
\end{alignedat}
\label{eq:Iinter=00003D+cos}
\end{equation}

\paragraph{Demodulacja obrazu.}

Amplitudę i fazę ostatniego członu można odzyskać jeśli na przekroju
$h(\mathbf{r_{\bot}})$ mamy do dyspozycji wiele prążków, a tak ustawiamy
interferometr. Wówczas transformata Fouriera zmierzonego profilu $I_{\text{inter}}(\mathbf{r_{\bot}})$
zawiera trzy wyraźnie rozdzielone składowe: dwa pierwsze człony równania
\ref{eq:Iinter=00003D+cos} są wolnozmienne i wobec tego ich transformata
będzie skupiona wokół zerowych częstości, natomiast człon interferencyjny
rozdzieli się na składową wokół $\mathbf{K}_{0}$ oraz sprzężenie
zespolone wokół $\mathbf{-K}_{0}$. Po przeprowadzeniu numerycznej
transformaty Fouriera każdego obrazu z kamery można wyciąć fragment
odpowiadający $A_{0}|h(\mathbf{r_{\bot}})|\exp(i\mathbf{K}_{0}\cdot\mathbf{r_{\bot}}+i\varphi(\mathbf{r_{\bot}}))$
i odzyskać cała tę funkcję zespoloną dokonując odwrotnej transformaty.
Tym sposobem można wyznaczyć amplitudę $2A_{0}|h(\mathbf{r_{\bot}})|$
oraz fazę sygnału odczytu $\varphi(\mathbf{r_{\bot}})$ \cite{Bone86}.
Wykonując pomiar z modulacją oraz bez niej można odzyskać fazę wprowadzaną
przez modulację niezależnie od ewentualnych aberracji układu.

\paragraph{Profil schodkowy.}

Aby zademonstrować możliwości metody i wydajność modulatora przestrzennej
fazy, wykonajmy modulację fazy z płaskim profilem wzdłuż $x$ i stopniem
o wysokości $\pi$ przy wybranym $y$ (rysunek \ref{fig:Interferencyjny-pomiar-fazy}).
Wstawka (i) przedstawia przykładowe prążki interferencyjne zarejestrowane
na kamerze dalekiego pola. Wstawka (ii) przedstawia fazę $\varphi(x,y)$
odzyskaną z $10^{4}$ zarejestrowanych klatek. Ciągła niebieska linia
na głównym wykresie przedstawia fazę uśrednioną wzdłuż osi $x$, $\langle\varphi(x,y)\rangle_{x}$,
podczas gdy obszar zacieniony na niebiesko odpowiada jego odchyleniu
standardowemu. Czerwona linia przerywana przedstawia natomiast uśredniony
wzdłuż osi $x$ profil natężenia wiązki starkowskiej, obserwowanej
przez dodatkową kamerę kalibracyjną, przeskalowaną w celu dopasowania
do obserwowanych profili fazowych. Wierność odwzorowania pomiędzy
zmierzonym $\varphi(x,y)$, a oczekiwanym profilem fazowym $\varphi_{\mathrm{0}}(x,y)$
jest zdefiniowana analogicznie do wyrażenia \ref{eq:fidelity} z $\sqrt{I}\rightarrow\varphi$
i wynosi $\mathscr{F}=98\%$. Wydajność zdefiniowana jako stosunek
całkowitej energii sygnału odczytu zmodulowanego do niezmodulowanego
wynosi $\eta=77\%$.

\paragraph{Profile cylindryczne.}

Panel (b) rysunku \ref{fig:Interferencyjny-pomiar-fazy} przedstawia
profile fazowe kilku cylindrycznych soczewek Starkowskich, uśrednionych
wzdłuż osi $x$. Obszar zacieniony odpowiada jednemu odchyleniu standardowemu.
By odzyskać ogniskową soczewki, do odzyskanego profilu fazy $\varphi(x,y)$
dopasowana została powierzchnia paraboliczna, a jej średnia wzdłuż
osi $x$ została przedstawiona przez różne typy linii przerywanych.
Niepewność odzyskanej ogniskowej uzyskano jako odchylenie standardowe
z 10 niezależnych pomiarów. Wierność odwzorowania waha się od $\mathscr{F}=95\%$
dla $f=82\;\text{mm}$ do $\mathscr{F}=98\%$ dla $f=163\;\text{mm}$
i $f=401\;\text{mm}$. Dla wszystkich ogniskowych wydajność $\eta$
pozostała powyżej $80\%$.

\section{Zanik dopasowana fazowego\label{sec:SSM---dekoherencja}}

\begin{figure}
\centering\includegraphics{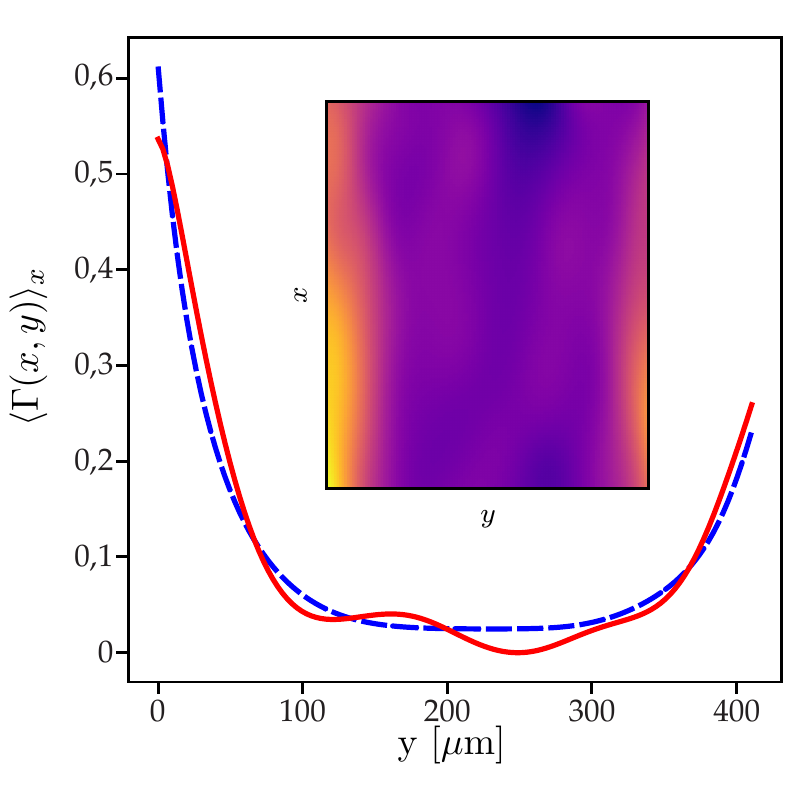}

\caption{Obserwowana interferencyjnie dekoherencja fal spinowych. Wstawka przedstawia
stosunek amplitud sygnału odczytu zmodulowanego $h(x,y)$ oraz niezmodulowanego
$h_{0}(x,y)$. Wykres główny przedstawia obliczony czynnik zaniku
$2\Gamma=\ln h/h_{0}$ (czerwona linia ciągła) oraz dopasowanie funkcji
proporcjonalnej do kwadratu nałożonej fazy $\gamma\varphi(y)^{2}$
(niebieska linia przerywana)\label{fig:Dekoherencja-fal-spinowych}}
\end{figure}

Charakteryzacja interferometryczna umożliwia dokładniejsze zbadanie
zagadnienia osłabienia natężenia światła odczytywanego po dokonaniu
modulacji fal spinowych. Eksperyment pokazał, że zanik ten skaluje
się z kwadratem nałożonej fazy. Okazuje się, że można to całkowicie
wyjaśnić jako utratę dopasowania fazowego wywołaną przez niejednorodne
oświetlenie modulujące.

Podobnie jak w poprzednim rozdziale (sekcja \ref{sec:Revival-koncepcja-i-wyniki})
zapiszmy sygnał odczytywany z pamięci w postaci analogicznej jak w
przy rozważaniu dopasowania fazowego przy odczycie fali spinowej (sekcja
\ref{sec:Dopasowanie-fazowe}). Faza nabyta przez atomy jest zależna
od położenia $z$ i wynosi $\varphi(x,y,z)=\alpha TI(y,z)$, gdzie
$\alpha$ jest czynnikiem proporcjonalności, a $T$ jest czasem przez
jaki wiązka Starkowska oddziałuje na atomy. Amplituda w wybranym punkcie
kamery $A(x,y)$ stanowi sumę wkładów od poszczególnych plasterków
chmury atomowej:
\begin{equation}
A(x,y)\propto\intop n\mathrm{e}^{i\varphi(x,y,z)}\text{d}z.\label{eq:4:S=00003Ddecoh}
\end{equation}

Przy doskonale działającym oświetleniu laserowym rozkład natężenia
wiązki Starkowskiej $I$ oświetlającej atomy powinien w rozważanym
eksperymencie zależeć jedynie od $y$, oraz nie zależeć od $z$.\footnote{Oś $x$ jest ustawiona wzdłuż biegu wiązki modulującej i na grubości
chmury \textasciitilde 0,5mm zaniedbujemy wszelką zależność natężenia} Jednakże, w rzeczywistości natężenie to ma pewne odstępstwa od tego
ideału $I(y,z)=I_{0}(y)+\Delta I(z,y)$. W efekcie dla każdej wysokości
$y$ na obrazie wyjściowym będziemy uzyskiwać sumę wkładów od atomów
które nie są ze sobą zgodne w fazie. Przyjmijmy, że odstępstwa $\Delta I(z,y)$
są losowe, gaussowskie i o średniej zero. Założenie o gaussowkim szumie
potwierdza obraz wiązki Starkowskiej zarejestrowanej na kamerze kalibracyjnej.
Co więcej odchylenie standardowe szumu jest proporcjonalne do lokalnego
natężenia światła w stosunku $\sigma_{I}/I_{0}\approx6\%$. Obserwowaną
zależność można zrozumieć jako efekt interferencji właściwej wiązki
zmodulowanej z przypadkowymi rozproszeniami tej wiązki na różnych
elementach układu. Ponownie spójrzmy na równanie \ref{eq:Iinter=00003D+cos}
- niech $h$ gra przez chwilę rolę rozproszeń. Rozproszenia o bardzo
małym udziale energetycznym $|h|^{2}\simeq0,4\%|A_{0}|^{2}$ wystarczą,
aby spowodować obserwowane efekty interferencyjne $hA_{0}^{*}\simeq6\%|A_{0}|^{2}$.
Rozproszenia od dowolnych elementów metalowych, krawędzi szkła itp.
dają fale padające na chmurę pod bardzo dużym katem w stosunku do
wiązki zasadniczej, czyli bardzo geste prążki. Mieszanina takich prążków
w czasie kilku mikrosekund jest nieruchoma i stanowi omawiany szum.

Wykorzystajmy wzór \ref{eq:4:S=00003Ddecoh} aby obliczyć natężenie
na kamerze $|A(x,y)|^{2}$. Ponieważ szum natężenia $\Delta I(z,y)$
zmienia się na odległościach dużo mniejszych niż długość chmury $L$,
to całkowanie wzdłuż $z$ możemy zastąpić uśrednianiem po rozkładzie
prawdopodobieństwa wystąpienia odchylenia: 
\[
A(x,y)\propto\exp\left[i\alpha TI_{0}(y)\right]\left\langle \exp(i\alpha T\Delta I)\right\rangle =\exp\left[i\alpha TI_{0}(y)\right]\exp\left(-\frac{\alpha^{2}T^{2}\sigma_{I}^{2}}{2}\right).
\]

Warto zwrócić uwagę na to, że o ile faza nałożona na spójność atomową
rośnie liniowo z energią impulsu modulującego $\varphi(x,y,z)=\alpha TI(y,z)$,
to zanik dopasowana fazowego $\alpha^{2}T^{2}\sigma_{I}^{2}$ skaluje
się już kwadratowo. Natężenie odczytu spada jak $\exp[-\gamma\varphi^{2}]$
z $\gamma=(\sigma_{I}/I_{0})^{2}$.

Kwadratowa zależność wykładnika zaniku od fazy została potwierdzona
przez pomiar interferencyjny opisany w poprzedniej sekcji. Pomiar
ten umożliwia jednoczesne odzyskanie nadrukowanej fazy oraz przestrzennego
rozkładu amplitudy sygnału odczytu $h(x,y)$. Porównując sygnały odczytu
z modulacją $h(x,y)$ oraz bez modulacji $h_{0}(x,y)$, można odzyskać
$2\Gamma=\ln h/h_{0}$. Rysunek \ref{fig:Dekoherencja-fal-spinowych}
przedstawia wielkość czynnika $\Gamma$(y), uśrednionego wzdłuż osi
$x$, dla Soczewki Starkowskiej o ogniskowej $f=82\pm2$ mm (profil
fazy przedstawiony na rysunku \ref{fig:Interferencyjny-pomiar-fazy}
(b)). Z dopasowania otrzymaliśmy $\gamma=(4,2\pm1,4)\cdot10^{-2}$.
Odpowiada to względnemu odchyleniu standardowemu szumu natężenia na
poziomie $\sigma_{I}/I_{0}=(29\pm5)\%$. Wynik ten jest znacznie wyższy,
niż to wynika z bezpośrednich pomiarów na kamerze kalibracyjnej (około
6\%). Jest to spowodowane prawdopodobnie zbyt niską rozdzielczością
kamery i precyzją jej ustawienia, co utrudnia obserwowanie szybko
zmiennych profili, takich jak gęste plamki.\footnote{Analogiczny pomiar przeprowadzony w najnowszym układzie optycznym
wykazał już dobrą zgodność.} Dekoherencja spowodowana utratą dopasowania fazowego na skutek losowej
modulacji fazy przez przypadkowy szum na obrazie optycznym jest główną
przyczyną dekoherencji fal spinowych przechowywanych w pamięci, znacznie
istotniejszą niż absorpcja wiązki Starkowskiej, czy poszerzenie natężeniowe.

\section{Stabilność fazy\label{sec:Stabilno=00015B=000107-fazy}}

\begin{figure}
\centering\includegraphics{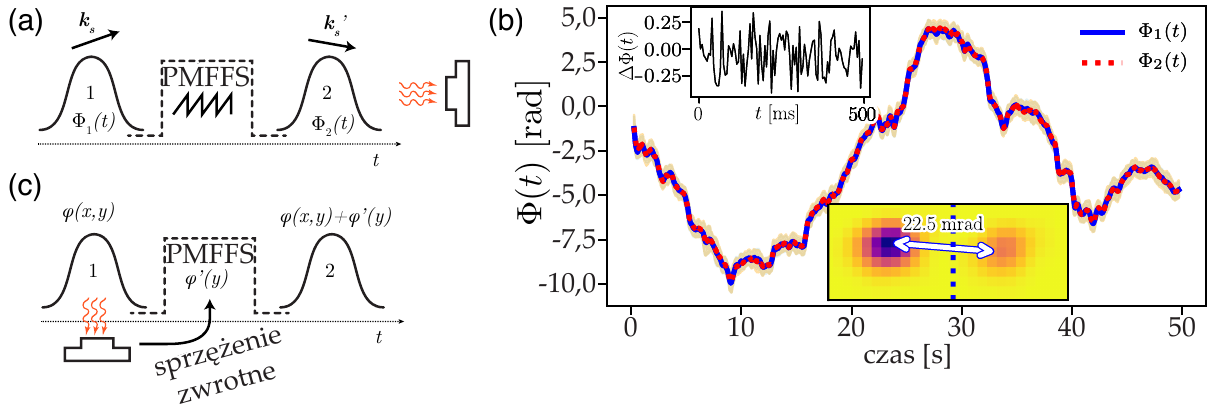}

\caption{(a), (b) Pomiar stabilności fazy pomiędzy dwoma sygnałami odczytu
wykonanymi w różnych momentach czasu. (a) Sekwencja operacji na pamięci.
Drugi odczyt jest przesunięty w przestrzeni fourierowskiej poprzez
nałożenia na spójność atomową dodatkowej liniowej fazy. (b) Odzyskana
globalna faza sygnałów odczytu ($\Phi_{1}$(t) i $\Phi_{2}(t)$).
Górna wstawka przedstawia różnicę faz $\Delta\Phi(t)=\Phi_{1}(t)-\Phi_{2}(t)$
dla pierwszych 500 ms pomiaru. Dolna wstawka przedstawia natomiast
moduł transformacji Fouriera obrazu zarejestrowanego na kamerze. Dwa
widoczne maksima odpowiadają odpowiednio sygnałowi z pierwszego i
drugiego odczytu. (c) Idea pomiarów adaptacyjnych z pętlą sprzężenia
zwrotnego. Modulacja przestrzennej fazy przed drugim pomiarem może
zależeć od wyniku pierwszego.\label{fig:(a),-(b)-Pomiar}}

\end{figure}
Spójność atomowa może być odczytana jako kilka czasowo oddzielonych
od siebie impulsów świetlnych poprzez zastosowanie serii krótkich
impulsów wiązki sprzęgającej. Dzięki modulacji możliwa jest zmiana
fazy przestrzennej kolejnych odczytów. Modulacja $n$-tego impulsu
może zależeć od wyników wszystkich poprzednich pomiarów (np. poprzez
pętlę sprzężenia zwrotnego). W ten sposób, modulacja fazy przestrzennej
może być stosowana do realizacji pomiarów adaptacyjnych z rozdzielczością
przestrzenną. Idea takiego pomiaru została pokazana na rysunku \ref{fig:(a),-(b)-Pomiar}
(c). Można sobie wyobrazić dokonanie w czasie rzeczywistym takiej
modulacji, która spowoduje przejście od pomiaru w bliskim polu do
pomiaru w polu dalekim. W przypadku gdyby emitowane światło podlegało
detekcji homodynowej, możnaby w każdym punkcie zmieniać podlegającą
detekcji kwadraturę. Otwierają się zatem nowe możliwości w optycznych
protokołach komunikacji i przetwarzania informacji.

Rysunek \ref{fig:(a),-(b)-Pomiar} (a) przedstawia schemat zastosowanej
sekwencji pomiarowej. Pierwszy odczyt wykonywany jest po 1 $\mu$s
przechowywania w pamięci przy użyciu krótkiego impulsu wiązki sprzęgającej
o długości 200 ns. Dalej, na spójność atomową zostaje nadrukowana
liniowa faza Fresnelowska (profil piłokształtny z zębami o wysokości
2$\pi$ rad), która przesuwa pozostałą spójność w domenie Fourierowskiej.
Jest ona odczytywana po 10 $\mu$s. Dzięki modulacji przestrzennej
fazy spójności atomowej odczyty są rozdzielone przestrzennie w polu
dalekim. Oba odczytywane impulsy interferują z wiązką odniesienia,
a prążki interferencyjne są rejestrowane na pojedynczej klatce kamery
bliskiego pola.

Stosując analizę fourierowską, odczytane sygnały mogą zostać łatwo
odseparowane (rysunek \ref{fig:(a),-(b)-Pomiar} (b)). Dla każdego
z nich odzyskiwana była faza globalna $\Phi(t)$, metodą opisaną w
sekcji \ref{sec:charakteryzacja-interferometrycz}. Wstawka w lewym
górnym rogu rysunku \ref{fig:(a),-(b)-Pomiar} (b) przedstawia ich
różnicę $\Delta\Phi(t)=\Phi_{1}(t)-\Phi_{2}(t)$ przez pierwsze 500
ms. Średnie odchylenie standardowe $\Delta\Phi(t)$ wynosi około 0,2
rad. Może to wynikać z niestabilności fazy pomiędzy wiązką sprzęgającą,
a sygnałem. Sygnał jest wytwarzany poprzez zmodulowanie lasera wiązki
sprzęgającej za pomocą modulatora elektrooptycznego na częstości 6,8
GHz, a następnie odfiltrowanie odpowiedniego pasma za pomocą aktywnie
stabilizowanej wnęki Fabry-Pérota. Zbyt wolna pętla sprzężenia zwrotnego
powoduje oscylacje częstości rezonansowej wnęki. Wiedząc, że finezja
wnęki wynosi 100, a zakres między kolejnymi jej rezonansami wynosi
10 GHz, można obliczyć, że każdy MHz odsunięcia od rezonansu wnęki
nadaje wiązce dodatkową fazę 20 mrad. Obserwując sygnał błędu użyty
do stabilizacji wnęki, oszacowaliśmy, że jej niestabilność powinna
generować średnie odchylenie standardowe $\Delta\Phi(t)$ na poziomie
około 0,1 rad. Uwzględniając to, że wiązka sprzęgająca, sygnał jak
i wiązka odniesienia muszą przepropagować się przez 10 m światłowodu,
odchylenie to może dodatkowo wzrosnąć do zmierzonego poziomu 0,2 rad.

\section{Podsumowanie}

W rozdziale tym opisana została technika modulacji przestrzennej fazy
fali spinowej przy pomocy efektu ac-Starka. Zademonstrowane zostało
nakładanie na sygnał odczytu dowolnie wybranej, jednowymiarowej przestrzennej
fazy. Najważniejsze wyniki przedstawione w tym rozdziale to:
\begin{itemize}
\item Demonstracja kompensacji fazy nałożonej na sygnał odczytu za pomocą
soczewki cylindrycznej z wiernością $\mathcal{\mathscr{F}}=95\%$
oraz wydajnością definiowaną jako stosunek mocy sygnału kompensowanego
do niemodulowanego na poziomie 80\%,
\item Interferencyjny pomiar nadrukowanej fazy. Wierność pomiędzy fazą zmierzoną,
a żądaną na poziomie powyżej 95\%.
\item Wykorzystanie analizy dopasowania fazowego do potwierdzenia, że najważniejszą
przyczynę dekoherencji jest ziarnista strukturą obrazu wyświetlanego
na chmurze atomów
\item Demonstracja dwuetapowego odczytu z modulacją fazy pomiędzy etapami.
Zmierzona została stabilność fazy pomiędzy odczytami oddzielonymi
czasowo o 10 $\mu$s. Średnie odchylenie standardowe różnicy faz między
poszczególnymi odczytami $\Delta\Phi(t)$ wynosiło 0,2 rad i jej głównym
źródłem były zewnętrzne elementy układu eksperymentalnego.
\end{itemize}
Przedstawione wyniki dają obiecującą perspektywę na dalsze zastosowania
przestrzennej modulacji fazy. Wysoka wierność nakładania fazy na fale
spinowe umożliwia stworzenie elementu optycznego z kontrolowalnym
i dowolnym profilem fazowym. Może mieć to zastosowanie w kompensacji
aberracji w układzie obrazującym, co wpływa pozytywnie na liczbę modów
dostępnych w pamięci kwantowej.

Dzięki przestrzennej modulacji fazy możliwe jest w umieszczenie w
układzie dodatkowej soczewki o dowolnej ogniskowej. Można to wykorzystać
m.in do szybkiego zmienienia układu obrazującego pomiędzy blisko,
a dalekopolowym. Aktywny wybór bazy pomiarowej jest niezbędny w protokołach
komunikacji kwantowej opartej na paradoksie Einsteina-Podolskiego-Rosena
\cite{Edgar2012,Aspden2013,Dabrowski2018}.

Zademonstrowany wieloetapowy odczyt z pamięci kwantowej może pozwolić
w dalszej przyszłości na tworzenie sekwencji ze sprzężeniem zwrotnym,
w którym modulacja fazy fali spinowej zależy od wyniku wcześniejszego
pomiaru. Otwiera to nowe możliwości prowadzenia pomiarów adaptatywnych
\cite{Wiseman2009} a także superaddytywnych \cite{Klimek2016}.

Ponieważ modulacja fazy opiera się na rozsuwaniu poziomów energetycznych,
pomiędzy którymi przechowywana jest spójność atomowa, przedstawiony
schemat można adaptować, poprzez zmianę długości fali lasera modulującego,
do innych ośrodków materialnych np. na ciepłych par atomowych \cite{Reim2010,Hosseini2011,Bashkansky2012}
(konieczne znaczne zwiększenie odstrojenia i mocy), centrów barwnych
w diamencie \cite{Maurer2012,Tsukanov2013,Pfender2017}, czy domieszek
metali ziem rzadkich w ciele stałym \cite{Saglamyurek2011,DeRiedmatten2008}

Modulacja fazy fal spinowych wzdłuż osi kwantyzacji pozwala na modyfikowanie
struktury czasowej sygnału odczytu. Takie użycie modulacji przestrzennej
fazy fal spinowych zostanie zaprezentowane w następnym rozdziale.

\chapter{Czasowo-częstotliwościowa transformacja Fouriera dla światła\label{chap:Czasowo-cz=000119stotliwosciowa-transf}}

W toku prowadzonych badań okazało się, że naturalnym rozwinięciem
koncepcji soczewki przedstawionej w poprzednim rozdziale są soczewki
czasowe, które można zrealizować przy wykorzystaniu protokołów GEM
(gradient echo memory). Soczewkę można wykorzystać do realizacji przestrzennej
transformacji Fouriera pola optycznego. Naturalnie wobec tego nasuwa
się użycie soczewki czasowej do konstrukcji spektrometru. Realizacja
spektrometru w pamięci kwantowej okazuje się skutkować nadzwyczajną
rozdzielczością, z którą konkurować mogą jedynie bardzo dobre wnęki.
Aby zrozumieć przyczynę tego faktu przypomnijmy sobie, że rozdzielczość
spektrometru $\delta_{\omega}$ jest ograniczona maksymalnym opóźnieniem
$\tau$ pomiędzy najkrótszą, a najdłuższą drogą optyczną jaka się
w nim realizuje (skrajne pozycje na siatce, skrajne pozycje lustra
w interferometrze lub finezja wnęki razy jej długość), $\delta_{\omega}=1/\tau$.
W przypadku pamięci kwantowej opóźnienie $\tau$ jest czasem życia
pamięci, ograniczonym ostatecznie w naszym przypadku termicznym ruchem
atomów. W opracowanym protokole do pamięci zapisuje się badany impuls
świetlny, a po przetworzeniu odczytujemy po kolei poszczególne jego
częstości.

\begin{figure}
\centering\includegraphics{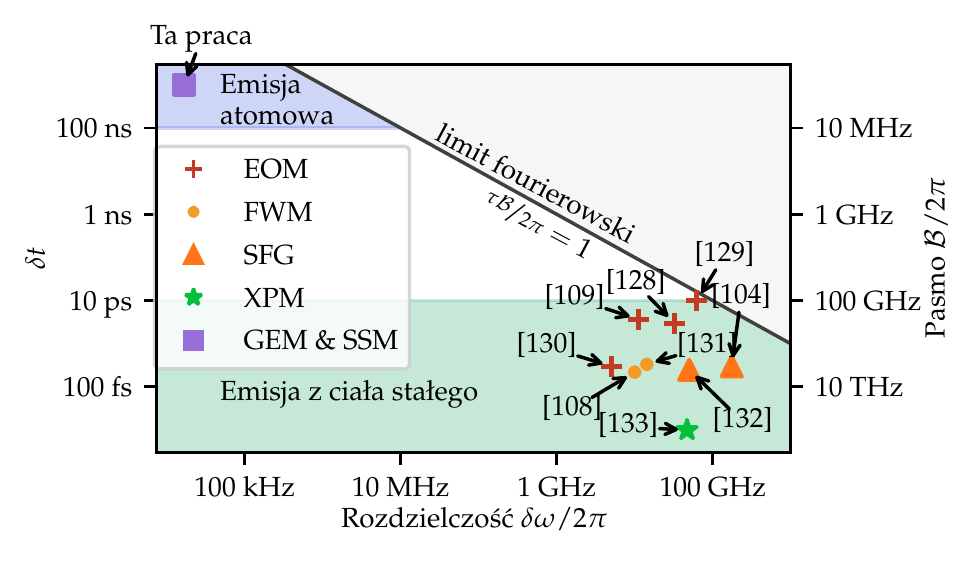}

\caption{Istniejące schematy obrazowania w domenie czasowej scharakteryzowane
przez rozdzielczość spektralną $\delta_{\omega}$ oraz dostępne pasmo
pomiarowe $\mathcal{B}=1/\delta_{t}$. Wiele implementacji bazujących
na ciele stałym (EOM - modulator elektrooptyczny \cite{foster_ultrafast_2009,Kauffman1994,Azana2004,Babashah2019},
FWM - mieszanie czterech fal \cite{foster_silicon-chip-based_2008,Salem2009},
SFG - generacja sumy częstości \cite{bennett_upconversion_1999,Suret2016},
XPM - skrośna modulacja fazy \cite{Mouradian2000}) jest dobrze dopasowanych
do impulsów piko- a nawet femtosekundowych o szerokim widmie, osiągając
rozdzielczość nie lepszą niż 1 GHz z ilością niezależnych punktów
pomiarowych$\mathcal{B}/\delta_{\omega}$ na poziomie $2\pi\cdot2000$.
Nasz system (GEM \& SSM) osiąga rozdzielczość $\delta_{\omega}\sim2\pi\times20$
KHz, czyli 6 rzędów wielkości lepszą, przy wciąż dobrym $\mathcal{B}/\delta_{\omega}$.
Wyszarzony region oznacza niefizyczny obszar w którym $\mathcal{B}/\delta_{\omega}<2\pi$.\label{fig:Istniej=000105ce-schematy-obrazowania}}

\end{figure}

Protokół zachowuje relacje fazowe i jest odwracalny - może służyć
do syntezy impulsu o zadanym widmie z podanych po kolei składników
spektralnych. Najkrótszy impuls jaki możnaby uzyskać poprzez spójne
dodanie częstości z pełnego pasma spektrometru $\mathcal{B}$ będzie
miał długość $\delta_{t}\sim1/\mathcal{B}$. Wobec tego dla spektrometru
który potrafi rozdzielić pasmo na $\mathcal{B}/\delta_{\omega}$ punktów
pomiarowych możemy liczbę punktów przepisać na wiele sposobów$\mathcal{B}/\delta_{\omega}=\mathcal{\tau B}=1/(\delta_{t}\delta_{\omega})$.
Rysunek \ref{fig:Istniej=000105ce-schematy-obrazowania} przedstawia
porównanie naszego spektrometru z istniejącymi implementacjami pod
względem dostępnego pasma częstości $\mathcal{B}$, z którego spektrometr
przetwarza światło oraz rozdzielczości spektralnej $\delta_{\omega}$.

Struktura rozdziału jest następująca:
\begin{itemize}
\item Sekcja \ref{sec:Teoria-funkcje-Wignera} - Wprowadzona zostanie funkcja
Wignera, która umożliwia przedstawienie modulacji czasu i częstości
sygnału świetlnego w języku optyki geometrycznej.
\item Sekcja \ref{sec:Uk=000142ad-obrazuj=000105cy-w} - Przedstawiona zostanie
koncepcja soczewki oraz propagacji czasowej.
\item Sekcja \ref{sec:Spektrometr} - Zaprezentowana zostanie idea spektrometru
oraz ograniczenia jego rozdzielczości i pasma wynikające bezpośrednio
z rozmiaru chmury atomowej oraz oraz natężenia wiązki sprzęgającej.
\item Sekcja \ref{sec:Eksperyment} - Omówiona zostanie budowa układu eksperymentalnego
oraz sekwencja pomiarowa. Zestawione ze sobą zostaną również wyniki
pomiarów wraz z wynikami symulacji numerycznych na bazie równań Maxwella-Blocha
wyprowadzonych w rozdziale \ref{chap:interfejs-=00015Bwiat=000142o-atomy}. 
\item Sekcja \ref{sec:Rozdzielczo=00015B=000107-i-pasmo,} - Pokazuje zależność
wydajności spektrometru od jego rozdzielczości oraz dostępnego pasma.
\end{itemize}
Wyniki rozdziału zostały opublikowane w pracy \cite{Mazelanik2020}.

\section{Teoria-funkcje Wignera \label{sec:Teoria-funkcje-Wignera}}

Funkcja Wignera jest kwazirozkładem prawdopodobieństwa w przestrzeni
fazowej położeń i pędów \cite{Wigner1932}. Dla naszych celów użyjemy
funkcji Wignera dla obwiedni pola $A(x)$, którą można zapisać w postaci
\begin{equation}
\begin{alignedat}{2}W(x,k) & = &  & \frac{1}{2\pi}\intop_{-\infty}^{\infty}A(x+\frac{y}{2})A^{*}(x-\frac{y}{2})\mathrm{e}^{-iky}\mathrm{d}y\\
 & = &  & \frac{1}{2\pi}\intop_{-\infty}^{\infty}\widetilde{A}(k-\frac{k'}{2})\widetilde{A}^{*}(k+\frac{k'}{2})\mathrm{e}^{ik'x}\mathrm{d}k'.
\end{alignedat}
\label{eq:wigner}
\end{equation}
Funkcja Wignera jest znormalizowana jak rozkład prawdopodobieństwa
$\intop_{-\infty}^{\infty}W(x,k)\mathrm{d}k\mathrm{d}x=1$. Jej marginale
są rozkładami natężenia w położeniach (bliskopolowy) i pędach (dalekopolowy):
\begin{equation}
\begin{gathered}\intop_{-\infty}^{\infty}W(x,k)\mathrm{d}k=|A(x)|^{2},\\
\intop_{-\infty}^{\infty}W(x,k)\mathrm{d}x=|\widetilde{A}(k)|^{2}.
\end{gathered}
\end{equation}
Oznacza to, że gdy funkcja Wignera przyjmuje wartości dodatnie, może
być interpretowana jako rozkład prawdopodobieństwa w przestrzeni położeń
i kierunków, stanowiąc analogię do zespołu promieni optyki geometrycznej
o określonym rozkładzie położeń jak i kierunków.

Używając funkcji Wignera ewolucję wiązki w układzie optycznym można
przedstawić za pomocą narzędzi optyki geometrycznej. W przypadku soczewki
o ogniskowej $f$, która w przestrzeni położeń nadaje wiązce dodatkową
fazę kwadratową $A'(x)=A(x)\exp\left(-ik_{0}/(2f)x^{2}\right)$ można
z równania \ref{eq:wigner} obliczyć transformacje, której podlega
funkcja Wignera $W(x,k)\rightarrow W'(x,k)$. Funkcja Wignera pola
za soczewką $W'(x,k)$ może być przedstawiona jako początkowa funkcja
Wignera w nowych współrzędnych $W'(x,k)=W(x',k')$. Transformacja
współrzędnych jest nam znana z kursu optyki geometrycznej:
\begin{equation}
\begin{gathered}\left(\begin{array}{c}
x'\\
\frac{k'}{k_{0}}
\end{array}\right)=\left(\begin{array}{cc}
1 & 0\\
-\frac{1}{f} & 1
\end{array}\right)\left(\begin{array}{c}
x\\
\frac{k}{k_{0}}
\end{array}\right)\end{gathered}
\end{equation}

Podobnie w przypadku propagacji o odległość d, która zgodnie z równaniem
\ref{eq:ewolucja pola} nadaje obwiedni dodatkową fazę kwadratową
$\widetilde{A}'(k)=\widetilde{A}(k)\exp\left(-id/(2k_{0})k^{2}\right)$
w przestrzeni fourierowskiej
\begin{equation}
\begin{gathered}W(x,k)\xrightarrow[\textrm{propagacja}]{\widetilde{A}(k)\rightarrow\widetilde{A}(k)\mathrm{e^{-i\frac{d}{2k_{0}}k^{2}}}}W'(x,k)=W(x',k')\\
\downarrow\\
\left(\begin{array}{c}
x'\\
\frac{k'}{k_{0}}
\end{array}\right)=\left(\begin{array}{cc}
1 & d\\
0 & 1
\end{array}\right)\left(\begin{array}{c}
x\\
\frac{k}{k_{0}}
\end{array}\right)
\end{gathered}
\end{equation}
Analogicznie do ewolucji wiązki przestrzeni położeń i wektorów falowych,
funkcji Wignera można użyć do opisu światła w domenie czasowej i częstościowej.
Wtedy działanie soczewki czasowej o ogniskowej $f_{t}$ sprowadza
się do nałożenia na światło czasowej fazy $\exp\left(-i\omega_{0}/(2f_{t})t^{2}\right)$,
gdzie $\omega_{0}$ to jego częstość nośna, a propagacja o odcinek
czasowy $d_{t}$ jest związana z nałożeniem dodatkowej fazy spektralnej
$\exp\left(-id_{t}/(2\omega_{0})\omega^{2}\right)$

\section{Układ obrazujący w domenie czasowej \label{sec:Uk=000142ad-obrazuj=000105cy-w}}

\subsection{Pamięć gradientowa\label{subsec:Pami=000119=000107-gradientowa}}

\begin{figure}
\centering\includegraphics{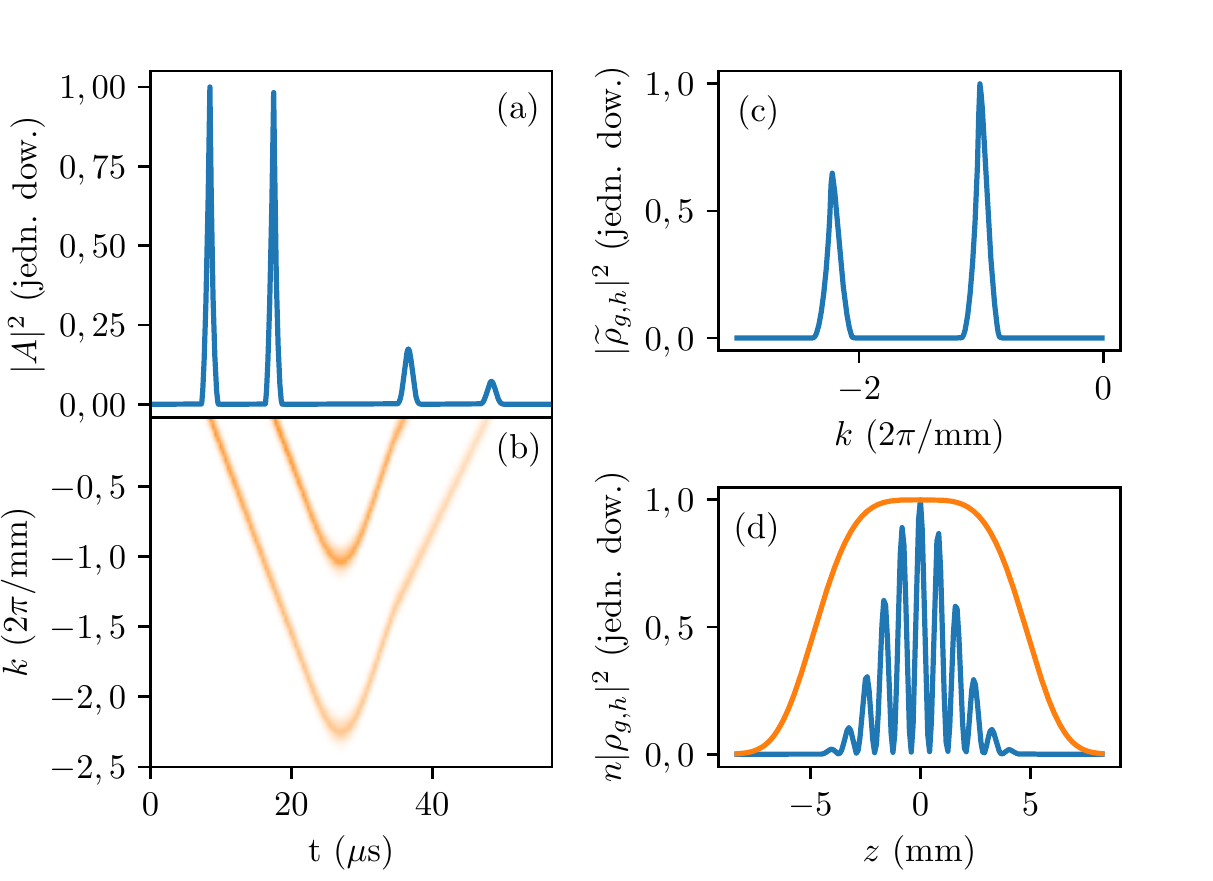}\caption{Symulacja pamięci GEM. Do 25 $\mu$s następuje proces zapisu do pamięci.
W 25 $\mu$s następuje odwrócenie gradientu pola magnetycznego i rozpoczyna
się proces odczytu. (a) Natężenie sygnału zapisywanego i odczytywanego
z pamięci GEM w funkcji czasu. (b) Diagram modułu kwadrat spójności
atomowej $|\widetilde{\rho}_{g,h}|^{2}$ w przestrzeni fourierowskiej
w funkcji czasu. (c) Moduł kwadrat spójności atomowej $|\widetilde{\rho}_{g,h}|^{2}$
w przestrzeni fourierowskiej w 25 $\mu$s. (d) N liczba wzbudzeń przechowywanych
w atomach w 25 $\mu$s w funkcji $z$ (niebieska linia). Pomarańczowa
linia przedstawia rozkład koncentracji atomowej w $z$. \label{fig:Symulacja-pami=000119ci-GEM.}}

\end{figure}

Pamięć gradientowa (gradient echo memory, GEM) jest pamięcią kwantową,
w której dodatkowo używa się cewek generujących jednorodny przestrzennie
gradient pola magnetycznego wzdłuż chmury atomowej $B=B_{0}+\beta z/\mu_{B}$\cite{Sparkes2013}.
Sprawia to, że przesunięcie Zeemanowskie, a co za tym idzie przerwa
energetyczna pomiędzy stanami $|g\rangle$ i $|h\rangle$, gdzie przechowywane
są fale spinowe, zależy liniowo od położenia na osi $z$. W równaniu
ewolucji spójności $\rho_{g,h}$ \ref{eq:spojnosc} pojawia się liniowo
zmienne w $z$ odstrojenie dwufotonowe $\delta(z)=\beta z$. Dla różnych
warstw atomów $\rho_{g,h}(z)$ rezonans dwufotonowy zachodzi z różnymi
składowymi $\widetilde{A}(\omega)$ słabego pola. Dzięki temu pamięć
odwzorowuje poszczególne składowe spektralne przychodzącego pola $\widetilde{A}(\omega)$
w różnych miejsca $z=(\omega-\omega_{0})/\beta$.

Równoważny wniosek można uzyskać analizując proces zapisu w czasie.
Ewolucja w czasie spójności wywołana gradientem pola $\dot{\rho}_{g,h}=i\beta z\rho_{g,h}$
może być przepisana jako równanie przesuwania w przestrzeni pędów
$\dot{\widetilde{\rho}_{g,h}}(K)=\beta\partial_{z}\widetilde{\rho}_{g,h}(K)$.
Zapis krótkiego impulsu powoduje powstanie fali spinowej o zerowym
wektorze falowym $K$ (por. równanie \ref{eq:rho=00003Dexp(ikr)}).
Dlatego przebieg czasowy pola $A(t)$ odwzoruje się jako rozkład amplitud
spójności w przestrzeni pędów $\widetilde{\rho}_{g,h}(K=\beta t)\sim A(t)$.
Transformując tą relacje obustronnie odzyskujemy zależność z poprzedniego
paragrafu.

Rysunek \ref{fig:Symulacja-pami=000119ci-GEM.} przedstawia symulację
działania pamięci GEM. Przez cały okres działania pamięci atomy są
oświetlane wiązką sprzęgającą o częstości Rabiego $\Omega=2\pi\cdot0,75$
MHz. Do 25 $\mu$s gradient pola przesunięcia Zeemanowskiego wynosi
$\beta=1,25$ MHz/cm. Następnie gradient jest zmieniany na przeciwny.
Rysunek \ref{fig:Symulacja-pami=000119ci-GEM.} (c) przedstawia spójność
atomową w przestrzeni Fourierowskiej w 25 $\mu$s, czyli po procesie
zapisu. Widać tutaj odtworzenie struktury czasowej przychodzącego
sygnału. Pewna niezgodność w amplitudzie wynika z tego, że sygnał
zapisany wcześniej uległ większej dekoherencji z powodu oddziaływania
z wiązką sprzęgającą. Rysunek \ref{fig:Symulacja-pami=000119ci-GEM.}
(d) przedstawia natomiast tę samą spójność atomową, ale w przestrzeni
$z$. Rozkład ten dobrze odtwarza widmo sygnału zapisanego w pamięci
GEM.

Podsumowując, dodanie pola magnetycznego o stałym gradiencie do pamięci
kwantowej sprawia, że zapamiętuje ona jego strukturę czasową odtwarzając
jego funkcję Wignera $W_{t}(t,\omega)\sim W_{\rho}(k,z)$.

W 25 $\mu$s gradient pola magnetycznego został odwrócony i rozpoczął
się proces odczytu. Tam gdzie długość wektora falowego fali spinowej
$K$ wynosi zero, zaczynają być spełnione warunki dopasowania fazowego,
co skutkuje wygenerowaniem makroskopowego sygnału świetlnego przez
atomy. W efekcie sygnał odczytany z pamięci GEM odtwarza profil czasowy
sygnału zapisywanego (z dokładnością do dekoherencji wywołanej wiązką
sprzęgającą), przy czym jest on odwrócony. Pierwszy zapisany impuls
został odczytany jako ostatni, a ostatni zapisany odczytany jako pierwszy.
Ze względu na przeciwny gradient pola przesunięcia Zeemanowskiego
podczas odczytu, odwrócone zostało również widmo sygnału. Komponenty
o najniższych częstościach zapisane w pamięci GEM mają przy odczycie
najwyższą częstość i odwrotnie, komponenty o najwyższych częstościach
mają przy odczycie najniższą częstość.

\subsection{Soczewka czasowa}

Soczewka czasowa o ogniskowej $f_{t}$ powinna nadawać wiązce dodatkową
fazę kwadratową $\exp\left(-i\omega_{0}/(2f_{t})t^{2}\right)$. Odpowiada
to liniowej w czasie zmianie częstości nośnej $\omega(t)=\omega_{0}+\alpha t$,
gdzie $\alpha=-\omega_{0}/f_{t}$. Można tego dokonać w łatwy sposób,
używając modulatora akusto lub elektrooptycznego \cite{foster_ultrafast_2009,Kauffman1994,Azana2004,Babashah2019}.
Z perspektywy oddziaływania atomów ze światłem (równanie \ref{eq:spojnosc}),
zakładając że $\alpha t\ll\Delta$ w trakcie trwania całego impulsu,
istotne jest jedynie odstrojenie dwufotonowe $\delta$. Sprawia to,
że soczewka czasowa może zostać zrealizowana również bezpośrednio
w trakcie zapisu w pamięci GEM. Zamiast sygnału należy jedynie zmodulować
częstość wiązki sprzęgającej przy czym z odwrotnym przyrostem częstości
$\omega_{S}(t)=\omega_{0S}-\alpha t$

\subsection{Propagacja czasowa}

Propagacja czasowa na odległość $d_{t}$ wiąże się z nałożeniem dodatkowej,
kwadratowej fazy spektralnej $\exp\left(-id_{t}/(2\omega_{0})\omega^{2}\right)$.
Przy zapisie w obecności gradientu przesunięcia Zeemanowskiego $\beta$
spektralne komponenty sygnału są rzutowane na przestrzenne komponenty
spójności atomowej $\rho(z)\sim\widetilde{A}(\beta z)$. Oznacza to,
że modulacja fazy spektralnej światła jest równoważna modulacji przestrzennej
fazy spójności atomowej przetrzymywanej w pamięci. Aby nadać spójności
atomowej fazę $\exp\left(-id_{t}/(2\omega_{0})\beta^{2}z^{2}\right)$
można posłużyć się modulatorem przestrzennej fazy fali spinowej \cite{Leszczynski2018,Parniak2019,Lipka2019,mazelanik_coherent_2019}.

\section{Spektrometr\label{sec:Spektrometr}}

Spektrometr jest urządzeniem, które wykonuje transformację Fouriera
na przychodzącym sygnale. Można go zrealizować przy pomocy dwóch soczewek
czasowych o ogniskowej $f_{t}$ odseparowanych od siebie odległością
czasową $d_{t}=f_{t}$. W języku optyki geometrycznej transformacja
ta jest opisywana równaniem
\begin{equation}
\begin{alignedat}{1}\left(\begin{array}{c}
t'\\
\frac{\omega'}{\omega_{0}}
\end{array}\right) & =\left(\begin{array}{cc}
1 & 0\\
-\frac{1}{f_{t}} & 1
\end{array}\right)\left(\begin{array}{cc}
1 & f_{t}\\
0 & 1
\end{array}\right)\left(\begin{array}{cc}
1 & 0\\
-\frac{1}{f_{t}} & 1
\end{array}\right)\left(\begin{array}{c}
t\\
\frac{\omega}{\omega_{0}}
\end{array}\right)\\
 & =\left(\begin{array}{cc}
0 & f_{t}\\
-\frac{1}{f_{t}} & 0
\end{array}\right)\left(\begin{array}{c}
t\\
\frac{\omega}{\omega_{0}}
\end{array}\right).
\end{alignedat}
\label{eq:wigner_obrot}
\end{equation}
Spodziewamy się zatem, że amplituda pola wyjściowego będzie wynosić
\begin{equation}
A'(t)=\widetilde{A}(\alpha t).
\end{equation}
Jest to wynik idealny, zakładający stuprocentową wydajność zapisu
i odczytu z pamięci kwantowej oraz nieskończenie długą chmurę atomową. 

\subsection{Rozdzielczość i pasmo spektrometru}

Projektując spektrometr w efekcie końcowym chcemy otrzymać określoną
dostępną szerokość pasma $\mathcal{B}$ oraz rozdzielczość $\delta\omega$.

Jak to było wspomniane w podsekcji \ref{subsec:Pami=000119=000107-gradientowa},
przy gradiencie przesunięcia Zeemanowskiego $\beta,$ poszczególne
komponenty spektralne zapisywanego sygnału są mapowane na komponenty
przestrzenne spójności atomowej $z=(\omega-\omega_{0})/\beta$. Przy
założeniu prostokątnego rozłożenia atomów na odcinku o długości $L$,
maksymalny zakres częstości dostępnych w pamięci będzie miał szerokość
$\mathcal{B}=\beta L$. Oznacza to, że szerokość pasma jest proporcjonalna
do gradientu pola magnetycznego jakie było obecne podczas zapisu.

Rozdzielczość spektrometru zależy natomiast od tego jak długi impuls
jesteśmy w stanie zapisać w pamięci. Nazwijmy maksymalną długość $\tau_{\text{max}}$.
Po zadziałaniu na monohromatyczny sygnał o długości $\tau_{\text{max}}$
soczewki czasowej o ogniskowej $f_{t}=-\omega_{0}/\alpha$ jego widmo
ma szerokość $|\alpha|\tau_{\text{max}}$. Widmo to musi być tak szerokie
jak szerokość pasma, skąd dostajemy wartość $\tau_{\text{max}}=\mathcal{B}/|\alpha|$.
Widmo prostokątnego monohromatycznego impulsu $A(t)$ o długości $\tau_{\text{max}}$,
zdefiniowane jako szerokość $|\widetilde{A}(\omega)|^{2}$ w połowie
wysokości, wynosi $\delta\omega\approx2\pi\cdot0,89/\tau_{\text{max}}=2\pi\cdot0,89|\alpha|/\mathcal{B}$.

Czynnikiem, który może dodatkowo ograniczyć rozdzielczość jest dekoherencja
spójności atomowej pod wpływem oddziaływania z wiązką sprzęgającą.
Średni czas życia fali spinowej w polu o częstości Rabiego $\Omega$
wynosi $\tau=4\Delta^{2}/\left(\Gamma\Omega^{2}\right)$. W przypadku,
gdy $\tau\ll\tau_{\text{max}}$ rozdzielczość spektrometru zdefiniowana
jako szerokość widma $|\widetilde{A}(\omega)|^{2}$ w połowie wysokości
impulsu o amplitudzie $A(t)=\Theta(t)\exp(-t/(2\tau))$ wynosi $\delta\omega=1/\tau$.

\section{Eksperyment\label{sec:Eksperyment}}

\begin{figure}
\centering\includegraphics{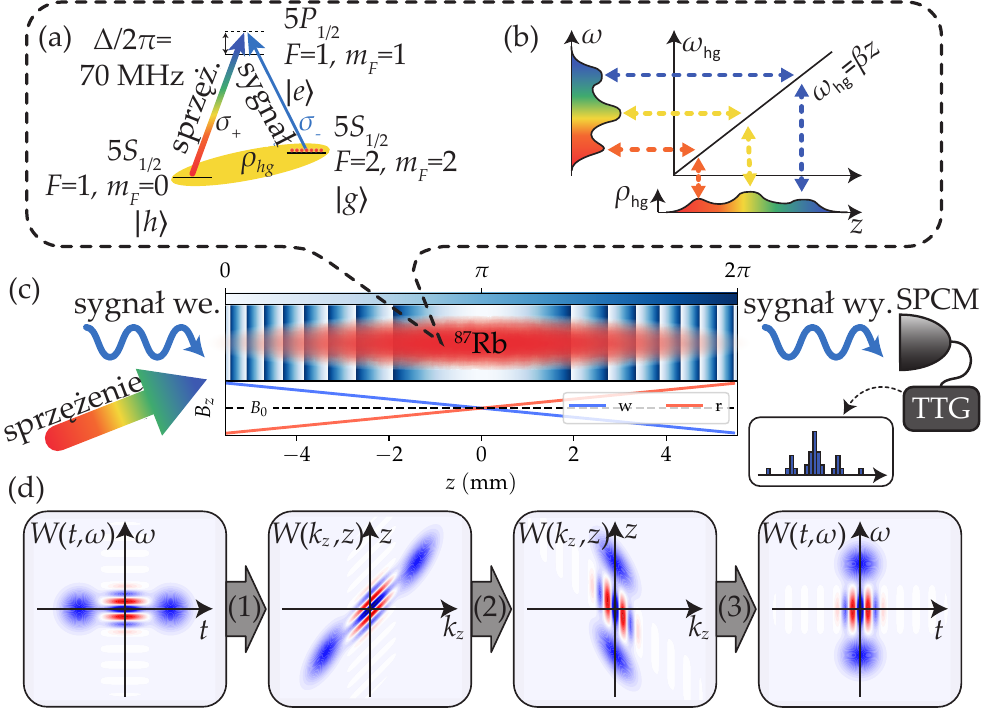}

\caption{(a) Interfejs światło-atomy. Zmodulowane pole sprzęgające pozwala
na jednoczesne mapowanie sygnału wejściowego na spójność atomową $\rho_{g,h}$
i realizację soczewki czasowej. (b) Projekcja składowych spektralnych
sygnału na składowe przestrzenne spójności atomowej w pamięci gradientowej
z gradientem przesunięcia Zeemana $\beta$. (c) Podczas procesu zapisu
atomy są umieszczane w ujemnym gradiencie pola magnetycznego wzdłuż
chmury (w). Po zakończeniu zapisu, na spójność atomową nakładana jest
przestrzenna faza o Fresnelowskim profilu parabolicznym, która realizuje
czasowy odpowiednik propagacji w pustej przestrzeni. Na końcu, gradient
jest przełączany na dodatni (r), a spójność jest z powrotem przekształcana
na światło, które jest dalej rejestrowane przez moduł zliczający pojedyncze
fotony. (d) Ewolucja czasowo-spektralnej funkcji Wignera na kolejnych
etapach obrazowania czasowego dalekiego pola: (1) soczewka czasowa,
(2) propagacja w pustej przestrzeni, (3) soczewka czasowa. Całkowita
transformacja efektywnie obraca początkową funkcję Wignera dwóch impulsów
(odpowiednik funkcji Wignera stanu kota Schrödingera w przestrzeni
fazowej) o $\pi/2$, w przestrzeni fazowej jak podano w równaniu \ref{eq:wigner_obrot}.}

\end{figure}
\begin{figure}
\centering\includegraphics{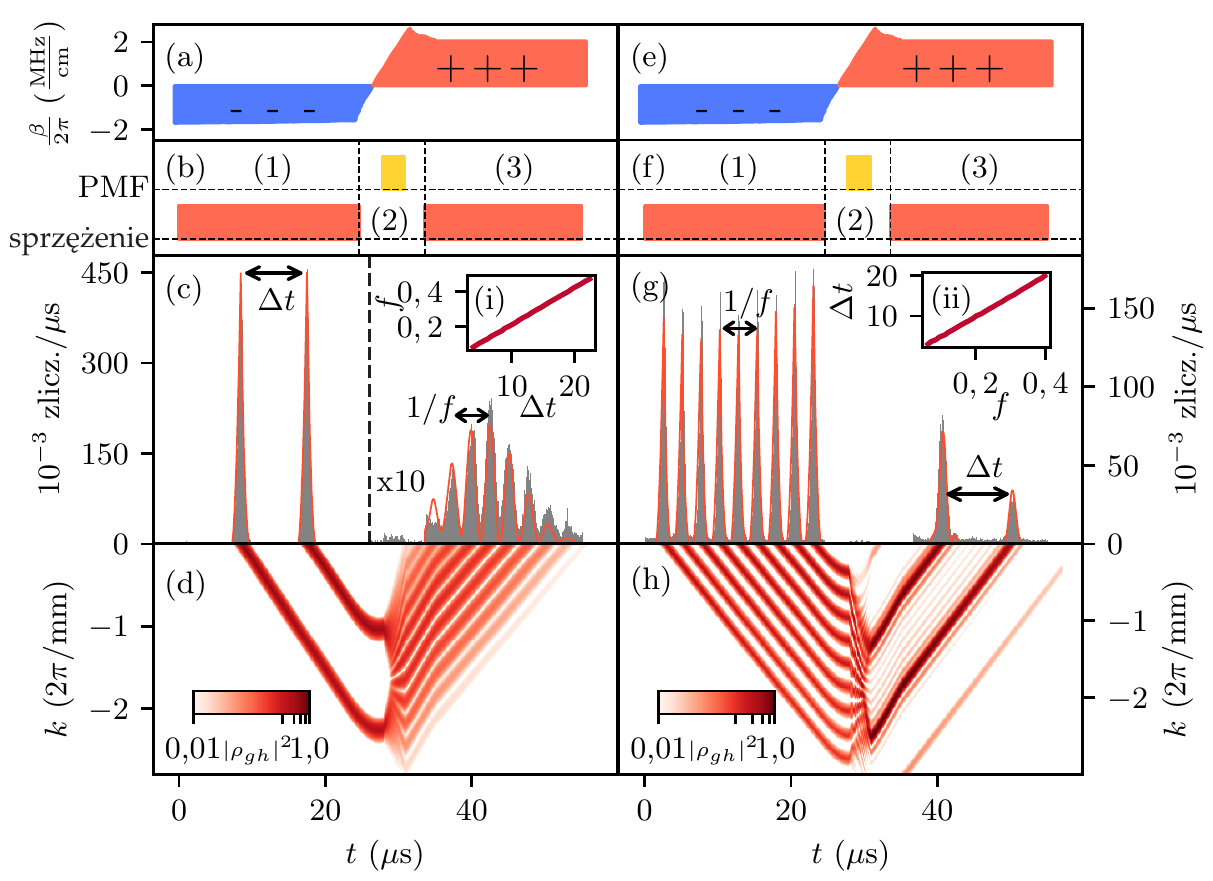}

\caption{Sekwencja czasowa spektrometru. (a, e) Gradient przesunięcia Zeemanowskiego
w czasie. (b, f) Sekwencja włączania pola sprzęgającego na przejściu
$|h\rangle\rightarrow|e\rangle$ (kolor czerwony) oraz wiązki modulatora
przestrzennej fazy fali spinowej (kolor żółty). (b, f) Cały proces
można podzielić na 3 następujące po sobie części. (1) Wiązka sprzęgająca
ze zmodulowaną częstością umożliwia zapis sygnału przychodzącego implementując
jednocześnie soczewkę czasową. (2) Następnie jest ona wyłączana, a
włączona na 3 $\mu$s zostaje wiązka modulująca o parabolicznym rozkładzie
natężenia, nadrukowując spójności atomowej dodatkową kwadratową fazę
przestrzenną i tym samym realizując propagację czasową. Równolegle
gradient pola magnetycznego jest przełączany na przeciwny. (3) Na
koniec wiązka sprzęgająca zostaje ponownie włączona umożliwiając konwersję
fal spinowych z powrotem na światło. Jeżeli byłaby ona zmodulowana,
tak jak przy procesie zapisu, to realizowała by ona drugą soczewkę
czasową. Jednakże, ponieważ wpływa to jedynie na fazę wychodzącego
sygnału, która i tak nie jest rejestrowana przez fotodiodę lawinową,
dla uproszczenia użyta została wiązka o stałej częstości. (c, g) Przykładowe
rezultaty otrzymane sygnału wejściowego w kształcie odpowiednio dwóch
szpil, oraz fali sinusopodobnej. Szary kolor prezentuje liczbę zarejestrowanych
fotonów, a czerwona linia odpowiada wynikom symulacji numerycznej.
(d, h) Znormalizowany moduł kwadrat spójności atomowej w przestrzeni
fourierowskiej. Rysunki (i) oraz (ii) przedstawiają zmierzone relacje
pomiędzy parametrami $f$ i $\Delta t$ zdefiniowanymi na rysunkach
(c, g)\label{fig:Sekwencja-czasowa-spektrometru.}}

\end{figure}
 Przejście dwufotonowe $|g\rangle\rightarrow|h\rangle$ pomiędzy poziomami
stanu podstawowego opisane w poprzednim rozdziale jest tak zwanym
przejściem zegarowym, które jest niewrażliwe na działanie pola magnetycznego.
W konsekwencji nie byłoby możliwe skonstruowanie pamięci GEM. Z tego
powodu do niniejszego eksperymentu użyto układu $\Lambda$ z poziomami
podstawowymi $|g\rangle=5S_{1/2},F=2,m_{F}=2$, $|h\rangle=5S_{1/2},F=1,m_{F}=0$,
oraz poziomem wzbudzonym $|e\rangle=5P_{1/2},F=1,m_{F}=1$. Wiązka
sprzęgająca posiada polaryzację kołową $\sigma_{+}$ i jest odstrojona
o $\Delta=2\pi\times70$ MHz od przejścia $|h\rangle\rightarrow|e\rangle$,
natomiast sygnał ma ortogonalną polaryzację kołową $\sigma_{-}$ i
dostrojony do przejścia $|g\rangle\rightarrow|h\rangle$ z odstrojeniem
dwufotonowym $\delta\approx0$ (rysunek \ref{eq:wigner_obrot}). Chmura
atomowa ma kształt cygara o długości około 1 cm i średnicy około 0,5
mm. Dla ustalenia osi kwantyzacji atomy znajdują się w zewnętrznym
polu magnetycznym wzdłuż osi $z$ o natężeniu około 1 G. Ponadto sygnał
posiada średnicę jedynie 0,1 mm i oświetla środek chmury atomów, tam
gdzie gęstość optyczna jest największa i wynosi około $\mathrm{OD}=76$
na przejściu $|g\rangle\rightarrow|e\rangle$. Przed procesem zapisu
ośrodek atomowy jest pompowany za pomocą dwóch wiązek laserowych.
Pierwsza, niespolaryzowana jest dostrojona do przejścia $5S_{1/2},F=1\rightarrow5P_{1/2},F=2$,
przepompowując tym samym wszystkie atomy do poziomu $5S_{1/2},F=2$.
Druga zaś posiada polaryzację kołową $\sigma_{+}$ i jest dostrojona
do zamkniętego przejścia $5S_{1/2},F=2\rightarrow5P_{1/2},F=3$. Powoduje
to przepompowanie wszystkich atomów do podpoziomu Zeemanowskiego $|g\rangle$.
Wiązka modulująca fazę fal spinowych za pomocą efektu ac-Starka posiada
polaryzację $\pi$ i jest odstrojone o około 1 GHz ku niebieskiemu
od przejścia $5S_{1/2},F=1\rightarrow5P_{3/2}$. 

\subsection{Szum}

\begin{figure}
\centering\includegraphics{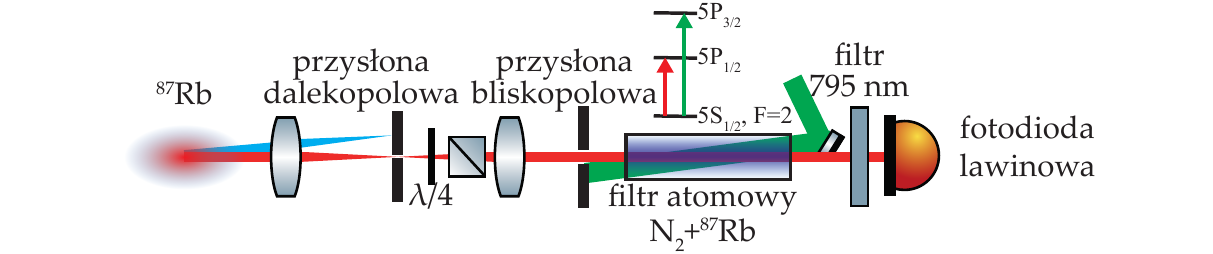}

\caption{Układ filtrujący sygnał wyjściowy. Przechodzi on kolejno przez przysłonę
dalekopolową, polaryzator Wollastona, przysłonę bliskopolową, optycznie
pompowany filtr atomowy oraz filtr interferencyjny na długość fali
795 nm\label{fig:Uk=000142ad-filtruj=000105cy.}}

\end{figure}

Zasadniczą kwestią w eksperymentalnej realizacji pamięci kwantowej
jest eliminacja wszelkiego innego światła niż odczytane z fal spinowych.
Aby uzyskać jak najlepszy stosunek sygnału do szumu, fotony emitowane
z pamięci kwantowej są przepuszczane przez wieloetapowy filtr (rysunek
\ref{fig:Uk=000142ad-filtruj=000105cy.}). Na początek, wiązka sprzęgająca
jest wstępnie usuwana przy pomocy przysłony dalekopolowej. Dalej,
korzystając z faktu, że ma ona ortogonalną polaryzację do sygnału,
jest ona dalej odfiltrowywana polaryzatorem Wollastona. Następnie,
by usunąć fotony nie pochodzące z chmury atomowej użyta jest przysłona
bliskopolowa. W kolejnym etapie, sygnał jest przepuszczany przez komórkę
z ciepłym $^{87}$Rb napompowanym do poziomu $5S_{1/2},F=1$ oraz
azotem jako gazem buforowym. Pozwala to usunąć resztę fotonów z lasera
sprzęgającego. Na samym końcu znajduje się filtr interferencyjny na
795 nm, który usuwa wszelkie fotony o innych długościach fali.

\begin{figure}

\centering\includegraphics{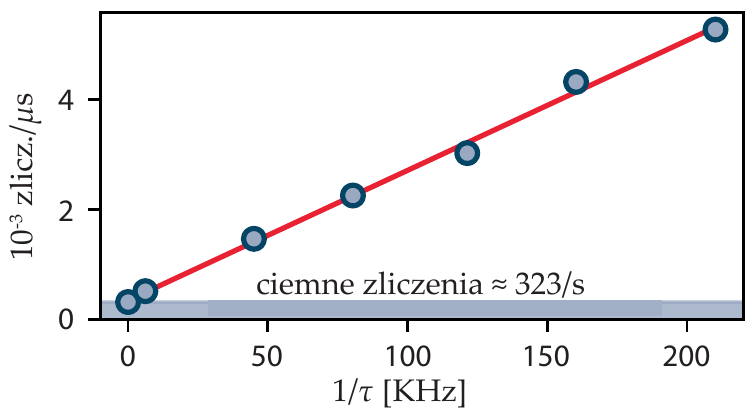}

\caption{Częstość rejestrowania fotonów szumu w funkcji stałej zaniku spójności
atomowej.\label{fig:szum}}

\end{figure}

Przejdźmy do wyznaczenia średniej liczby fotonów szumu rejestrowanych
w trakcie odczytu z pamięci. W tej fazie pamięć oświetlana jest wiązką
laserową, która powoduje zanik spójności atomowej ze stałą $1/\tau=\Gamma\Omega^{2}/(4\Delta^{2})$.
Przyjmijmy że proce odczytu trwa dokładnie $\tau$ --- przedłużanie
go byłoby i tak niecelowe. Dla kilku rożnych natężeń wiązki odczytującej
zmierzyliśmy średnią ilość fotonów odczytywanych z czystej pamięci
na jednostkę czasu przy stale włączonej wiązce sprzęgającej. Dla każdego
z próbnych natężeń wyznaczaliśmy $\tau$ z osobnego pomiaru całkowitej
energii odczytywanego sygnału w zależności od czasu pomiędzy zapisem,
a odczytem, przy ciągle włączonej wiązce sprzęgającej, a następnie
dopasowanie zaniku wykładniczego. Wyniki przedstawia rysunek \ref{fig:szum}.
Nachylenie prostej dopasowanej do zależności natężenia szumu od natężenia
lasera wyskalowanego jako stała zaniku $1/\tau$ ma interpretację
średniej liczby fotonów szumu zarejestrowanych w czasie odczytu $\tau$.
Wynosi ona jedynie 0,023, a więc foton szumu będzie rejestrowany średnio
raz na 43 pomiary.

\subsection{Demonstracja działania spektrometru}

Rysunek \ref{fig:Sekwencja-czasowa-spektrometru.} przedstawia wyniki
dwóch przykładowych pomiarów dokonanych przy pomocy opisanego wyżej
układu. Panele (a) oraz (e) pokazują zmienność czasową gradientu przesunięcia
Zeemanowskiego, którego wartość w trakcie zapisu wynosi około $\beta=-2\pi\times1,7$
MHz/cm. Panele (b) i (f) prezentują sekwencję uruchamiania wiązki
sprzęgającej oraz wiązki modulującej, którą można podzielić na 3 etapy.
(1) W pierwszym, wiązka sprzęgająca o częstości Rabiego $\Omega=2\pi\times4,7$
MHz jest używana do konwersji słabego sygnału (średnio 2,8 fotonu
na impuls) o amplitudzie $A(t)$ na spójność atomową $\rho_{g,h}$.
Dodatkowo częstość nośna wiązki sprzęgającej $\omega_{S}$ jest zmodulowana
za pomocą modulatora akustooptycznego $\omega_{S}(t)=\omega_{0}+\alpha t$
ze świergotem $\alpha=2\pi\cdot0,004$ MHz/$\mu$s, co odpowiada implementacji
soczewki czasowej o ogniskowej $f_{t}=9,6\cdot10^{3}$s. (2) W kolejnym
etapie, w czasie 3$\mu$s, za pomocą wiązki modulującej na spójność
atomową nadrukowywany jest paraboliczny fresnelowski profil fazy $\exp\left(-i\beta^{2}/(2\alpha)z^{2}\right)$,
co odpowiada propagacji na odległość czasową $d_{t}=f_{t}$. Równocześnie
gradient pola magnetycznego jest przełączany na przeciwny. (3) Na
sam koniec, wiązka sprzęgająca zostaje włączona i spójność atomowa
jest z powrotem konwertowana na sygnał świetlny. Ponieważ detekcja
za pomocą fotodiody lawinowej nie rejestruje fazy, częstość wiązki
sprzęgającej podczas odczytu nie jest modulowana, co upraszcza to
sterowanie.

Rysunki \ref{fig:Sekwencja-czasowa-spektrometru.}(c) oraz \ref{fig:Sekwencja-czasowa-spektrometru.}(g)
przedstawiają rezultaty dla dwóch różnych sygnałów wejściowych, odpowiednio:
dwa wąskie ucha lub sygnał prążkowany. Czerwona linia odpowiada wynikom
symulacji numerycznej pełnej interakcji między światłem a atomami,
zgodnie z równaniami \ref{eq:spojnosc} oraz \ref{eq:ewolucja pola}
wykonaną za pomocą skryptu napisanego na platformie XMDS2. Diagramy
(d) i (h) pokazują ewolucję czasową spójności atomowej w przestrzeni
fourierowskiej. Można zauważyć, jak struktura czasowa sygnału wejściowego
jest odwzorowana w przestrzeni fourierowskiej. Całkowita wydajność
konwersji światło wchodzące - odczytane dla obu typów sygnałów wejściowych
wynosiła około 7\%. Wstawki (i) oraz (ii) przedstawiają zależność
między parametrami $\Delta t$ i $f$ zdefiniowanymi na wykresach
(c) i (g) którą wyznaczyliśmy powtarzając pomiary dla wielu różnych
parametrów wchodzącego sygnału.

Pomiędzy wynikami symulacji (czerwone obwiednie na rysunku \ref{fig:Sekwencja-czasowa-spektrometru.}),
a pomiarami (szare histogramy zliczeń w czasie) istnieją pewne niewielkie
rozbieżności. Wynikają one z kilku czynników. Po pierwsze gradient
pola magnetycznego mógł nie być idealnie równy w całej objętości chmury
atomowej. Po drugie modulacja fazy powoduje pewną dekoherencję poprzez
niejednorodność natężenia wiązki oświetlającej atomy z boku. Po trzecie
w symulacji rozkład atomów w przestrzeni został przybliżony funkcją
supergaussowską.

\section{Rozdzielczość i pasmo, a wydajność\label{sec:Rozdzielczo=00015B=000107-i-pasmo,}}

\begin{figure}
\centering\includegraphics{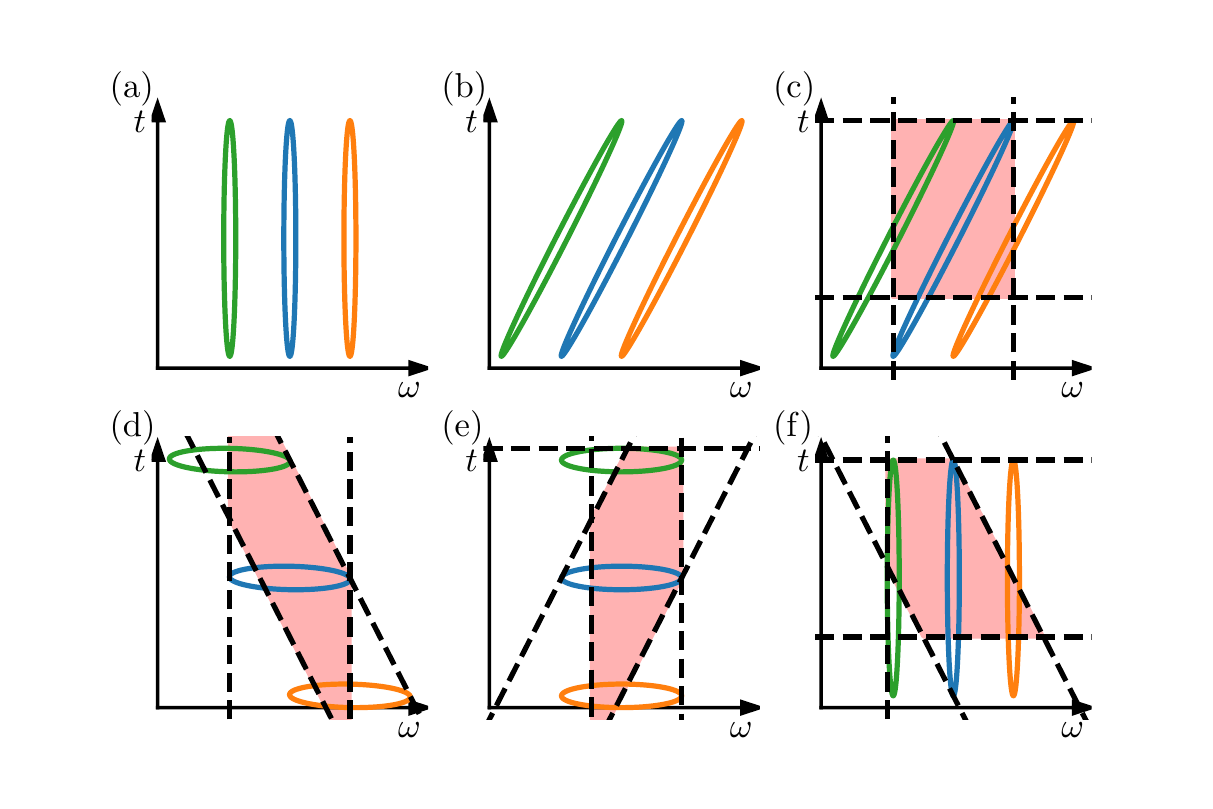}

\caption{Transformacje funkcji Wignera w kolejnych etapach działania spektrometru
oraz ograniczenia pasma i rozdzielczości wynikające z dekoherencji
i skończonego rozmiaru chmury atomowej. Trzy kolorowe elipsy symbolizują
funkcję Wignera trzech przykładowych sygnałów - długich impulsów o
wąskim widmie i trzech rożnych częstościach. (a) Funkcje Wignera sygnału
przychodzącego. (b) Funkcje Wignera po zastosowaniu soczewki czasowej
o ogniskowej $f$. (c) Zapis sygnału do atomów. Zacieniony obszar
to ta część funkcji Wignera, która zostanie zapisana w atomach. Ograniczenie
w częstościach $\omega$ wynika ze skończonego rozmiaru chmury, natomiast
w czasie $t$ z dekoherencji spójności atomowej pod wpływem wiązki
sprzęgającej. (d) Propagacja sygnału o odległość czasową $f$. (e)
Sygnał po zastosowaniu drugiej soczewki czasowej o ogniskowej $f$.
Ponieważ przy odczycie wiązka sprzęgająca zostaje ponownie uruchomiona,
pojawia się dodatkowe ograniczenie w $t$ z powodu dekoherencji. (f)
Nieprzetworzona funkcja Wignera sygnału przychodzącego (taka jak w
panelu (a)) z naniesionym obszarem roboczym spektrometru. \label{fig:Transformacje-funkcji-Wignera}}

\end{figure}
\begin{figure}
\centering\includegraphics{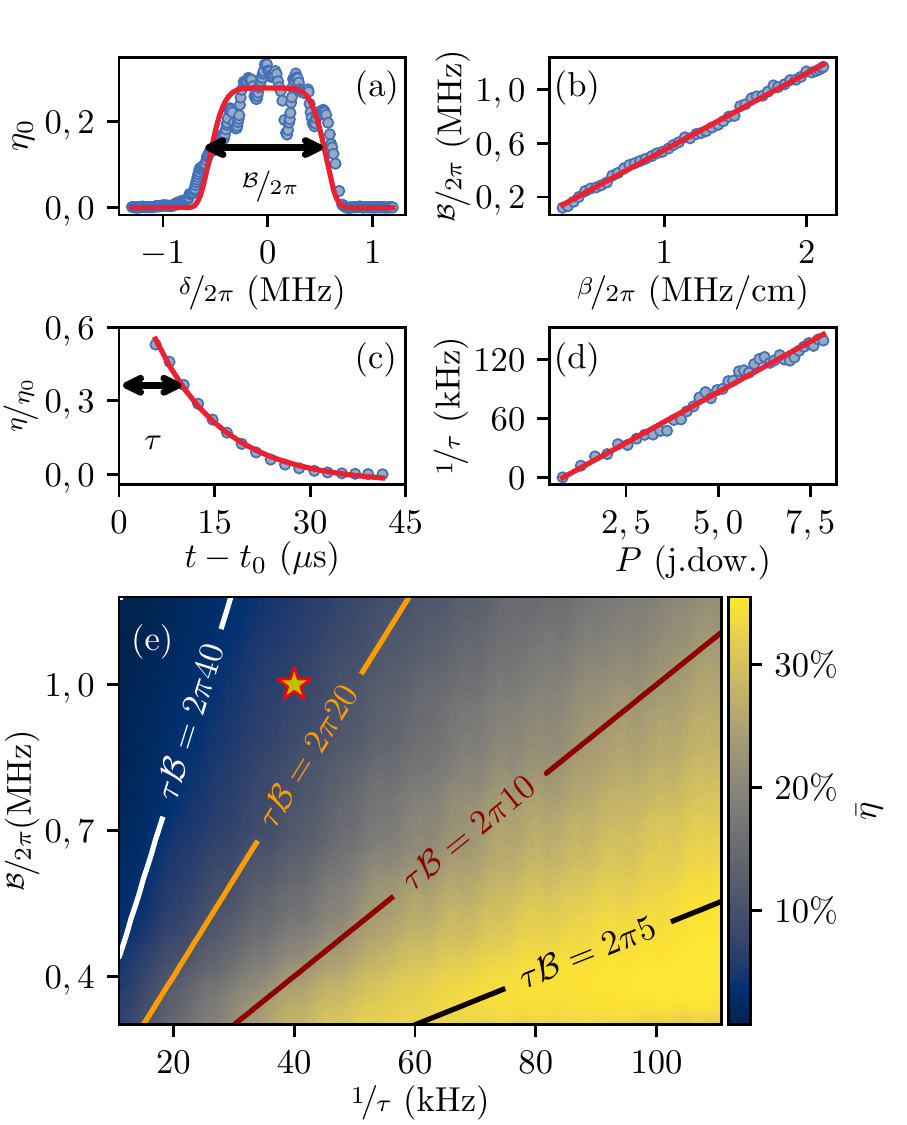}

\caption{(a) Wydajność spektralna w funkcji odstrojenia dwufotonowego $\delta=\omega-\omega_{0}$,
dla wybranego iloczynu $\tau\mathcal{B}=2\pi\cdot13$, z pasmem$\mathcal{B}$
zdefiniowanym jako szerokość $\eta_{0}(\omega)$ w połowie wysokości.
Czerwona linia jest przybliżeniem rozkładu atomów funkcją super-gaussowską
($\exp(-\delta^{4}/(4\sigma^{4}))$) używanego w symulacjach. (b)
Szerokość pasma $\mathcal{B}$ w funkcji gradientu przesunięcia Zeemana
$\beta$ razem z dopasowaną funkcją liniową (czerwona linia). (c)
Wydajność pamięci GEM w funkcji czasu przechowywania, przy ciągle
włączonej wiązce sprzęgającej. Średni czas życia $\tau$ otrzymany
jest z dopasowania zaniku wykładniczego (czerwona linia) - na rysunku
$\tau=10\text{ }\mu$s . (d) $1/\tau$ w funkcji mocy wiązki sprzęgającej
$P\sim|\Omega|^{2},$z dopasowaną funkcją liniową (czerwona linia).
(e) Obliczona mapa średniej wydajności $\bar{\eta}$ w funkcji pasma$\mathcal{B}$
oraz rozdzielczości $1/\tau$. Gwiazdka oznacza miejsce, w którym
wykonane były pomiary przedstawione na rysunku \ref{fig:Sekwencja-czasowa-spektrometru.}.
\label{fig:(a)-Wydajno=00015B=000107-spektralna}}
\end{figure}
Rysunek \ref{fig:Transformacje-funkcji-Wignera} przedstawia transformację
funkcji Wignera sygnału w kolejnych etapach działania spektrometru.
W pierwszej kolejności sygnał przychodzący (a) zostaje poddany działaniu
soczewki czasowej o ogniskowej $f$ (b). Przy zapisie (c) część sygnału
jest tracona. Wzdłuż osi $\omega$ ze względu na skończony rozmiar
chmury (\textbf{$\mathcal{B}=\beta L$}) oraz wzdłuż osi $t$ ze względu
na skończony czas życia fal spinowych $\tau$ pod wpływem dekoherencji
spowodowanej wiązką sprzęgającą. Zacieniony obszar oznacza przestrzeń
roboczą spektrometru, w której sygnał przychodzący zapisywany jest
w wydajny sposób. Panele (d) oraz (e) przedstawiają dalszą przekształcenie
przestrzeni roboczej w kolejnych etapach działania spektrometru, czyli
propagacji o odległość czasową $f$ (d) oraz ponowne zastosowanie
soczewki czasowej o ogniskowej $f$ (e). Przy odczycie (e) pojawia
się dodatkowe ograniczenie obszaru roboczego w $t$ ze względu na
ponowne uruchomienie wiązki sprzęgającej. Panel (f) przedstawia obszar
roboczy działania spektrometru w przestrzeni czas-częstość na nieprzetworzonym
sygnale przychodzącym. Rozmiar tego obszaru skaluje się z dobrym przybliżeniem
jak iloczyn $\tau\mathcal{B}=\mathcal{B}/\delta\omega$, który można
interpretować jako ilość pikseli dostępną w spektrometrze.

Szerokość pasma $\mathcal{B}=\beta L$ można zwiększać poprzez zwiększanie
gradientu przesunięcia Zeemanowskiego $\mathcal{B}$. Podobnie rozdzielczość
$\delta\omega=1/\tau=\Gamma\Omega^{2}/\left(4\Delta^{2}\right)$ można
zwiększać poprzez zmniejszenie natężenia wiązki sprzęgającej. Niestety
cena jaką należy zapłacić za zwiększenie iloczynu $\tau\mathcal{B}$
jest wydajność. W przybliżeniu atomów równo rozłożonych w przestrzeni
na odcinku o długości $L$ i gęstości optycznej OD, wydajność pamięci
GEM wyraża się jak \cite{Sparkes2013}
\begin{equation}
\eta_{0}=\left(1-\mathrm{e}^{-2\pi\frac{\mathrm{OD}}{\tau\mathcal{B}}}\right)^{2}.\label{eq:tb_eff_sim}
\end{equation}
Dla stałego iloczynu $\tau\mathcal{B}$ wydajność również jest stała.
Innymi słowy wydajność spektrometru jest monotoniczną funkcją ilości
dostępnych pikseli.

W realnym przypadku atomy nie są równo rozłożone w przestrzeni, więc
wydajność $\eta_{0}$ zależy od częstości $\omega$. Rysunek \ref{fig:(a)-Wydajno=00015B=000107-spektralna}
(a) przedstawia zmierzone $\eta_{0}$ w funkcji odstrojenia dwufotonowego
$\delta$. Szerokość tego rozkładu w połowie wysokości będzie traktowana
jako operacyjna definicja szerokości pasma $\mathcal{B}$. Jako średnią
wydajność przyjmiemy 
\begin{equation}
\bar{\eta}=\frac{1}{\mathcal{B}}\intop_{\mathcal{-B}/2}^{\mathcal{B}/2}\eta_{0}(\omega)\text{d}\omega.
\end{equation}
Rysunek \ref{fig:(a)-Wydajno=00015B=000107-spektralna} (c) przedstawia
natomiast natężenie sygnału odczytu w zależności od czasu przez jaki
wiązka sprzęgająca świeciła pomiędzy zapisem, a odczytem. Z tego pomiaru
możliwe było odzyskanie parametru $\tau$. 

Wartości $\tau$, $\mathcal{B}$ oraz $\bar{\eta}$ zostały zmierzone
dla różnych natężeń wiązki sprzęgającej $P\sim|\Omega|^{2}$ raz dla
różnych wartości gradientu przesunięcia Zeemanowskiego $\beta$. Diagram
na rysunku \ref{fig:(a)-Wydajno=00015B=000107-spektralna} (e) przedstawia
wartość $\bar{\eta}$ w funkcji $\tau$ oraz $\mathcal{B}$. Zmierzona
zależność dobrze odwzorowuje relację pomiędzy $\bar{\eta}$, a iloczynem
$\tau\mathcal{B}$ wynikającą z uproszczonej formuły \ref{eq:tb_eff_sim}. 

W przypadku zademonstrowanego spektrometru udało się osiągnąć wydajność
rzędu 20\% przy szerokości pasma około 1 MHz oraz rozdzielczości 100
KHz. Przy zejściu z wydajnością do poziomu około 5\% udało się uzyskać
rozdzielczość rzędu 20 kHz.

\section{Podsumowanie}

W tym rozdziale zaprezentowane zostało działanie modulacji przestrzennej
fazy fal spinowych wraz z pamięcią GEM, która odtwarza czasową strukturę
zapisywanego sygnału. Przedstawiona została idea czasowych odpowiedników
propagacji w przestrzeni oraz soczewki, a także za ich pomocą skonstruowany
został spektrometr. Najważniejsze wyniki uzyskane w tym rozdziale
to:
\begin{itemize}
\item demonstracja działania spektrometru oraz wysoka zgodność pomiędzy
danymi uzyskanymi z pomiarów, a wynikami symulacji numerycznych
\item zaprezentowanie wieloetapowego układu filtrującego, dzięki któremu
średnia liczba fotonów szumu rejestrowanych w trakcie odczytu z pamięci
wynosi 0,023.
\item charakteryzacja zależności pomiędzy wydajnością, szerokością pasma,
a rozdzielczością spektrometru. Przy wydajnościach rzędu 20\% osiągnięto
szerokość pasma około 1 MHz przy rozdzielczości 100 kHz, a przy obniżeniu
wydajności do około 5\% osiągnięto rozdzielczość rzędu 20 kHz.
\end{itemize}
Bardzo niski poziom rejestrowanego szumu pozwala na wykonywanie pomiarów
na stanach o bardzo małej liczbie fotonów. Jednocześnie bardzo wysoka
rozdzielczość sprawia, że układ ten doskonale nadaje się do charakteryzowania
wąskopasmowej emisji atomowej. Charakteryzacja w domenie czasowej
i manipulacja na poziomie pojedynczej fotonów jest stosowana w wielu
kwantowych protokołach przetwarzania informacji, architekturach sieci
kwantowych i metrologii. Nasze urządzenie pozwoli tym technikom zejść
do domeny bardzo wąskiego pasma. Ponieważ za pomocą efektu ac-Starka
możliwe jest nadrukowywanie dowolnych profili fazowych, możliwe jest
niemal dowolne manipulowanie stanami światła zapisanymi w GEM. Stwarza
to nowe możliwości w czasowym i spektralnym przetwarzaniu wąskopasmowych
kwantowych stanów światła pochodzących z emisji atomowej. Wraz z poprawą
gradientu pola magnetycznego oraz zwiększeniem gęstości optycznej
atomów, szerokość pasma GEM może osiągnąć dziesiątki MHz otwierając
nowe obszary zastosowań, takie jak półprzewodnikowe pamięci kwantowe
\cite{Hedges2010}, czy centra barwne \cite{Jeong2019}. Nasza technika
zastosowana w układach o większej szerokości pasma absorpcji \cite{Saglamyurek2018}
lub gęstości optycznej \cite{Cho2016} może wypełnić lukę w szerokości
pasma, umożliwiając zastosowanie w hybrydowych (ciało stałe - atomy)
sieciach kwantowych pracujących z pełnym czasowo-spektralnym stopniem
swobody.

\chapter{Sukcesywny odczyt-konwerter modów przestrzennych na czasowe\label{chap:Sukcesywny-odczyt-konwerter-mod=0000F3}}

Ostatnim zagadnieniem poruszonym w niniejszej rozprawie jest analiza
teoretyczna układu umożliwiającego sukcesywny odczyt wybranych przez
nas fal spinowych, przechowywanych w różnych modach przestrzennych
naszej pamięci kwantowej. Zaprezentowany pomysł pozwala nie tylko
na konwersję fali spinowej na foton, ale także wydajne sprzęgnięcie
go do światłowodu jednomodowego. Znacząco to zwiększy możliwości naszej
pamięci, ułatwiając jej wykorzystanie w protokołach przesyłu informacji
kwantowej. W opisie wykorzystana będzie pamięć kwantowa opisana w
\cite{Parniak2017}, modulacja fal spinowych za pomocą gradientu pola
magnetycznego (cewki GEM) \cite{Sparkes2013} oraz wnęka rezonansowa\cite{Cox2019}.
Struktura rozdziału jest następująca:
\begin{itemize}
\item Sekcja \ref{sec:Koncepcja} - Zaprezentowana zostanie konstrukcja
układu doświadczalnego, którego celem jest konwersja fal spinowych,
wytworzonych w procesie spontanicznego rozpraszania Ramana, na ciąg
fotonów sprzężonych do światłowodu jednomodowego. Omówiony zostanie
wpływ geometrii wiązek na kierunek generowania fotonów odczytu. Wprowadzona
zostanie również koncepcja korekcji dopasowania za pomocą gradientu
pola magnetycznego oraz wnęki rezonansowej jako narzędzia pozwalającego
na wydajny, selektywny odczyt z pamięci.
\item Sekcja \ref{sec:Odczyt-z-pier=00015Bcieniow=000105} - Wyprowadzone
zostaną równania oddziaływania światła i spójności atomowej w obecności
pierścieniowej wnęki rezonansowej. Dokładnie omówione zostaną wpływ
dopasowania fazowego na wydajność, a także straty związane z dekoherencją
oraz absorpcją światła, Na koniec przedstawiony zostanie również wynik
symulacji dla realnego zestawu parametrów, jakie oferuje nasza pamięć
kwantowa opisana w poprzednich rozdziałach niniejszej rozprawy.
\end{itemize}

\section{Koncepcja\label{sec:Projekt-uk=000142adu-do=00015Bwiadczalnego}}

\begin{figure}
\centering\includegraphics{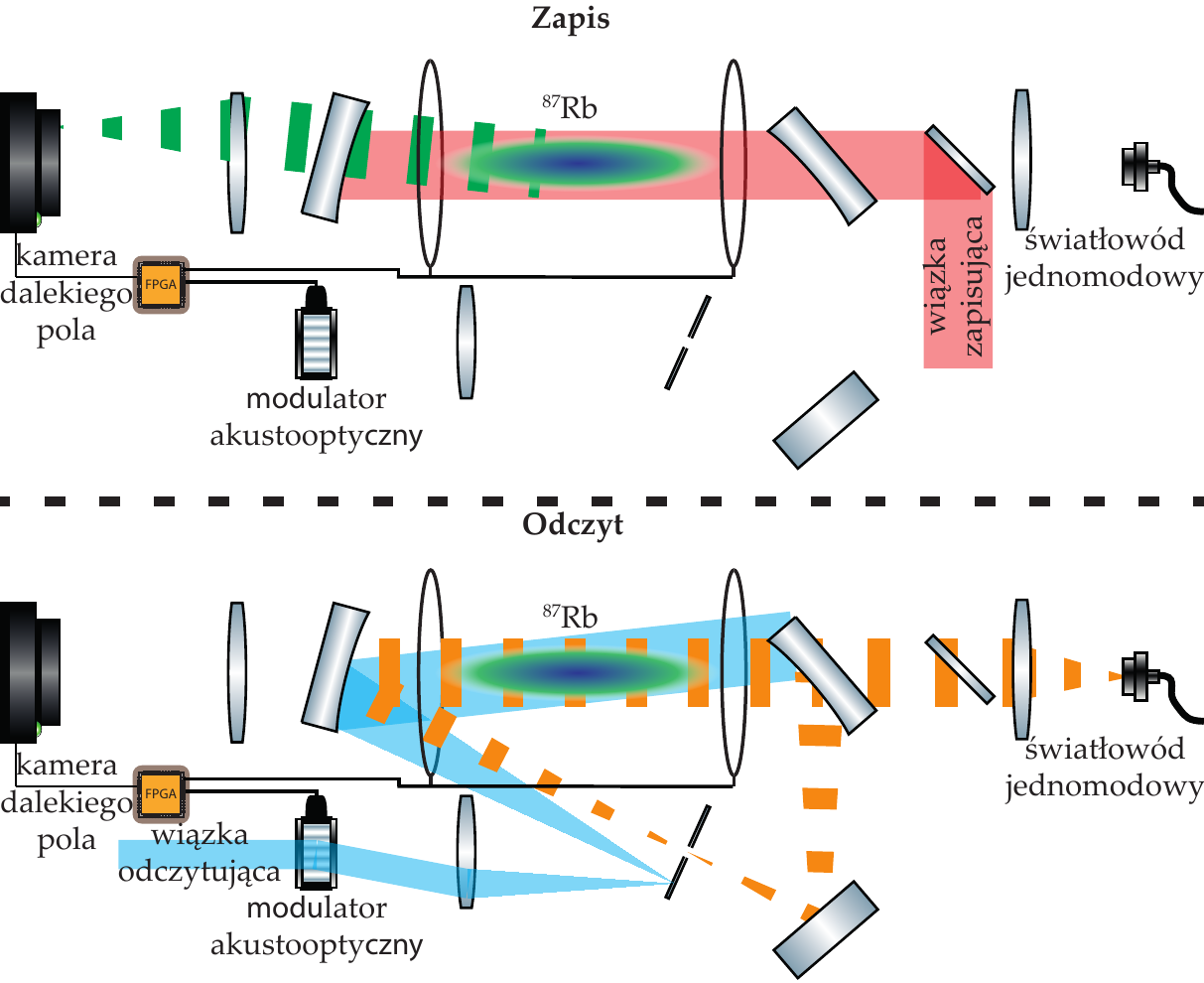}\caption{Sekwencja zapisu i sukcesywnego odczytu fal spinowych. Na początek
chmura atomów rubidu jest oświetlona wiązką zapisującą. Rozproszone
fotony ramanowskie są rejestrowane przez kamerę dalekiego pola ze
wzmacniaczem obrazu. Daje to informację o wektorach falowych wygenerowanych
fal spinowych. Następnie następuje seria konwersji kolejnych fal spinowych
na fotony i sprzęganie ich do światłowodu jednomodowego. Pojedynczy
akt odczytu składa się z dwóch etapów. Najpierw uruchamiane są cewki
generujące gradient pola magnetycznego celem korekcji wzdłużnej składowej
wektora falowego odczytywanej fali. Następnie na okres 1 $\mu$s uruchamiana
jest wiązka odczytująca, której kąt padania na chmurę atomów jest
kontrolowany za pomocą modulatora akustooptycznego. Kąt ten ustawiany
jest w taki sposób, by wyemitowany foton był sprzężony z modem wnęki.
Jednocześnie wnęka jest sprzężona ze światłowodem jednomodowym, do
którego owy foton trafia. \label{fig:Schemat-konwertera-modow-1}}
\end{figure}
Rysunek \ref{fig:Schemat-konwertera-modow-1} przedstawia schemat
układu doświadczalnego, którego zadaniem byłaby konwersja fal spinowych
wygenerowanych w procesie spontanicznego rozpraszania Ramana na ciąg
fotonów sprzężonych do światłowodu. Chmura atomów rubidu znajduje
się w pierścieniowej wnęce rezonansowej, ustawionej w taki sposób,
by sprzęgać się z fotonem wyemitowanym dokładnie wzdłuż osi chmury.
Wewnątrz wnęki, w dalekim polu względem atomów, znajduje się przysłona,
przez którą przedostaje się światło zgodne z jej modem podstawowym.
Pełni ona jednocześnie funkcję lustra, od którego odbija wiązka odczytująca.
Na drodze wiązki odczytującej znajduje się modulator akustooptyczny,
który jest obrazowany za pomocą teleskopu na chmurę atomową. Zmieniając
częstość podawanego sygnału można kontrolować kąt pod jakim oświetlany
jest rubid \cite{Cox2019}. Mod podstawowy wnęki optycznej jest sprzężony
z modem światłowodu jednomodowego. Pozwala to na bardzo wydajne wyłapywanie
światła wyciekającego z rezonatora. Lustra użyte w do budowy układu
są dichroiczne. Elementy z których skonstruowana jest wnęka optyczna
odbijają światło o długości fali 795 nm używane przy odczycie i są
jednocześnie przezroczyste dla światła o długości 780 nm, używanego
przy zapisie. Jednocześnie lustro, którym wiązka zapisująca jest kierowana
na atomy jest przezroczyste dla fotonów odczytu.

Cały eksperyment składał się będzie z kilku, następujących po sobie
faz. Na początek, atomy rubidu po uwolnieniu z pułapki magnetooptycznej
są pompowane optycznie do poziomu $|g\rangle$. W pierwszym kroku
atomy oświetlane są wiązką zapisującą. W wyniku tego dochodzi do spontanicznego
rozpraszania Ramana. Fotony wyemitowane w tym procesie są rejestrowane
przez kamerę dalekiego pola. Z otrzymanych pozycji można dokładnie
określić wektory falowe wszystkich powstałych fal spinowych. Następnie,
dane z kamery trafiają do modułu FPGA, który na tej podstawie przygotowuje
sekwencję odczytującą. Pojedynczy cykl tej sekwencji składa się z
dwóch etapów. Najpierw uruchamiane są cewki wytwarzające gradient
pola magnetycznego, aby zoptymalizować dopasowanie fazowe dla wybranej
do odczytu fali spinowej. Następnie uruchamiana jest wiązka odczytująca
i kierowana na chmurę atomową pod takim kątem, by wygenerowany foton
odczytu był sprzężony z podstawowym modem wnęki rezonansowej. Następnie
trafia on do światłowodu jednomodowego, którym może zostać przetransportowany
dalej. Każda kolejna fala spinowa podlega dokładnie tej samej procedurze.
Cewki naprawiają dopasowanie fazowe, a następnie wiązka odczytująca
oświetla atomy tak, by foton odczytu sprzągł się z modem wnęki i dalej
z modem światłowodu.

\paragraph{Analiza dopasowania fazowego\label{sec:Koncepcja}}

\begin{figure}
\includegraphics{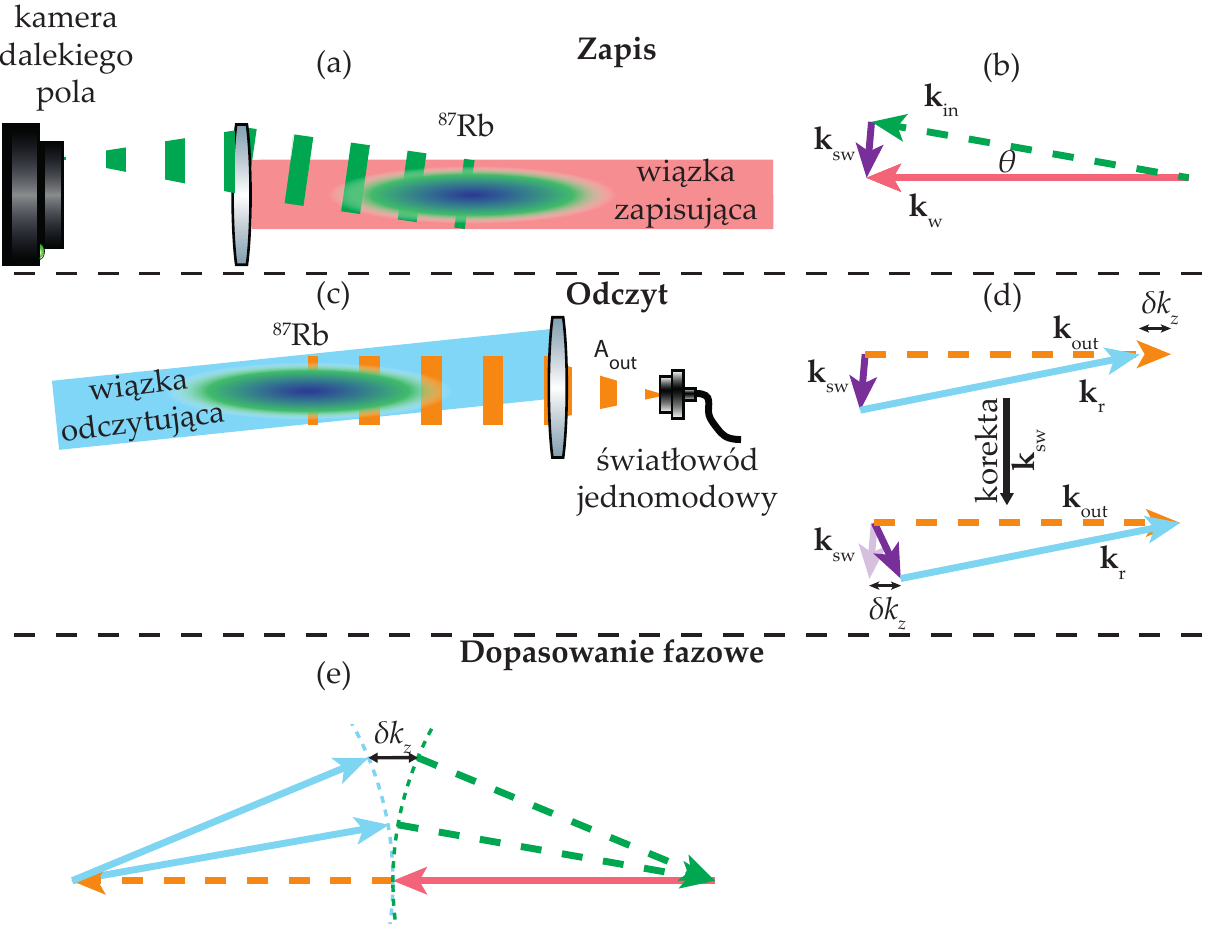}\caption{Analiza wektorów falowych. (a) Foton rozproszony w procesie spontanicznego
rozpraszania Ramana jest rejestrowany przez kamerę dalekiego pola.
(b) Pomiar ten pozwala obliczyć wektor falowy fali spinowej $\mathbf{k}_{\mathrm{sw}}=\mathbf{k}_{\mathrm{w}}-\mathbf{k}_{\mathrm{in}}$.
(c) Wiązka odczytująca oświetla atomy pod kątem $\theta$, (d) aby
spełnić zależność $\mathbf{k}_{\mathrm{in},\perp}=\mathbf{k}_{\mathrm{r,}\perp}$
i aby foton wygenerowany w procesie odczytu był emitowany wzdłuż osi
$z$. Panel (e) przedstawia wektory falowe fotonów biorących udział
w całym cyklu zapisu i odczytu. Brakujący do zamknięcia cyklu wektor
$\delta k_{z}$, zwany niedopasowaniem fazowym, uzupełniamy poprzez
np. modulację pomiędzy zapisem a odczytem.\label{fig:Schemat-zapisu-ze}}
\end{figure}

Pierwszym celem jest przekonwertowanie wybranej przez nas fali spinowej
na foton skierowany w kierunku $z$, czyli $\mathbf{k}_{\mathrm{out,}\perp}=0$
gdzie $\perp$ oznacz składowe prostopadłe do z. Przyjmijmy, że wiązka
lasera zapisującego jest skierowana wzdłuż osi z, czyli $\mathbf{k}_{\mathrm{w},\perp}=0$.
Aby uzyskać $\mathbf{k}_{\mathrm{out,}\perp}=0$, należy zapewnić
$\mathbf{k}_{\mathrm{r,}\perp}=\mathbf{k}_{\mathrm{in,}\perp}$. Ponieważ
długości fali wiązki zapisującej i odczytującej wynoszą odpowiednio
780 nm oraz 795 nm, to jeżeli foton zapisu został rozproszony pod
kątem $\theta$, wiązka odczytująca również powinna być pochylona
pod kątem $795/780\cdot\theta$ (rysunek \ref{fig:Schemat-zapisu-ze}
(c, d)).

Przy takiej konfiguracji pojawi się niedopasowanie fazowe $\delta k_{z}=(\mathbf{k}_{\mathrm{w}}-\mathbf{k}_{\mathrm{in}}+\mathbf{k}_{\mathrm{r}}-\mathbf{k}_{\mathrm{out}})_{z}$
(rysunek \ref{fig:Schemat-zapisu-ze} (d)). Rysunek \ref{fig:Schemat-zapisu-ze}
(e) przedstawia diagram wektorów falowych wszystkich fotonów biorących
udział w procesie zapisu i odczytu. Wynika z niego, że $\delta k_{z}=(1-\cos(\theta))(|\mathbf{k}_{\mathrm{r}}|+|\mathbf{k}_{\mathrm{in}}|)-0,045\text{ mm}^{-1}\approx(|\mathbf{k}_{\mathrm{r}}|+|\mathbf{k}_{\mathrm{in}}|)\theta^{2}/2$.
Dla $\delta k_{z}$ większego niż odwrotność długości chmury atomowej,
wydajność odczytu drastycznie spada. Z tego powodu ważne jest, by
czynnik ten skompensować.

\paragraph{Korekcja wzdłużnego wektora falowego.}

Aby skompensować niedopasowanie fazowe i uzyskać najlepszą wydajność
konwersji fali spinowej na foton po oświetleniu atomów wiązką odczytującą,
należy zmieniać $k_{z}$-ową składową wektora falowego fali spinowej
o wartość $\delta k_{z}$. Można tego dokonać, wytwarzając pole magnetyczne
o stałym gradiencie wzdłuż osi $z$. Aby jednak było to możliwe należy
odpowiednio wybrać poziomy stanu podstawowego, między którymi fale
spinowe będą przechowywane. Przejście dwufotonowe $5S_{1/2},F=1,m_{F}=-1\rightarrow5S_{1/2},F=2,m_{F}=1$
użyte w pracy \cite{Parniak2017}, nie jest do tego odpowiednie, ponieważ
jest to tak zwane przejście zegarowe, które jest niewrażliwe na działanie
pola magnetycznego. Jako stan początkowy dobrym wyborem jest poziom
$|g\rangle=5S_{1/2},F=2,m_{F}=2$. Wtedy fale spinowe przechowywane
będą między poziomami $|g\rangle$ oraz $|h\rangle=5S_{1/2},F=1,m_{F}=0$.
Z jednej strony umożliwia ono wykorzystanie pola magnetycznego, a
z drugiej fotony zapisu i odczytu posiadają ortogonalne polaryzacje
do wiązek zapisującej i odczytującej, co znacznie ułatwia ich odfiltrowanie.

\paragraph{Odczyt sukcesywny/rola wnęki.}

Drugim celem jest to, aby odczytowi uległa tylko wybrana przez nas
fala spinowa. Częściowo jest to zapewnione przez dopasowanie fazowe,
ponieważ przy jego braku fala spinowa jest praktycznie nienaruszona
\cite{Hosseini2012}. Jednakże, zwłaszcza dla małych kątów $\theta$,
w obszarze dobrego dopasowania może znajdować się więcej niż jedna
fala spinowa. Potrzebny jest więc dodatkowy mechanizm pozwalający
na odczytanie tylko jednej fali spinowej. Dobrym rozwiązaniem wydaje
się być pierścieniowa wnęka rezonansowa, która znacznie potrafi podnieść
efektywną gęstość optyczną dla fotonów, które są zgodne z jej modem.

\section{Model teoretyczny\label{sec:Odczyt-z-pier=00015Bcieniow=000105}}

Pierścieniowa wnęka rezonansowa, w której umieszczona jest chmura
atomów rubidu, znacząco podnosi szybkość konwersji fali spinowej na
foton \cite{Bao2012}. Jeżeli transmisja lustra wnęki, przez które
wycieka światło, wynosi $T$, to średnia liczba obiegów wykonanych
przez uwięziony w niej foton wyniesie $1/T$. Oznacza to, że efektywna
gęstość optyczna również wzrasta $1/T$ razy, co zgodnie z równaniem
\ref{eq:obroty_prawdziwe} przekłada się na odczyt $1/\sqrt{T}$-krotnie
szybszy, niż w przypadku braku wnęki. 

\paragraph{Ewolucja słabego pola.}

Ewolucja słabego pola o częstości Rabiego $\Omega=d_{g,e}A/\hbar$,
gdzie $A$ jest amplitudą pola elektrycznego, w obecności spójności
atomowej $\rho_{g,h}$, zgodnie z równaniem \ref{eq:ewolucja pola}
oraz pomijając dyfrakcję, jest opisana równaniem 
\begin{equation}
\frac{\partial}{\partial z}\Omega+\frac{1}{c}\frac{\partial}{\partial t}\Omega=-i\frac{k_{0}}{\hbar\epsilon_{0}}\frac{nd_{g,e}^{2}}{2\Delta+i\Gamma}(\Omega+\Omega_{r}\rho_{g,h}^{*}).\label{eq:pole-bez-dyfrakcji}
\end{equation}
Dla uproszczenia analizy można przeformułować zmienne o odpowiednie
fazy przestrzenne
\[
\begin{gathered}\rho_{g,h}\rightarrow\rho_{g,h}\exp(i(\mathbf{k}_{\mathrm{r}}-\mathbf{k}_{\mathrm{out}})\mathbf{r})\\
\Omega_{r}\rightarrow\Omega_{r}\exp(i\mathbf{k}_{\mathrm{r}}\mathbf{r})\\
\Omega\rightarrow\Omega\exp(i\mathbf{k}_{\mathrm{out}}\mathbf{r})
\end{gathered}
\]
Wtedy $\Omega_{r}$ jest rzeczywiste, $\Omega$ pozbywa się szybko
zmiennej fazy. Ponieważ spójność wygenerowana w procesie spontanicznego
rozpraszania Ramana jest falą płaską $\rho_{0}\exp(i(\mathbf{k}_{\mathrm{w}}-\mathbf{k}_{\mathrm{in}})\mathbf{r})$,
to po zmianie o podaną wyżej fazę ma ona postać $\rho_{0}\exp(i\delta k_{z}z)$
. Dla wnęki rezonansowej o długości $L$ i transmisji lustra $T\ll1$
możemy przyjąć, że częstość Rabiego $\Omega$ wewnątrz wnęki nie zależy
od $z$, czyli innymi słowy 
\begin{equation}
\Omega(z,t)=\Theta(z+L/2)\Theta(L/2-z)\Omega_{\mathrm{cav}}(t),\label{eq:plaskie-pole}
\end{equation}
gdzie $\Omega_{\mathrm{cav}}(t)$ jest częstością Rabiego pola we
wnęce zależną jedynie od czasu, a $\Theta$ jest thetą Heavside'a.
Podstawiając \ref{eq:plaskie-pole} do równania \ref{eq:pole-bez-dyfrakcji}
i całkując po $z$ oraz dodając czynnik odpowiedzialny za wyciek światła
z wnęki dostaniemy:
\begin{equation}
\frac{\partial}{\partial t}\Omega_{\mathrm{cav}}=-i\frac{c\Gamma\mathrm{OD}}{2L(2\Delta+i\Gamma)}\Omega_{\mathrm{cav}}-i\frac{ck_{0}}{L\hbar\epsilon_{0}}\frac{d_{g,e}^{2}}{2\Delta+i\Gamma}\Omega_{r}\intop_{-\infty}^{\infty}n\rho_{g,h}^{*}\mathrm{d}z-\frac{Tc}{2L}\Omega_{\mathrm{cav}}.\label{eq:pole-we-wnece}
\end{equation}
Jednocześnie równanie na ewolucję, zgodnie z równaniem \ref{eq:spojnosc}
oraz przy założeniu $\Omega_{\mathrm{r}}\gg\Omega_{\mathrm{cav}}$
oraz łącznym odstrojeniu dwufotonowym $\delta+\delta_{acS}=0$ ($\delta_{acS}$
- przesunięcia ac-Starka pod wpływem $\Omega_{\mathrm{r}}$):
\begin{equation}
\frac{\partial}{\partial t}\rho_{g,h}=\frac{i}{2}\frac{\Omega_{\text{cav}}^{*}\Omega_{r}}{2\Delta-i\Gamma}-\frac{\Gamma|\Omega_{r}|^{2}}{2\Gamma^{2}+8\Delta^{2}}\rho_{g,h}.\label{eq:spojnosc-z-wneka}
\end{equation}
Aby uprościć analizę zdefiniujmy rozkład
\begin{equation}
\begin{gathered}\sqrt{n}\rho_{g,h}=\sum_{j=0}^{\infty}c_{j}(t)u_{j}(z),\\
\intop_{-\infty}^{\infty}u_{j}(z)u_{k}^{*}(z)\text{d}z=\delta_{j,k}.
\end{gathered}
\end{equation}
Funkcje $u_{j}$ tworzą bazę ortonormalną. Załóżmy też, że $u_{0}=\sqrt{n}/N$,
gdzie $N$ to liczba atomów. Można zauważyć, że w równaniu \ref{eq:pole-we-wnece}
wkład do generowania światła ma jedynie czynnik $c_{0}=1/N\intop_{-\infty}^{\infty}n\rho_{g,h}\mathrm{d}z$.
Równanie na poszczególne $c_{j}$ można otrzymać, mnożąc obustronnie
równanie \ref{eq:spojnosc-z-wneka} przez $\sqrt{n}/N\cdot u_{j}^{*}$
oraz całkując wzdłuż osi $z$. Wtedy przyjmują one postać
\begin{equation}
\frac{\partial}{\partial t}c_{j}=-\frac{\Gamma|\Omega_{r}|^{2}}{2\Gamma^{2}+8\Delta^{2}}c_{j}+\begin{cases}
\frac{i}{2}\frac{\Omega_{\text{cav}}^{*}\Omega_{r}}{2\Delta-i\Gamma} & j=0\\
0 & j\neq0
\end{cases}\label{eq:cj_dt}
\end{equation}
Widać stąd, że jedynie człon $c_{0}$ oddziałuje z polem $\Omega_{\text{cav}}$.
Równanie na jego ewolucję wygląda dokładnie tak samo jak równanie
\ref{eq:spojnosc-z-wneka} w punkcie. Pozostałe człony nie biorą udziału
z procesie zapisu i odczytu. Zachodzi jedynie dekoherencja związana
z poszerzenie natężeniowym wywołanym polem $\Omega_{r}$. Ich zanik
jest opisany równaniem:
\begin{equation}
c_{j}(t)=c_{j}(0)\mathrm{e}^{-\frac{\Gamma|\Omega_{r}|^{2}}{2\Gamma^{2}+8\Delta^{2}}t}.
\end{equation}

\subsection{Dekoherencja}

\paragraph{Absorpcja jednofotnowa.\label{par:Absorpcja-jednofotnowa}}

\begin{figure}
\centering\includegraphics{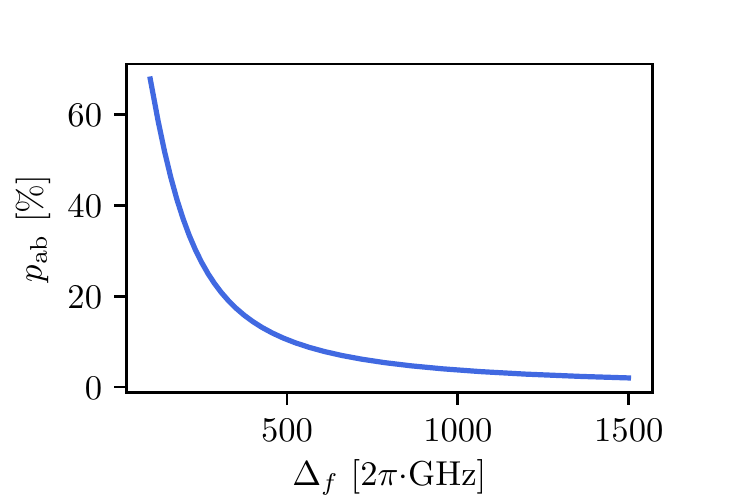}

\caption{Prawdopodobieństwo absorpcji fotonu we wnęce w funkcji odstrojenie
$\Delta_{f}$, przy wyłączonej wiązce odczytującej.\label{fig:Prawdopodobie=000144stwo-absorpcji-fot}}
\end{figure}

Z równania \ref{eq:pole-we-wnece} wynika, że istnieją dwa mechanizmy
zaniku pola $\Omega_{\mathrm{cav}}$ nie związane z oddziaływaniem
ze spójnością atomową: absorpcja jednofotonowa przez atomy ze średnim
czasem życia $\tau_{\mathrm{at}}=4L\Delta^{2}/(c\Gamma^{2}\mathrm{OD})$
(przy założeniu $\Delta\gg\Gamma$) oraz transmisja przez lustro wnęki
z czasem życia $\tau_{\mathrm{cav}}=L/(cT)$. Aby użycie wnęki miało
sens, $\tau_{\mathrm{cav}}$ musi być znacząco mniejsze od $\tau_{\mathrm{at}}$.
W innym przypadku absorpcja jednofotonowa będzie źródłem znacznych
strat. Prawdopodobieństwo absorpcji światła we wnęce, brzy braku wiązki
odczytującej wynosi $p_{\text{ab}}=\tau_{\mathrm{at}}\tau_{\mathrm{cav}}/(\tau_{\mathrm{at}}(\tau_{\mathrm{at}}+\tau_{\mathrm{cav}}))$.
Poziom $5P_{1/2}$, wykorzystywany w procesie odczytu, posiada 2 podpoziomy
z którymi może oddziaływać wiązka odczytująca: $|e\rangle=5P_{1/2}F=1,m_{F}=1$
oraz $|f\rangle=5P_{1/2}F=2,m_{F}=1$, odseparowane od siebie o $2\pi\cdot814$
MHz. Rysunek \ref{fig:Prawdopodobie=000144stwo-absorpcji-fot} przedstawia
prawdopodobieństwo absorpcji fotonu we wnęce w funkcji odstrojenia
$\Delta_{f}$ od poziomu $|f\rangle$, dla T=0,01, OD=70 na przejściu
$|g\rangle\rightarrow|e\rangle$ \cite{Mazelanik2020} oraz stosunku
momentów dipolowych $d_{g,f}/d_{g,e}=-1/\sqrt{3}$ \cite{Steck2001}. 

\paragraph{Rozmywanie fal spinowych.}

Ruch termiczny atomów powoduje powolną dekoherencję fali spinowych.
Można przyjąć, że jeżeli średnia droga pokonana przez atomy będzie
tego samego rzędu wielkości, co długośc fali, dojdzie do jej rozmycia.
Dla fali spinowej o większej długości potrzeba proporcjonalnie więcej
czasu, aby do takiego przemieszczenia atomów doszło. Z tego powodu
średni czas życia fali spinowej $\tau_{\text{term}}$ skaluje się
jak $1/|\mathbf{k}_{\text{sw}}|$. Dla niedużych kątów $\theta$,
$|\mathbf{k}_{\text{sw}}|\approx\mathbf{|k}_{\text{w}}|\theta$ .
Zakładając maxwellowski rozkład prędkości atomów, fale spinowe zanikają
jak $\exp(-t^{2}/\tau_{\text{term}}^{2})$, gdzie $\tau_{\mathrm{term}}=1/(\theta|\mathbf{k}_{\mathrm{w}}|\sqrt{k_{b}T)}$.
Przy temperaturze naszej chmury atomów wynoszącej około 20 $\mu$K
\cite{Parniak2017}, czas $\tau_{\text{term}}$ wynosi 80 $\mu$s
dla $\theta=1^{\circ}$, 40 $\mu$s dla $\theta=2^{\circ}$, a 20
$\mu$s dla $\theta=4^{\circ}$. Widać więc, że termiczna dekoherencja
znacząco ogranicza czas, w którym możliwy jest odczyt fal spinowych.
Przyjmijmy więc, że interesować nas będą tylko fale spinowe dla kątów
$\theta<1^{\circ}.$

\paragraph{Poszerzenia natężeniowe.}

\begin{figure}

\centering\includegraphics{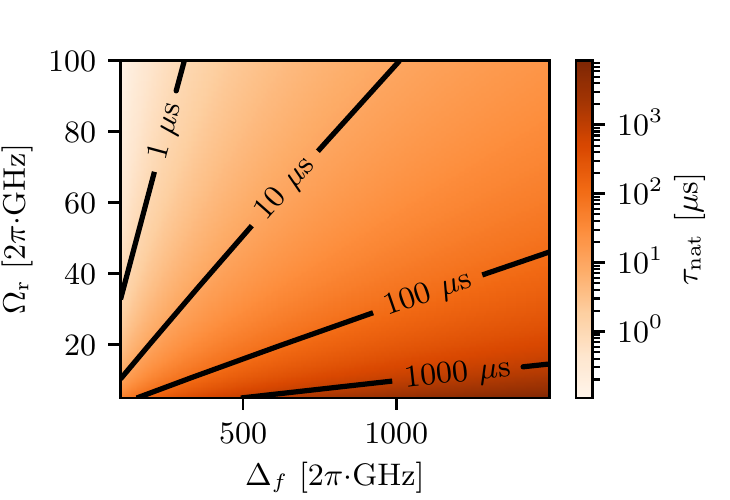}\caption{Średni czas życia fali spinowej pod wpływem poszerzenia natężeniowego
w funkcji $\Omega_{\text{r}}$ i $\Delta_{f}$.\label{fig:=00015Aredni-czas-=00017Cycia} }

\end{figure}
Dekoherencja wywołana poszerzeniem natężeniowym jest opisana członem
proporcjonalnym do $c_{j}$ w równaniu \ref{eq:cj_dt}. Dotyka ona
również w takim samym stopniu fale spinowe nie sprzężone z wnęką (równanie
\ref{eq:spojnosc}). Rysunek \ref{fig:=00015Aredni-czas-=00017Cycia}
przedstawia średni czas życia fali spinowej pod wpływem tego poszerzenia
w funkcji $\Omega_{\text{r}}$ oraz $\Delta_{f}$.

\subsection{Wyniki Symulacji\label{subsec:Symulacja}}

\paragraph{Wyniki symulacji.}

\begin{figure}
\centering\includegraphics{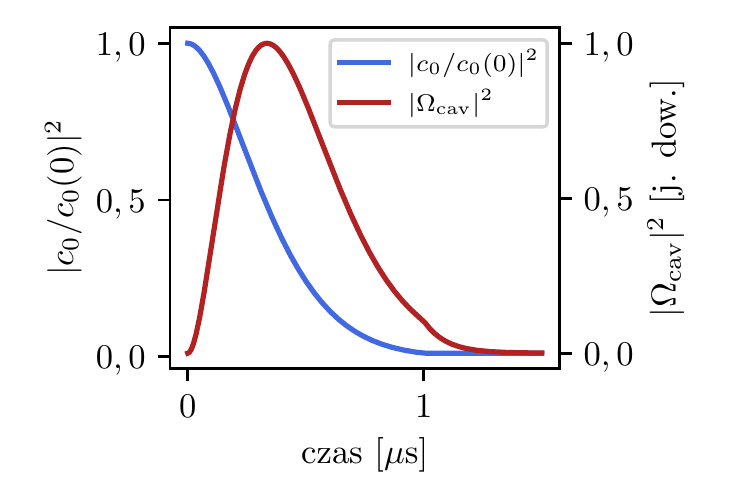}\caption{Symulacja konwersji spójności atomowej na światło we wnęce rezonansowej.
Niebieska linia przedstawia |$c_{0}/c_{0}(0)|^{2}$, które daje informację
o prawdopodobieństwie, że fala spinowa wciąż przechowywana jest w
atomach. Czerwona linia przedstawia $|\Omega_{\text{cav}}|^{2}$ proporcjonalne
do natężenia światła we wnęce. Wiązka odczytująca jest włączona na
okres 1 $\mu$s. Parametry wnęki oraz wiązek przedstawione są w tekście
głównym w sekcji \ref{subsec:Symulacja} \label{fig:symulacja_wneka}}
\end{figure}
Rysunek \ref{fig:symulacja_wneka} przedstawia prawdopodobieństwo
istnienia fali spinowej w atomach oraz natężenia światła we wnęce.
Symulacja została wykonana za pomocą skryptu w języku Python. Przyjęte
zostały parametry: długość wnęki $L=30$ cm, transmisja lustra wnęki
$T=0.01$, częstość Rabiego wiązki odczytującej $\Omega_{\text{r}}=2\pi\cdot30$
MHz na przejściu $|h\rangle\rightarrow|f\rangle$, odstrojenie $\Delta_{f}=2\pi\cdot1$
GHz oraz gęstość optycznej wynosząca 70 na przejściu |$g\rangle\rightarrow|e\rangle$
\cite{Mazelanik2020}. Dla przyjętych parametrów średni czas życia
fali spinowej pod wpływem poszerzenia natężeniowego wynosi około $\tau_{\text{nat}}=109$
$\mu$s, a więc przekracza $\tau_{\text{term}}$ dla $\theta=1^{\circ}$.
Wiązka odczytująca jest uruchamiana na okres 1 $\mu$s. Spójność początkowa
w symulacji posiada parametr $c_{0}(0)=1$ oraz pozostałe $c_{j}(0)=0$.
Prawdopodobieństwo odczytu, zdefiniowane jako stosunek liczby fotonów
w wyemitowanym świetle $n_{ph}=\epsilon_{0}/(2\hbar\omega_{0})\int|A(t)|\text{d}t$
do początkowej liczby atomów $n_{h}$ w stanie $|h\rangle$, wynosi
92\%.

Równocześnie, by określić niszczenie fal spinowych nie sprzężonych
z wnęką, przeprowadzona została symulacja bazująca na równaniach \ref{eq:spojnosc}
oraz \ref{eq:=00015Bwiatlo_biegnace}, za pomocą skryptu napisanego
w XMDS. Parametry takie jak $\Omega_{\text{r}},$$\Delta_{f}$ oraz
gęstość optyczna zostały takie same. Początkową spójność $\rho_{g,h,0}$
przyjęto równą 1 w całej objętości atomów. Rozkład atomów w przestrzeni
został przyjęty taki, jak opisano w \cite{Mazelanik2020}. Na podstawie
symulacji, prawdopodobieństwo zniszczenia fali spinowej zdefiniowane
jako $1-\int|\rho_{g,h}|\text{d}z/\int|\rho_{g,h,0}|\text{d}z$, wynosi
1,7\%. W przypadku fal spinowych, dla których warunek dopasowania
fazowego nie jest spełniony można przyjąć, że niszczenie jest spowodowana
jedynie poszerzeniem natężeniowym. Jego prawdopodobieństwo wynosi
wtedy $1-\exp(1/\tau_{\text{nat}}\cdot1\mu\text{s})=0,9\%$.

\section{Podsumowanie\label{sec:Wnioski}}

W tym rozdziale zaprezentowany został pomysł na konstrukcję konwertera
modów, który umożliwiłby odczyt fal spinowych przechowywanych w różnych
modach przestrzennych na ciąg fotonów sprzęgniętych ze światłowodem
jednomodowym. Do najważniejszych wyników przedstawionych w tym rozdziale
należą:
\begin{itemize}
\item Pokazanie, że w procesie odczytu biorą udział jedynie składowa fali
spinowej o całkowitym dopasowaniu fazowym.
\item Analiza strat podczas odczytu. Dla maksymalnej wydajności ważne jest
by średni czas absorpcji fotonu przez atomy $\tau_{\text{at}}$ był
jak największy w stosunku do średniego czasu życia fotonu we wnęce
$\tau_{\text{cav}}$. Jednocześnie do optymalnej pracy dekoherencja
związana ze świeceniem wiązką odczytującą nie powinna być mniejsza
od dekoherencji termicznej, która odgórnie ogranicza czas życia fali
spinowej w pamięci.
\item Symulacja numeryczna odczytu pojedynczej fali spinowej, dla parametrów
możliwych do osiągnięcia w pamięci kwantowej opisywanej w poprzednich
rozdziałach niniejszej pracy. Przewidywana wydajność odczytu wybranej
fali spinowej wyniosła 92\% w czasie 1 $\mu$s. Jednocześnie, w zależności
od dopasowania fazowego, prawdopodobieństwo zniszczenia innej fali
spinowej waha się pomiędzy 0,9\% do 1,7\%. Uwzględniając termiczną
dekoherencję zaprezentowana sekwencja pozwoliłaby na odczyt do kilkudziesięciu
fal spinowych.
\end{itemize}
Przedstawiony schemat jest kompatybilny z przestrzenną modulacja fazy.
Dzięki temu możliwe jest interferowanie ze sobą wielu fal spinowych
\cite{Parniak2019}. W połączeniu z możliwością dokonywania pomiarów
na wybranych modach umożliwi to np. generowanie na żądanie stanów
n-falospinowych (przekładających się na stany n-fotonowe w odczycie),
czy też możliwość projektowania bramek kwantowych działających na
falach spinowych. Wydajne wprzęganie wygenerowanych fotonów prosto
do światłowodu jednomodowego sprawia, że wzrasta łatwość przesyłania
wygenerowanej informacji na duże odległości. Daje to więc szerokie
pole do wykonywania w przyszłości obliczeń kwantowych, czy też realizowania
kolejnych protokołów przesyłu informacji kwantowej.

\chapter{Podsumowanie i wnioski}

W niniejszej rozprawie doktorskiej zaprezentowana została idea wykorzystania
przestrzennej modulacji fazy fal spinowych do modyfikacji dopasowania
fazowego. Przedstawiono szereg idei które umożliwiają wykorzystanie
wielomodowych pamięci kwantowych na nowe sposoby i do nowych celów.

\section{Otrzymane rezultaty}

\paragraph{Przestrzenna modulacja fazy.}

W rozdziale \ref{chap:Przestrzenna-modulacja-fazy} przedstawiona
została idea modulowania fazy spójności przechowywanej w atomach za
pomocą efektu ac-Starka wywołującego fikcyjne pole magnetyczne \cite{Cohen-Tannoudji1972}.
Zaprezentowana była konstrukcja umożliwiająca kontrolę przestrzennego
rozkładu natężenia wiązki światła. przekłada się to na kontrolę przesunięcia
ac-Starka z rozdzielczością przestrzenną. Modulując przestrzenną fazę
spinów precesujących w zewnętrznym polu magnetycznym zademonstrowano
ideę kontrolowania wydajności odczytu z pamięci kwantowej poprzez
modyfikowanie dopasowania fazowego. Jednocześnie pokazane zostało,
że niejednorodność nadrukowanej fazy jest głównym źródłem dekoherencji.

\paragraph{Modyfikacja fazy poprzecznej.}

W rozdziale \ref{chap:Kompensacja-aberracji} pokazana została koncepcja
poprzecznej fazy fal spinowych w pamięci kwantowej, co przekłada się
bezpośrednio na przestrzenną fazę światła wygenerowanego w procesie
odczytu. Zademonstrowana została kompensacja z wysoką wydajnością
i wiernością aberracji wprowadzonych sztucznie przez dodatkową soczewkę
cylindryczną. Za pomocą pomiarów interferometrycznych potwierdzono
wysoką wierność fazy nakładanej spójność atomową z profilami natężenia
wiązki starkowskiej oświetlającej atomy. Pokazane również zostało,
że dekoherencja fal spinowych pod wpływem modulacji fazy jest proporcjonalna
do kwadratu nałożonej fazy, a współczynnik proporcjonalności skaluje
się kwadratowo ze względnym odchyleniem standardowym natężenia wiązki
modulującej fazę od zadanego profilu. 

\paragraph{Spektrometr o rekordowej rozdzielczości.}

W rozdziale \ref{chap:Czasowo-cz=000119stotliwosciowa-transf} zaprezentowana
była modulacja podłużnej fazy fal spinowych w pamięci kwantowej typu
GEM, która zapamiętuje strukturę czasową zapisywanego sygnału. Zademonstrowany
został czasowy odpowiednik propagacji w przestrzeni oraz soczewki.
Posłużyły one do konstrukcji spektrometru o szerokości pasma rzędu
1MHz i rozdzielczości rzędu kilkudziesięciu kHz. Dzięki możliwości
nadrukowywania dowolnych profili fazowych, możliwe jest niemal dowolne
manipulowanie stanami światła zapisanymi w GEM. Stwarza to nowe możliwości
w czasowym i spektralnym przetwarzaniu wąskopasmowych kwantowych stanów
światła pochodzących z emisji atomowej.

\paragraph{Selektywny odczyt.}

W rozdziale \ref{chap:Sukcesywny-odczyt-konwerter-mod=0000F3} przedstawiona
została koncepcja konwertera modów, który umożliwia konwersję fal
spinowych przechowywanych w różnych modach przestrzennych pamięci
kwantowej na ciąg fotonów sprzęgniętych do światłowodu jednomodowego.
W tym celu została użyta wnęka optyczna zwiększająca wydajność odczytu,
gradient pola magnetycznego kompensujący niedopasowanie fazowe oraz
kontrolowanie kąta świecenia wiązki odczytującej na atomy, by wybrać
kierunek odczytu zgodny z modem wnęki. Dobrano zestaw realnych parametrów,
które pozwalają na odczyt wybranej fali spinowej z wysoką wydajnością
(92\%) przy jednoczesnym niewielkim niszczeniu pozostałych fal spinowych
(0,9\% - 1,7\%).

\section{Perspektywy}

Przetwarzanie informacji kwantowej w przestrzennych stopniach swobody
\cite{Pu2017,Parniak2017,mazelanik_coherent_2019,Seri2019} i w czasie/częstości
\cite{Humphreys2014,Brecht2015a,Reimer2016,Lu2019} stanowi ważny
aspekt współczesnych badań. Modulacja fazy przestrzennej przechowywanych
stanów daje w tym względzie nowe możliwości.

Po pierwsze otrzymujemy narzędzie pozwalające na wprowadzenie na żądanie
dodatkowego elementu optycznego o dowolnym profilu fazowym. Z jednej
strony umożliwia to skompensowanie aberracji, co pozytywnie wpływa
na zwiększenie liczby modów przestrzennych w pamięci, a także może
zwiększać sprzężenie odczytywanego sygnału z modem światłowodu jednomodowego
\cite{Lipka2019}. Z drugiej strony możliwa jest szybka zmiana bazy
pomiarowej, co jest potrzebne w protokołach komunikacji kwantowej
opartych o paradoks EPR \cite{Edgar2012,Aspden2013,Dabrowski2018}
i może być wykorzystane do pomiarów adaptatywnych \cite{Wiseman2009}.
Ponadto uzyskujemy zdolność do oddziaływania ze sobą pojedynczych
fal spinowych przechowywanych w różnych modach przestrzennych \cite{Parniak2018}.

Po drugie pełna kontrola podłużnej fazy stanów w pamięci GEM \cite{Sparkes2010,Chaneliere2015}
daje możliwość czasowego i spektralnego przetwarzaniu kwantowych stanów
światła pochodzących z wąskopasmowej emisji atomowej \cite{Zhao2014,Guo2017,Farrera2016},
który to obszar był wcześniej niedostępny.

Połączenie pamięci typu GEM z wnęką rezonansową, oraz aktywnie sterowanym
ustawieniem kąta padania wiązki sprzęgającej \cite{Cox2019} umożliwia
wykonanie selektywnego odczytu fal spinowych wytworzonych w procesie
spontanicznego rozpraszania Ramana w różnych modach przestrzennych.

Wreszcie pełna kontrola nad odczytem z wielomodowej przestrzenie pamięci
kwantowej w połączeniu z przestrzenną modulacją fazy \cite{mazelanik_coherent_2019}
otwiera nowe pola do tworzenia bramek kwantowych \cite{Parniak2018}
i przeprowadzania obliczeń kwantowych z użyciem kilku kubitów. Jest
to ważny krok, który w przyszłości umożliwi konstrukcję procesora
kwantowego.

\footnotesize\bibliographystyle{26C__Repozytoria_aldoktorat_osajnl}
\bibliography{25C__Repozytoria_aldoktorat_aldoktorat}

\end{document}